\documentclass[twocolumn]{aastex631}
\usepackage{framed}
\usepackage{multirow}
\usepackage{makecell}
\usepackage{amsmath, bm}
\usepackage{graphicx}
\usepackage{amssymb}
\usepackage{amsmath}
\usepackage{times}
\usepackage{subfigure}
\usepackage[T1]{fontenc}

\received{Jun 26, 2024}
\revised{July 25, 2024}
\accepted{July 29, 2024 }

\submitjournal{ApJ}

\shorttitle{Three types of galaxy alignment in simulations}
\shortauthors{Y. Lan et al.}

\begin{document}

\title{Mock Observations: Three Different Types of Galaxy Alignment in TNG100 Simulations}

\correspondingauthor{Weipeng Lin, Lin Tang}
\email{linweip5@mail.sysu.edu.cn, tanglin23@cwnu.edu.cn}

\author{Yanyao Lan}
\affiliation{School of Physics and Astronomy, Sun Yat-sen University, DaXue Road 2, 519082, Zhuhai, China}

\author[0000-0001-6395-2808]{Lin Tang}
\affiliation{School of Physics and Astronomy, China West Normal University, ShiDa Road 1, 637002, Nanchong, China}
\affiliation{CSST Science Center for the Guangdong-Hong Kong-Macau Greater Bay Area, DaXue Road 2, 519082, Zhuhai, China}

\author[0000-0003-2204-2474]{Weipeng Lin}
\affiliation{School of Physics and Astronomy, Sun Yat-sen University, DaXue Road 2, 519082, Zhuhai, China}
\affiliation{CSST Science Center for the Guangdong-Hong Kong-Macau Greater Bay Area, DaXue Road 2, 519082, Zhuhai, China}

\author{Junyu Gong}
\affiliation{School of Physics and Astronomy, Sun Yat-sen University, DaXue Road 2, 519082, Zhuhai, China}


\begin{abstract}

In this study, galaxy samples have been generated using mock observation techniques based on the results of TNG100-1 simulations to investigate three forms of intrinsic alignment: satellite-central alignment between the orientation of the brightest group galaxies (BGG) and the spatial distribution of their satellites, radial alignment between the satellites' orientation and the direction toward their BGG, as well as direct alignment between the orientation of BGG and that of its satellites. Overall, the predictions of galaxy alignment generally align with observations, although minor discrepancies have been identified.
For satellite-central alignment, the alignment strength and color-dependence trends are well replicated by the mock observations. Regarding radial alignment, the signals are weak but discernible, with no apparent color dependence.  As for direct alignment, no signal is detected, nor is there any color dependence. We also investigate the alignment dependencies on halo or the BGG properties, and proximity effect. For satellite-central alignment, the predicted alignment signal shows a positive correlation with halo and BGG mass, consistent with observations and previous predictions. Similar correlations have also been observed with the BGG age and metallicity, which merit future observational analysis for confirmation. Proximity effects have been observed for all three types of alignment, with satellites closer to the BGG exhibiting stronger alignment signals. The influence of galaxy definition and shape determination on alignment studies is also analyzed. This study underscores the importance of employing mock observation techniques for a fair comparison between predictions and observations.

\end{abstract}


\section{Introduction} \label{sec:intro}
The hierarchical clustering model of galaxy formation, as proposed by \cite{white1978core} and further developed by subsequent studies \citep[e.g.,][]{White1991, Lacey&Cole1993, Ghigna2000, dolag2009}, suggests that galaxies form within dark matter halos through a process of hierarchical merging. This model posits that smaller halos merge to form larger ones, with the dense cores of smaller halos persisting as substructures within the larger halos, contributing to the complexity of galaxy assembly.
The assembly histories of dark matter structures play a significant role in determining the properties of galaxies, such as their color, size, metallicity, angular momentum, satellite number density, and spatial distribution. Understanding the connection between the assembly history of dark matter structures and the properties of galaxies is crucial for unraveling the mysteries of galaxy formation and evolution \citep[e.g.,][]{Shen2003, yang2005galaxy,  Yang2006, mo2010galaxy, Taylor2015, Wang2018b, Tang2020, Wang2021}.
Among these, the spatial distribution of galaxies can provide important clues to understanding the merging history of dark matter structures \citep[e.g.,][]{Faltenbacher2009, Li2013, Wang2020, Zhang2021} and the formation of galaxies \citep[e.g.,][]{Faltenbacher2007, Agustsson&Brainerd2010, Debattista2015}.

Galaxy alignment is a critical tool for characterizing the anisotropy distribution of galaxies and has been the subject of extensive research, with notable studies providing in-depth reviews on the topic \cite[see, e.g.,][for review]{Joachimi2015, Kiessling2015}.
Based on previous classifications \citep[see][]{Faltenbacher2007}, there are primarily three distinct types of galaxy alignment observed within groups\footnote{Here, groups actually include galaxy groups and clusters, introduced from the FoF clustering analysis.}: satellite-central alignment\footnote{We will not adhere to the original term ``halo alignment'' as used in \cite{Faltenbacher2007}, but will refer to it as ``satellite-central alignment'', as there is a difference in the orientation of the halo and the BGG.} between the orientation of the brightest group galaxies (BGG) and the distribution of its satellite galaxies, radial alignment between the orientation of a satellite galaxy and the direction toward its BGG, and direct alignment between the orientation of the BGG and that of its satellites.

The investigation of satellite-central alignment has been a topic of interest for several decades. In a historical context, \cite{Sastry1968} found that there is a strong tendency for the major axis of the galaxy distribution to be oriented along the major axis of cD galaxies,  while \cite{Holmberg1969} reported that the satellites of local galaxies distribute peculiarly to the Galaxy disk.
Since then, numerous studies have shown robust evidence of satellite-central alignment using large samples of galaxy surveys \citep[e.g.,][]{Sales&Lambas2004, Brainerd2005, Agustsson2006b, Weinmann2006, Yang2006, Faltenbacher2007,  Wang2018a, Rodriguez2022} and simulations \citep[e.g.,][]{Kang2007, Faltenbacher2008, Bailin2008, Agustsson&Brainerd2010, Bahl2014, Dong2014, Velliscig2015, Zhang2019, Tang2020, Tenneti2021}.
It has indeed been demonstrated that satellite galaxies often align with the major axes of their central galaxies, and the strength of this alignment is affected by the color of the galaxies, with red galaxies showing stronger alignment signals compared to blue galaxies \citep[e.g.,][]{Yang2006, Faltenbacher2007, Wang2018a}.
Additionally, studies have revealed that the strength of satellite-central alignment also depends on the radial distance of satellite galaxies from the BGG. Specifically, galaxies located in the inner regions of clusters exhibit stronger alignment with the major axis of their BGG compared to galaxies in the outer regions \citep[e.g.,][]{Faltenbacher2007, Dong2014}.

While the phenomenon of satellite-central alignment has been established through various studies, there are discrepancies between theoretical predictions and observational findings. For instance, simulations often predict stronger satellite-central alignment compared to what is observed in actual data. Additionally, the predicted relationship between satellite-central alignment and galaxy color has not been accurately replicated in observations \citep[e.g.,][]{Kang2007,Wang2014}.
So far, most of the studies have focused on predicting the 3D alignment of halos rather than considering the 2D alignment that is relevant for observational comparisons. Additionally, lots of studies have defined friends-of-friends (FoF) groups as individual ``galaxies'', which may not accurately represent the real galaxies in observations. These methodological limitations can contribute to the discrepancies observed between theoretical predictions and actual measurements of satellite-central alignment in galaxy clusters.
Therefore, mock observations are highly demanded and have been practiced. For instance, the study by \cite{Tang2020} provides an example of how mock observations can help improve the alignment signal in galaxy clusters by constructing a mock galaxy sample from a Gadget-2 hydrodynamical simulation \citep{Jing2006ApJ, Lin2006ApJ}, \cite{Tang2020} were able to more accurately capture the distribution and properties of galaxies within clusters compared to previous studies that defined ``galaxies'' as FoF stellar groups, such as \cite{Dong2014}.
However, as the galaxy colors in the simulations used by \cite{Tang2020} do not follow those of observations,  the predicted color dependence of satellite-central alignment can not fit with the observed one.
Though it is indicated that a fair comparison between simulation predictions and observations must be made and state-of-the-art simulations are needed to better understand the nature of satellite-central alignment.

The initial observation of radial alignment was made  by \cite{Hawley&Peebles1975}, highlighting the tendency for galaxies to be aligned in a radial manner within their host halos.
Approximately 30 yr later, using the galaxy catalog from the Sloan Digital Sky Survey (SDSS) data, \cite{Agustsson2006a} found significant evidence for radial alignment on scales of $70\ h^{-1}$ kpc, confirmed by \cite{Wang2019} using the SDSS DR13 data.
In a study by \cite{Faltenbacher2007} a radial alignment signal was observed on distance scales of  $r<0.7R_{\rm{vir}}$ for red satellites within galaxy clusters. However, no significant radial alignment signal was found for blue satellites in the same study, suggesting a potential color dependence.
Radial alignment signal has also been detected in other surveys. For example, \cite{Pereira&Kuhn2005} found radial alignment in a sample of 85 X-ray-selected clusters, while \cite{Schneider2013} confirmed the existence of radial alignment signal by analyzing a group catalog from the GAMA survey.
Several studies have successfully reproduced the observed signal of radial alignment using numerical simulations \citep[e.g.,][]{Knebe2008, Pereira2008, Rong2015, Knebe2020, Tenneti2021}.
For instance, by using a cosmological $N$-body simulation, \cite{Pereira2008}  found a strong tendency for substructures to align radially with their host halos and
suggested that tidal torque is the most plausible cause of the formation of radial alignment.
Also, using a cosmological $N$-body simulation, \cite{Knebe2008} found that, in the innermost region of galaxy clusters, the observed signal of radial alignment can be reproduced and independent of substructure mass.
Lately, \cite{Knebe2020} represented the dependence of radial alignment signal on the distance to the center of the galaxy cluster and the dynamical state of galaxies by using hydrodynamical simulations.

The phenomenon of direct alignment has just been revealed in this century. \cite{Plionis2003} analyzed a sample of 303 Abell clusters to show evidence of direct alignment in high-density environments, and found that the alignment signal is dependent on the velocity dispersion of the clusters.
\cite{Ferrari2006} specifically reported that Abell 521 exhibited considerable red-sequence galaxy direct alignment out to about $\sim 2 \ h^{-1}$ Mpc from the cluster core.
However, subsequent studies have found that the direct alignment signal on the outer regions of clusters is absent. For example, \cite{Agustsson2006a} showed that the signal of direct alignment is weak but noticeable on the innermost scales of clusters, while no signal is observed on large scales \cite[see also][]{Mandelbaum2006}.
\cite{Faltenbacher2007} analyzed SDSS DR4 data and found a faint but significant indication for direct alignment on scales $\rm r<0.1R_{\rm{vir}}$ for red satellites,
and subsequently verified by a $\Lambda$CDM N-body simulation \citep{Faltenbacher2008}.
So far, the evidence for direct alignment and its dependence on galaxy properties are still controversial and need further investigation.

In this study, in order to make an appropriate comparison between simulation predictions and observations, we utilize a mock galaxy sample constructed from the state-of-the-art TNG100-1 simulations
to quantify signals for the three types of alignment and explore their dependence on galaxy or halo properties.
The paper is organized as follows.
Section 2 introduces the methodology, including the simulation data,  mock sample construction, and definition of alignment.
The main results will be presented in Section 3.
Conclusions and a discussion will be given in Section 4.

\section{Methodology}\label{Methodology}
\subsection{simulations}\label{simulation}
The TNG100-1 simulations utilized here are provided by the IllustrisTNG Project \footnote{https://www.tng-project.org} and described in \cite{pillepich2018simulating}, \cite{springel2018first}, and \cite{weinberger2016simulating}, while the public data release has been described in \cite{Nelson2019}.
These simulations incorporate a comprehensive model of gas physics and star formation, aiming to realize structure evolution from redshift  $z = $ 127  to 0.
The simulated cubic periodic box has a side length of $75\ h^{-1}$ Mpc, and contains $1820^3$ dark matter particles and the same number of gas particles, such that
each dark matter and gas particle has a mass of $7.5\times 10^6 M_{\odot}$ and $1.4\times10^6 M_{\odot}$, respectively.
A force softening length of $0.74\ h^{-1}$ kpc is adopted.
The FoF algorithm \citep{Davis1985} is used to identify dark matter halos
and the SUBFIND algorithm \citep{Springel2001} is employed to define substructures.
The cosmological model for the simulations is the concordance ${\Lambda}$CDM model, with parameters adopted from Planck \citep{Planck2016},
where $\Omega_{\Lambda}=0.6911$,  $\Omega_m=0.3089$,  $\sigma_8=0.8159$, $h =0.6774$.

\subsection{construction of mock galaxy samples}\label{galaxy samples}
The mock galaxy samples utilized in this study are sourced from \cite{Tang2021},
and the galaxy definition is based on the surface brightness limit segmentation procedure (SBLBP) algorithm, as detailed in \cite{Tang2020,Tang2021}.
The SBLBP algorithm comprises three main steps.
 Initially, the FoF groups of particles are projected onto $x$-$y$, $y$-$z$, and $z$-$x$ planes to produce raw images, assuming a pixel size similar to that of the detector and convolved with a Moffat point spread function (PSF) with parameters consistent with the typical seeing of the SDSS survey, as described by \cite{Moffat1969}.
Subsequently, potential `galaxy' candidates are identified based on a series of surface brightness thresholds that differentiate them from the background and diffuse stellar components.
Finally, the third step involves reconstructing the galaxy samples following the procedure outlined in \cite{Tang2020} and elaborated in \cite{Tang2021}.
In essence, to replicate observations of galaxies, the SBLBP algorithm considers several factors, including the PSF, CCD pixel size, surface brightness threshold, redshift dimming, cosmological geometry effects, and dust extinction\citep{Baes2011}.

The utilization of these mock galaxy samples has enabled the successful reproduction of observed galaxy color bimodality, the galaxy star-forming rate - stellar mass relationship, and the galaxy size dependence on stellar mass \citep{Tang2021}. Furthermore, the study by \cite{Tang2021} also predicted the galaxy's stellar mass function and luminosity function, showcasing good agreement with observational data from studies such as {\cite{Yang2009} and \cite{Bernardi2013}}, particularly at the massive or bright end of the galaxy population.
This is in contrast to results derived from substructures by setting an upper limit of galaxy size with an arbitrary radius of $30\ h^{-1}$ kpc, as discussed by \cite{Pillepich2018}.
Overall, these findings underscore the reliability of the mock observation techniques and highlight their potential for application in studying galaxy alignment phenomena.

The specific selection criteria for the galaxy sample used in this study of alignment are as follows:
\begin{enumerate}

\item Surface brightness threshold: To compare with galaxy alignment study using SDSS data \citep[e.g.,][]{Yang2006},  a surface brightness threshold of $25.0 \ \rm mag \ arcsec^{-2}$ in the $r$ band has been adopted to define galaxies in the simulation.

\item Redshifts: The galaxy groups are generated from snapshots for three redshifts ($z \sim$  0.01, 0.1 and 0.2), to mimic SDSS galaxy redshift distribution.

\item Galaxy groups: Only galaxy groups with a virial mass, $M_{200}$, greater than $10^{12}\ h^{-1} M_{\odot}$ within a radius of 200 times the mean density of the Universe are included.

\item Central and satellite galaxies: Only centrals and satellites with stellar mass greater than $1.4\times10^{8}\ h^{-1} M_{\odot}$ (more than 200 particles) and semi-major axes longer than $2.96\ h^{-1}$ kpc (4 times the softening length) are included in the samples, to alleviate resolution effect. Here, the central galaxy is defined as the BGG.

\item Apparent magnitude: The galaxies of interest are limited to have an apparent magnitude, $m_{r}$, brighter than 17.77 mag in the SDSS$-r$ band \citep[see][]{Yang2006}.
\end{enumerate}

Applying the above selection criteria, we extracted in total 12,987 `satellite-BGG' pairs from the mock galaxy samples. The sample of these pairs will be referred to ``mock sample''  in the following studies.
To compare with previous studies, we also analyze the spatial distribution of  substructures defined by the SUBFIND algorithm applying the same criteria.


\subsection{definition of galaxy alignment}\label{definition of galaxy alignment}
\begin{figure}[ht!]
\centering
\includegraphics[width=7 cm]{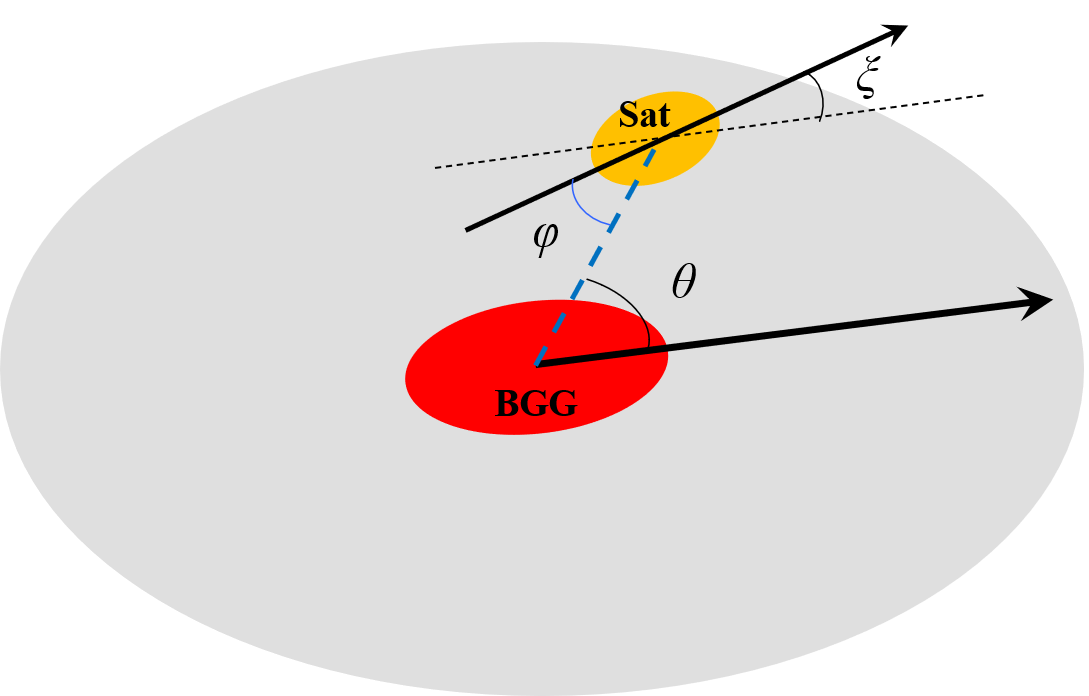}
\caption{The illustration of three types of alignment and the representing angles between the BGGs and satellites (or substructures). The solid lines with arrows represent the major axes of the satellite galaxy (yellow) and BGG (red). The long-dashed line connects the centers of the satellite and the BGG, while the dotted line is parallel to the major axis of the BGG. Here, $\theta$, $\varphi$ and $\xi$ denote the satellite-central, radial, and direct alignments, respectively.}
\label{figure 3angle}
\end{figure}

Figure \ref{figure 3angle} illustrates three types of satellite alignment related to their host BGGs,
where the solid lines with arrows represent the major axes of the satellite galaxy (yellow) and BGG (red).
To calculate the shape of the galaxy or substructure and determine the major axis, two methods will be applied.
In ``method-{mock}'',  the ellipticity of the mock galaxy is determined by analyzing its image and fitting ellipses to the isophotes.
This method leverages the brightness contours to extract the major axis direction of the galaxy, providing valuable information on its shape and orientation.
On the other hand, ``method-{sub}'' focuses on the 2D shape of substructures by calculating the moments of inertia of the stellar mass distribution. By computing the principal axes of the inertia tensor, the major axis direction of the substructure can be identified. This approach offers insights into the intrinsic structure and alignment of substructures within the galaxy system.
The inertia tensor, $I_{i,j}=\sum_{k=1}^{N} x_{i,k}x_{j,k}m_{k}/\sum_{k}m_{k}$, where $x_{i,k}$ and $x_{j,k}$ are the distance of $k$th particle from the subhalo center in $i$ and  $j$-dimension respectively, and $m_{k}$ is the mass of $k$th particle. Here, $N$ is the particle number of the stellar substructures.
By summing over the particle masses and distances, the inertia tensor provides a quantitative measure of the mass distribution and enables the determination of the major axis direction.

In Figure \ref{figure 3angle}, BGG is placed on the coordinate origin,
and $\theta$, $\varphi$ and $\xi$ are indicators for satellite-central, radial, and direct alignment, respectively.
This is also illustrated in Figure 1 of \cite{Faltenbacher2007} and \cite{Kiessling2015}.
In the context of satellite-central alignment, a significant alignment signal is indicated by a distribution of small angles or an average angle $\langle \theta \rangle< 45^{\circ}$.
This suggests that satellite galaxies tend to be preferentially distributed along the major axis of the central galaxy. The strength of the alignment signal is quantified by the average angle, with smaller average values indicating a stronger alignment signal.
Similar concepts apply to radial and direct alignment signals, which are characterized by the average angles $\langle \varphi \rangle$ and $\langle \xi\rangle$, respectively.
In the absence of any alignment, the average angle would be $45^\circ$.
However, it is important to note that an average angle of $45^{\circ}$ does not necessarily imply an isotropic distribution of satellite galaxies around central galaxies.

\section{Results}
In this section, results for the three types of alignment and their dependence on galaxy or halo properties, including galaxy color, halo mass, stellar mass, age, and metallicity, as well as the distance from the BGG, will be given.
To quantify the strength of the alignment signal, a normalized number density distribution is utilized.
The normalized number density distribution provides a measure of the relative density of aligned galaxies, taking into account variations in galaxy number counts.
It is computed by dividing the number of galaxy pairs within a specific angle range by the same number in a random sample, as follows:
\begin{equation}
\centering
P(\text{angle})=\frac{N(\text{angle})}{\langle N_r(\text{angle})\rangle} \label{cm}
\end{equation}
where $N$(angle) is the count of satellite galaxies in angle bins, while $\langle N_r(\text{angle})\rangle$ is the average count of random samples in the same bins.
This calculation considers the area or volume covered by each measurement, ensuring that the distribution accounts for the spatial coverage.
$P(\text{angle}) = 1$ means the isotropic distribution, while $P(\text{angle})>1$ at any angle implies a distribution with preferred alignment.

\subsection{alignment for the whole samples}

\begin{figure*}[htp!]
\centering
\includegraphics[width=5.9 cm]{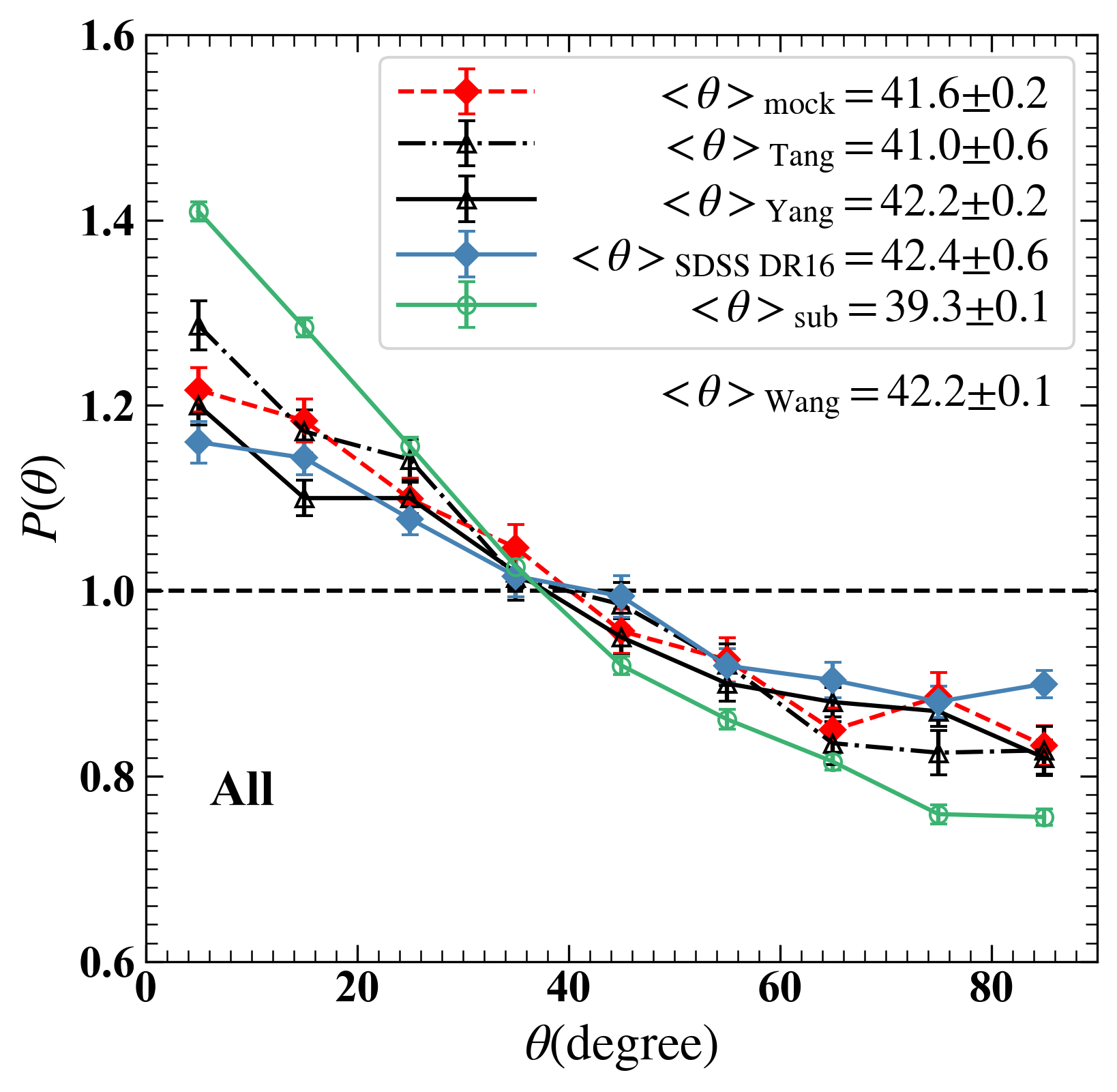}
\includegraphics[width=5.9 cm]{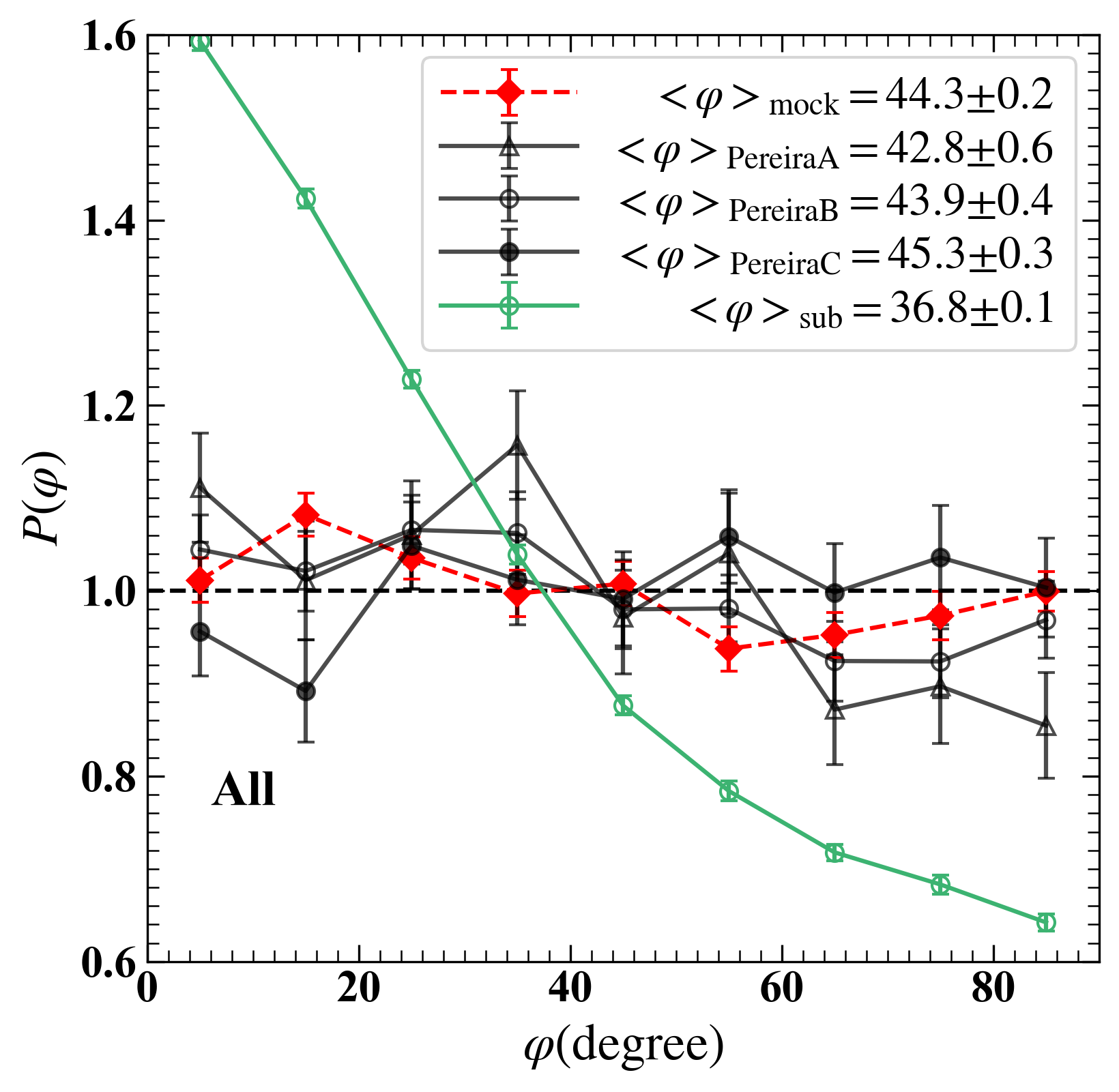}
\includegraphics[width=5.9 cm]{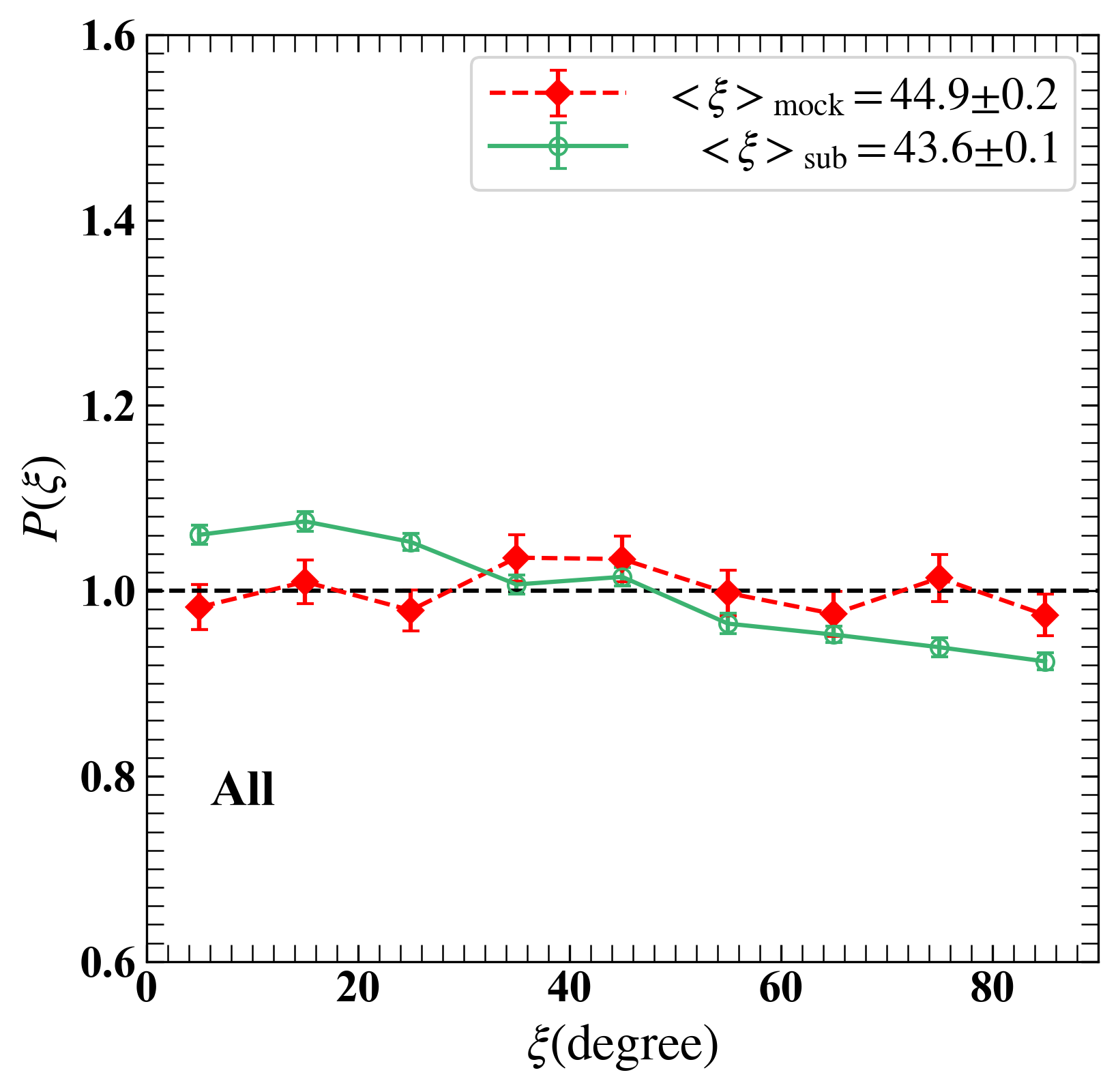}
\caption{Results for the three types of alignment, i.e., satellite-central, radial, and direct alignment, are plotted in the left, middle, and right panels, respectively.
In all panels, the red long-dashed line with filled diamonds represents the results for the mock galaxy sample from the TNG100-1 simulation, while the green solid lines with open circles represents the results for stellar substructures.
The error bars are computed by $\sigma_r(\text{angle})/N_r(\text{angle})$, where $\sigma_r(\text{angle})$ is the standard deviation of $N_r(\text{angle})$ obtained from the 100 random samples.
The legends in each panel provide average values for $\langle \theta \rangle$, $\langle \varphi \rangle$, and $\langle \xi \rangle$ with 1 $\sigma$ standard deviation,
while the subscript ``mock'' and ``sub'' refer to results derived by ``method-{mock}'' and ``method-{sub}'' respectively.
Left panel: results for satellite-central alignment.  The dotted-dashed line with an open triangle, the solid line with an open triangle, and the blue solid line with a filled diamond, represent results for the mock galaxies from Gadget-2 simulation \citep{Tang2020}, observational results from SDSS DR2 \citep{Yang2006}, and SDSS DR16 \citep{Rodriguez2022}, respectively. 
In this plot, the result derived from SDSS DR7 (Wang et al. 2018) is also listed. 
Middle panel: results for radial alignment. The other three solid lines with triangle, open circle, and filled circle are for observational results from three different samples  \citep{Pereira&Kuhn2005}. Note that the result denoted as ``PereiraC'' represents an isotropic orientation of satellite galaxies for a control sample, with a null alignment signal.
Right panel: results for direct alignment.
}
\label{figure theta_All_Ms}
\end{figure*}

The predictions of satellite-central, radial, and direct alignment from the mock sample and stellar substructures, are plotted as the red long-dashed and green solid lines  in Figure \ref{figure theta_All_Ms}.
Observational results from SDSS data \citep{Pereira&Kuhn2005, Yang2006, Wang2018a, Rodriguez2022}, and the prediction from Gadget-2 simulation \citep{Tang2020}, are also shown for comparison.

In Figure \ref{figure theta_All_Ms}, it is evident that the stellar substructures exhibit prominent alignment signals across all three types, with a particularly notable strength in radial alignment. These predicted alignment signals surpass those observed in real data and in predictions from the mock sample, thus confirming and reinforcing earlier findings \citep{Tang2020}.

The left panel of Figure \ref{figure theta_All_Ms} illustrates results for satellite-central alignment. The normalized probability distribution of satellite-central alignment angles for the mock sample, represented by the red long-dashed line with filled diamonds, closely matches the observations with only a small difference. The average angle for the mock sample (${\langle \theta \rangle}_{\rm{mock}}=41^\circ.6\pm 0^\circ.2$) is marginally larger than a previous prediction of $41^\circ.0\pm 0.6^\circ$ \citep{Tang2020}, and is approaching the observational results, specifically $42^\circ.2\pm 0^\circ.2$ \citep{Yang2006} and $42^\circ.4\pm 0^\circ.6$ \citep{Rodriguez2022}.

The middle panel of Figure \ref{figure theta_All_Ms} displays the findings for radial alignment. The red long-dashed line indicates the results for the mock sample, while the observational results for various SDSS galaxy samples \citep{Pereira&Kuhn2005} are depicted by three solid black lines with distinct symbols. The plot reveals that our predictions show a modest yet noticeable radial alignment, with an average angle of ${\langle \varphi \rangle}_{\rm{mock}} =44^\circ.3 \pm 0^\circ.2$. There are tendencies toward excess or deficiency in small or large angles, respectively, broadly consistent with the observational outcomes for galaxies in clusters \citep{Pereira&Kuhn2005}, namely $42^\circ.8 \pm 0^\circ.6$ for a spectroscopic sample (``Sample A'') and $43^\circ.9 \pm 0^\circ.4$ for a photometric sample (``Sample B'').

In the right panel of Figure \ref{figure theta_All_Ms}, the outcomes for direct alignment are represented. As there are no observational data points available for this plot, we solely rely on the results from the mock sample. The findings suggest an almost isotropic distribution between the orientation of the BGG and those of satellites. Specifically, we find that ${\langle \xi \rangle}_{\rm{mock}} =44^\circ.9\pm0^\circ.2$, with $P(\xi) \sim 1$.
This means that there is little direct alignment between the central and satellite galaxies in our mock sample, indicating a random distribution of orientations.

\subsection{dependence on galaxy or halo properties}
\label{section properties}
In further analysis, we delve deeper into exploring the relationship between galaxy alignment and various properties of mock galaxies or halos, including galaxy color, halo mass, BGG stellar mass/age/metallicity, as well as the radial distance of satellites toward the BGG.
By examining how galaxy alignment may vary with these diverse characteristics, we aim to uncover the underlying mechanisms that drive galaxy alignments.

\subsubsection{galaxy color}\label{color}

The galaxy color is defined as $^{0.1}(g-r)$, which represents the $(g-r)$ color k-corrected to a redshift of $z=0.1$.
This color is calculated using the formula $^{0.1}(g-r) = 0.7188+1.3197[(g-r)-0.6102]$, as detailed in \citet{blanton2007k}.
For the star particles within the galaxies, their magnitudes are computed in each band using the simple stellar population model by \cite{bruzual2003stellar}.

In this approach, we will divide the mock galaxy sample into red and blue categories based on the $^{0.1}(g-r)$ color criterion of 0.7.
This choice is motivated by the need to accurately classify galaxies in the high-mass range, as the fitting line used in previous studies \citep[e.g.,][]{Tang2020} may misclassify a significant fraction of high-mass mock galaxies as blue.
The decision to use $0.7$ instead of $0.83$, the value adopted in \citep{Yang2006}, is supported by the color bimodality observed in TNG100 simulations. The distribution of galaxy colors in TNG100 simulations, as shown in Figure 3 of \cite{Nelson2017}, reveals a dip between the two peaks of the color distribution around $\sim 0.6$ in $(g-r)$ color, which corresponds to approximately $^{0.1}(g-r) \sim 0.7$.
Furthermore, the stellar mass-color map derived from our mock sample (refer to the left panel of Figure 3 in \cite{Tang2021}) demonstrates that the horizontal line at $^{0.1}(g-r) = 0.7$ effectively separates the red and blue clouds in our simulated data. This criterion provides a clear distinction between red and blue galaxies in our sample, aligning with the color classification used in previous studies by \cite{Yang2006} and \cite{Wang2018a}, which serve as the main comparative targets for our analysis.
Other color thresholds, such as $^{0.1}(g-r)$ values of 0.73 and 0.76, were tested, but they did not significantly impact the alignment predictions. Additionally, galaxies with $\log_{10}{(M/h^{-2}M_\odot)}>11.5$ were classified as red galaxies to align with the SDSS survey's distribution of spiral galaxies (Xiaohu Yang, private communication). This classification did not notably affect the predictions.

The mock sample, consisting of satellite-BGG pairs, was then divided into four color-based subgroups: red satellite-BGG pairs, blue satellite-BGG pairs, red BGG-satellite pairs, and blue BGG-satellite pairs, labeled as ``Red Satellites,'' ``Blue Satellites,'' ``Red Centrals,'' and ``Blue Centrals,'' respectively. The number distribution of these pairs is presented in Table \ref{table1}. These subgroups are referred to as single-color subgroups, as they are based on the colors of either centrals or satellites. Similarly, four subgroups of substructure-BGG pairs were also created using the same method for comparison.

\begin{figure*}[htp!]
\centering
\includegraphics[width=4.2 cm]{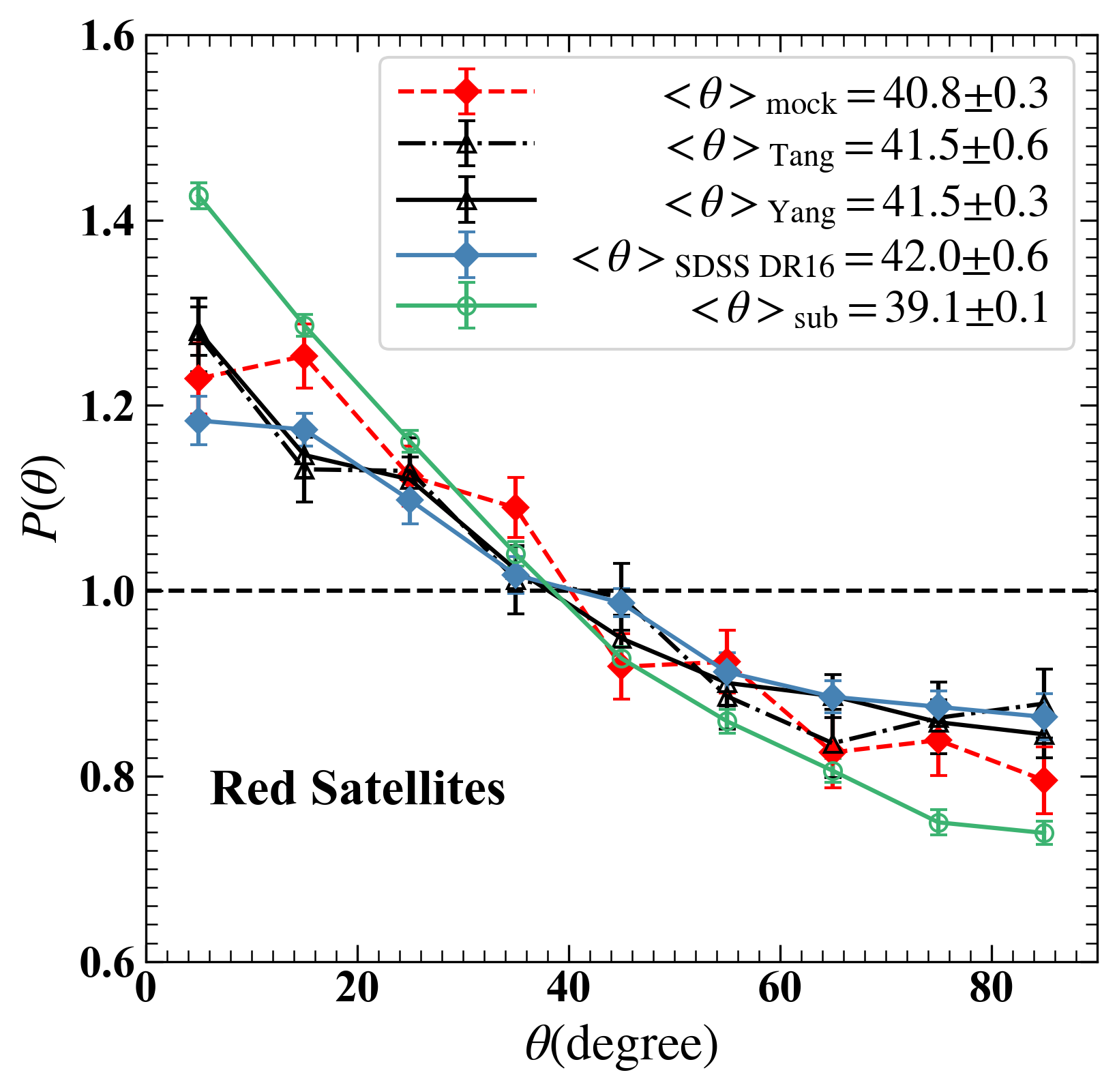}
\includegraphics[width=4.2 cm]{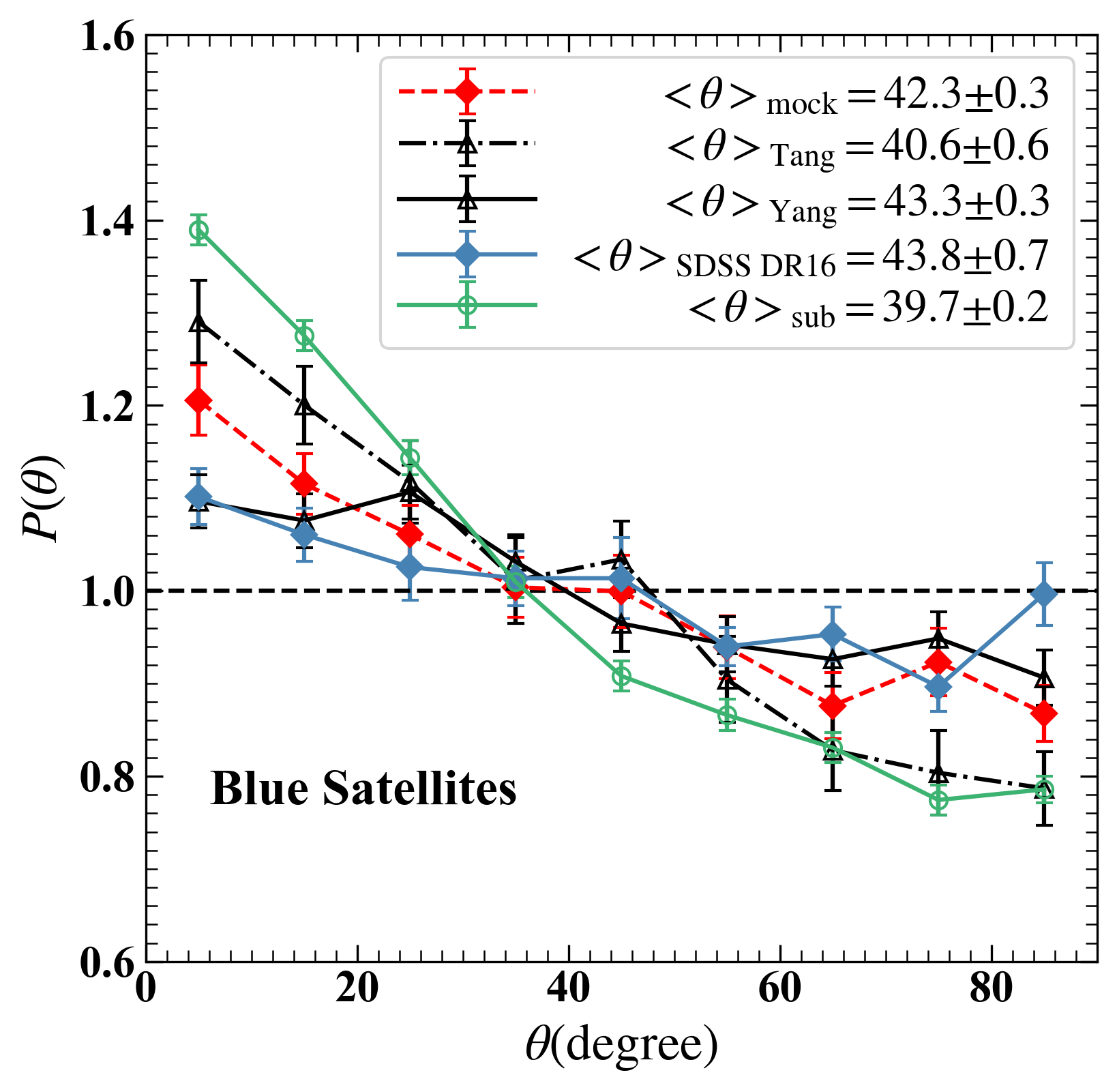}
\includegraphics[width=4.2 cm]{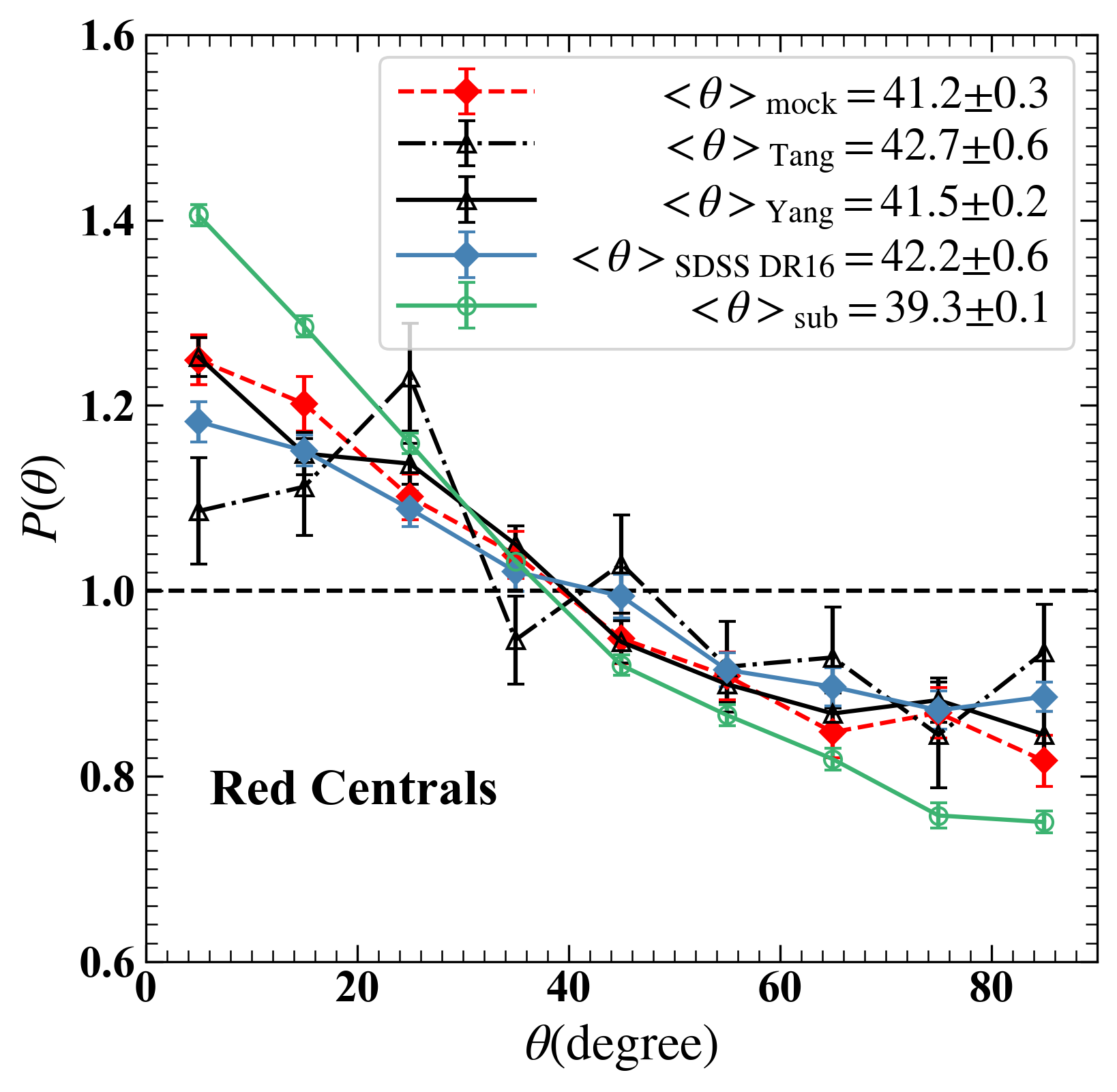}
\includegraphics[width=4.2 cm]{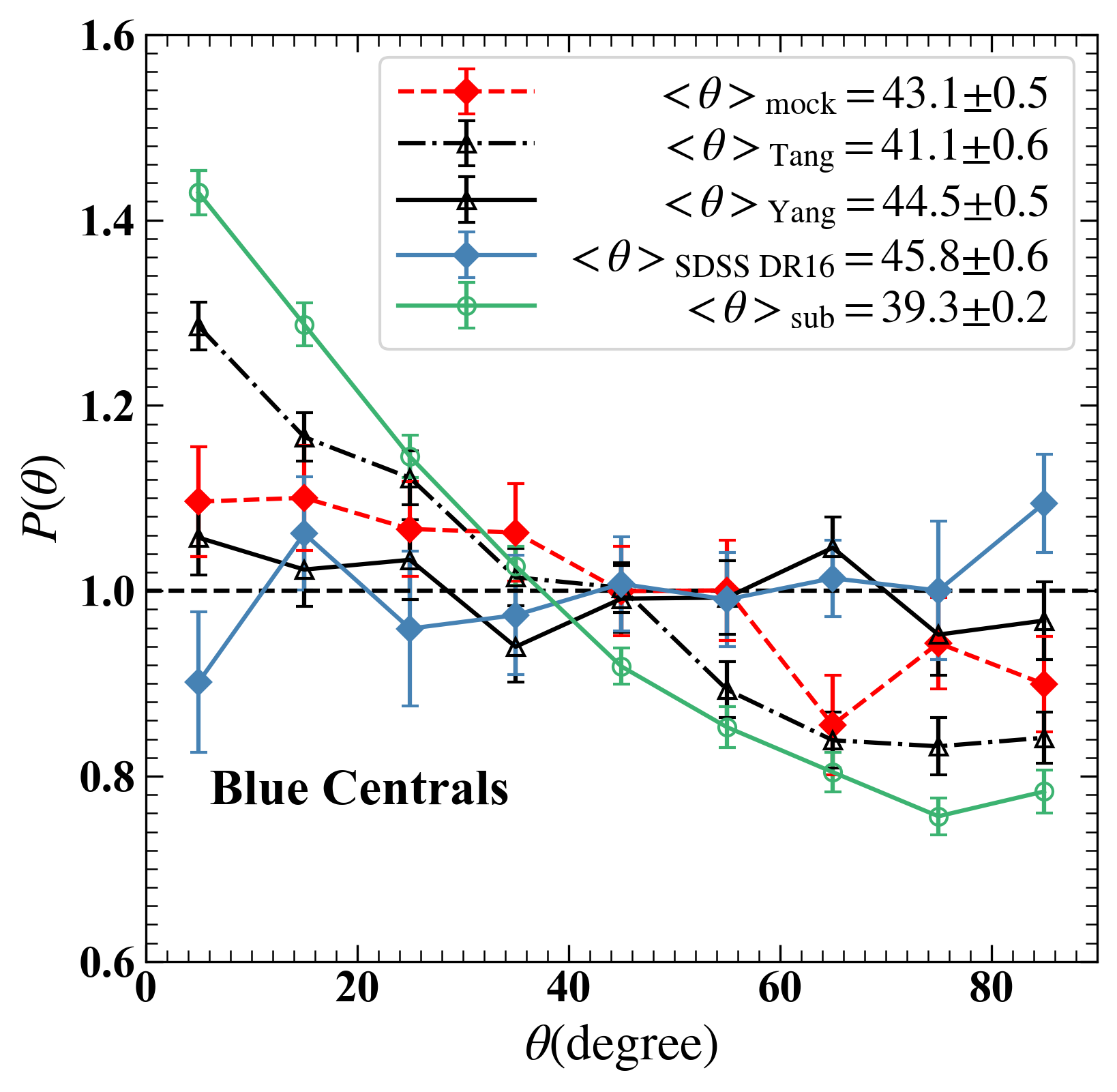}
\includegraphics[width=4.2 cm]{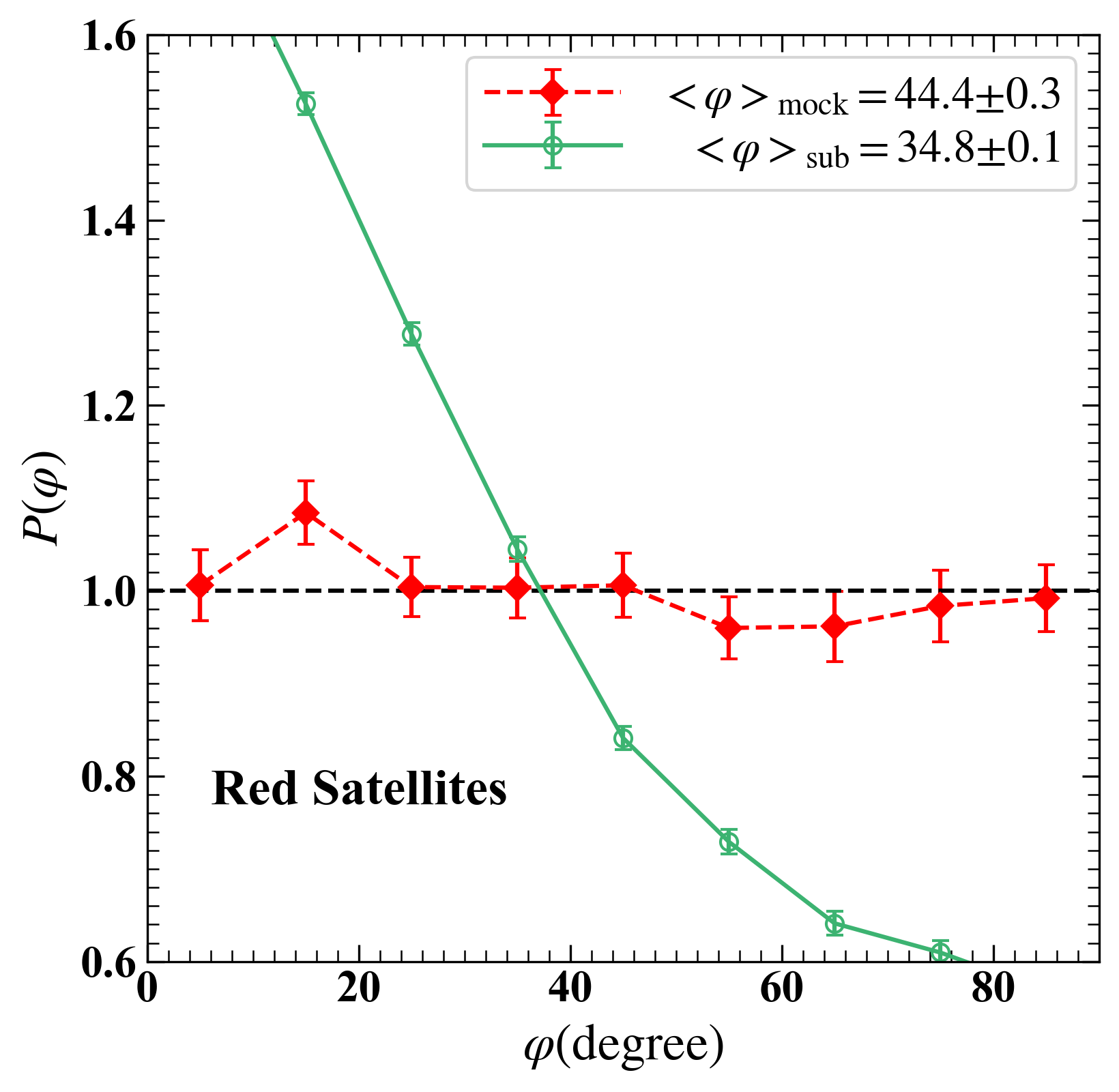}
\includegraphics[width=4.2 cm]{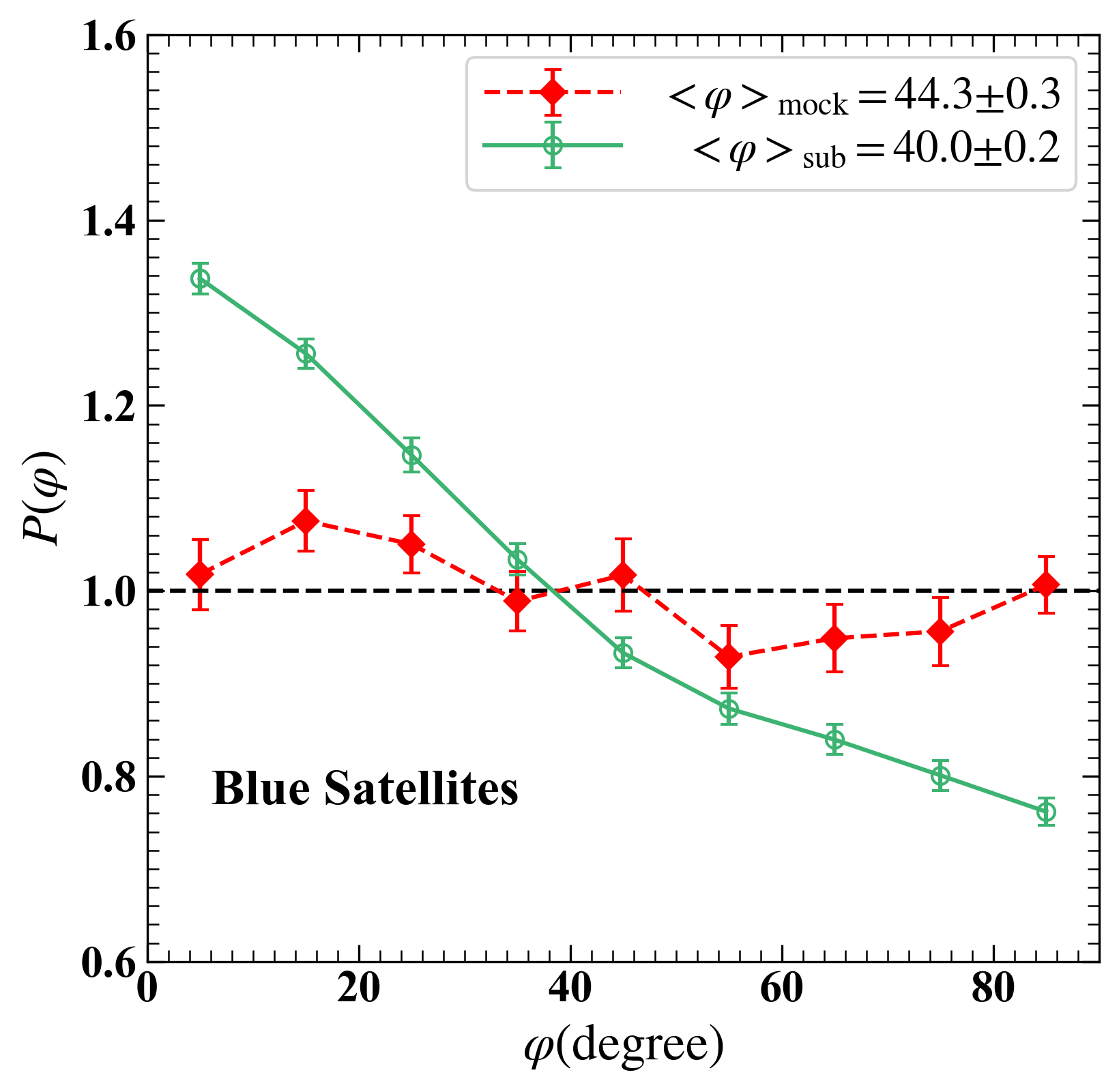}
\includegraphics[width=4.2 cm]{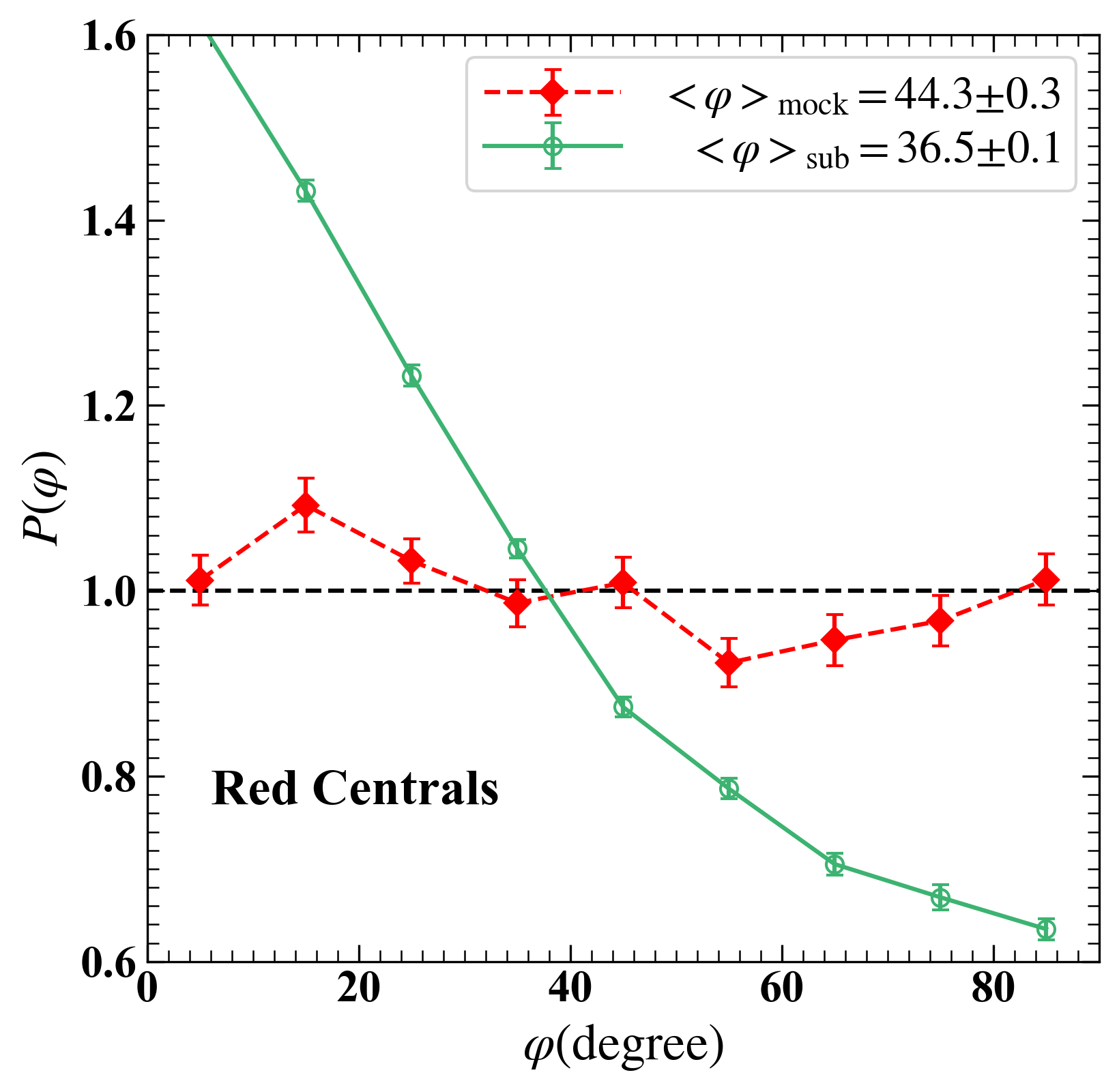}
\includegraphics[width=4.2 cm]{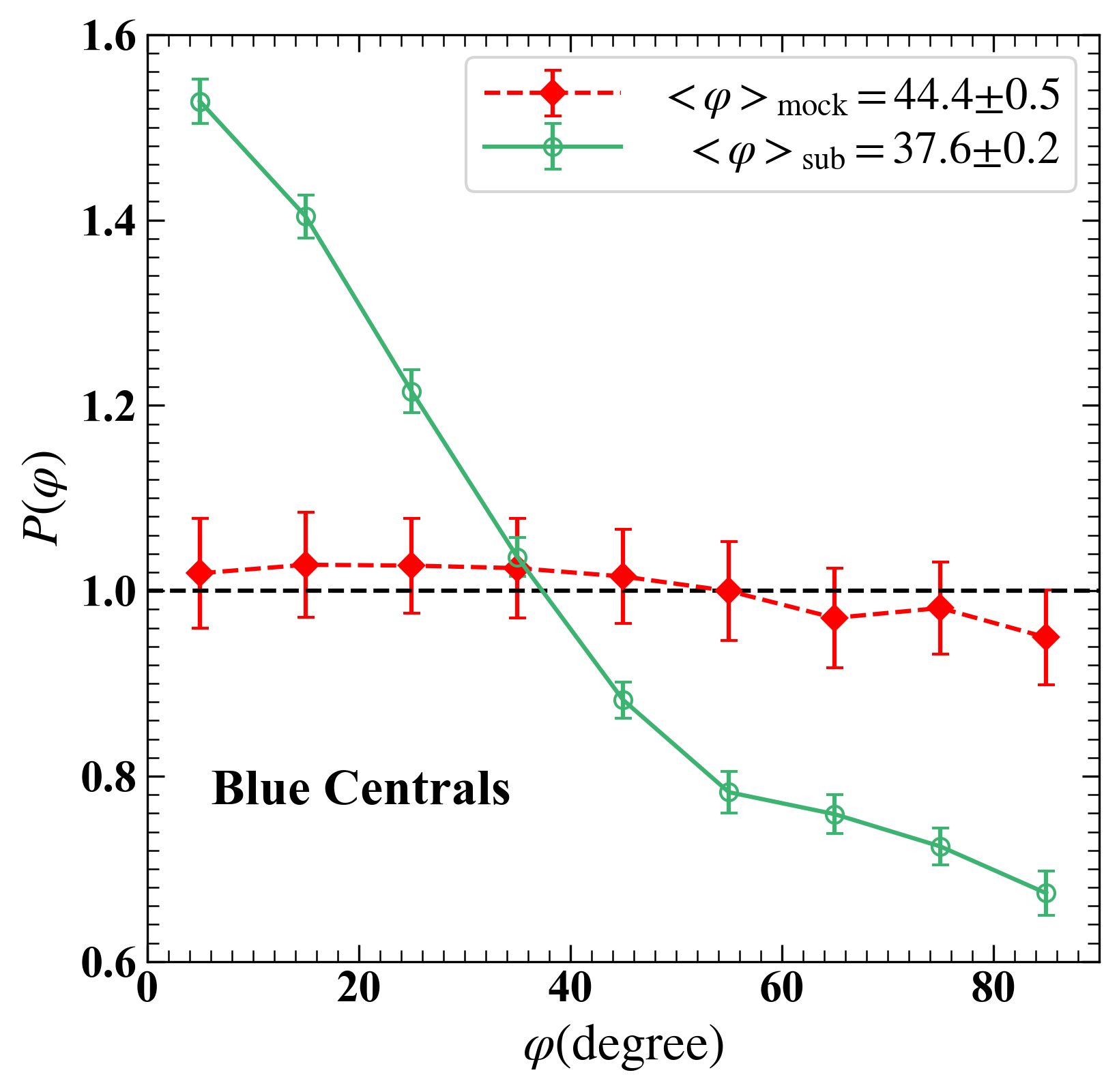}
\includegraphics[width=4.2 cm]{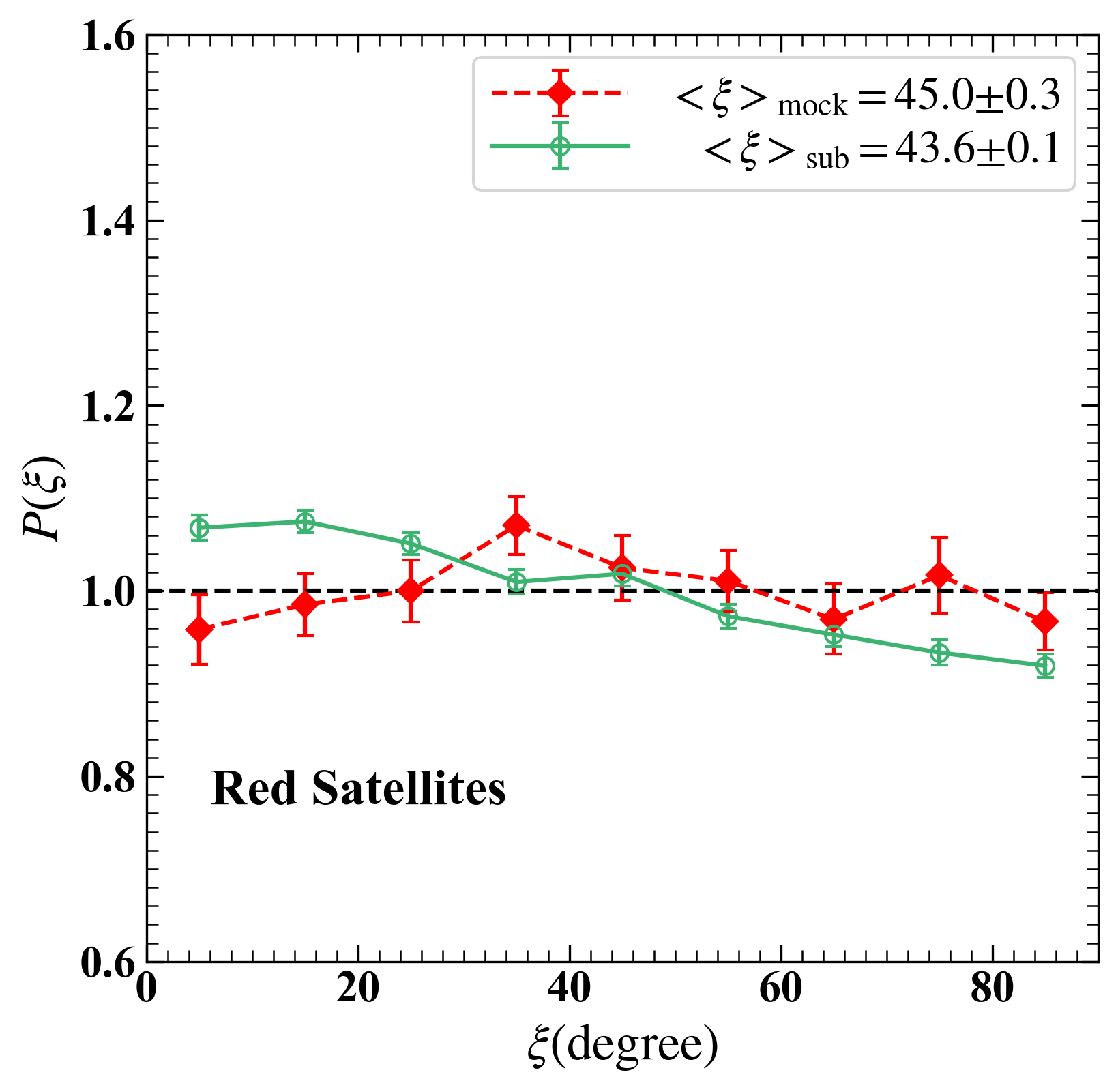}
\includegraphics[width=4.2 cm]{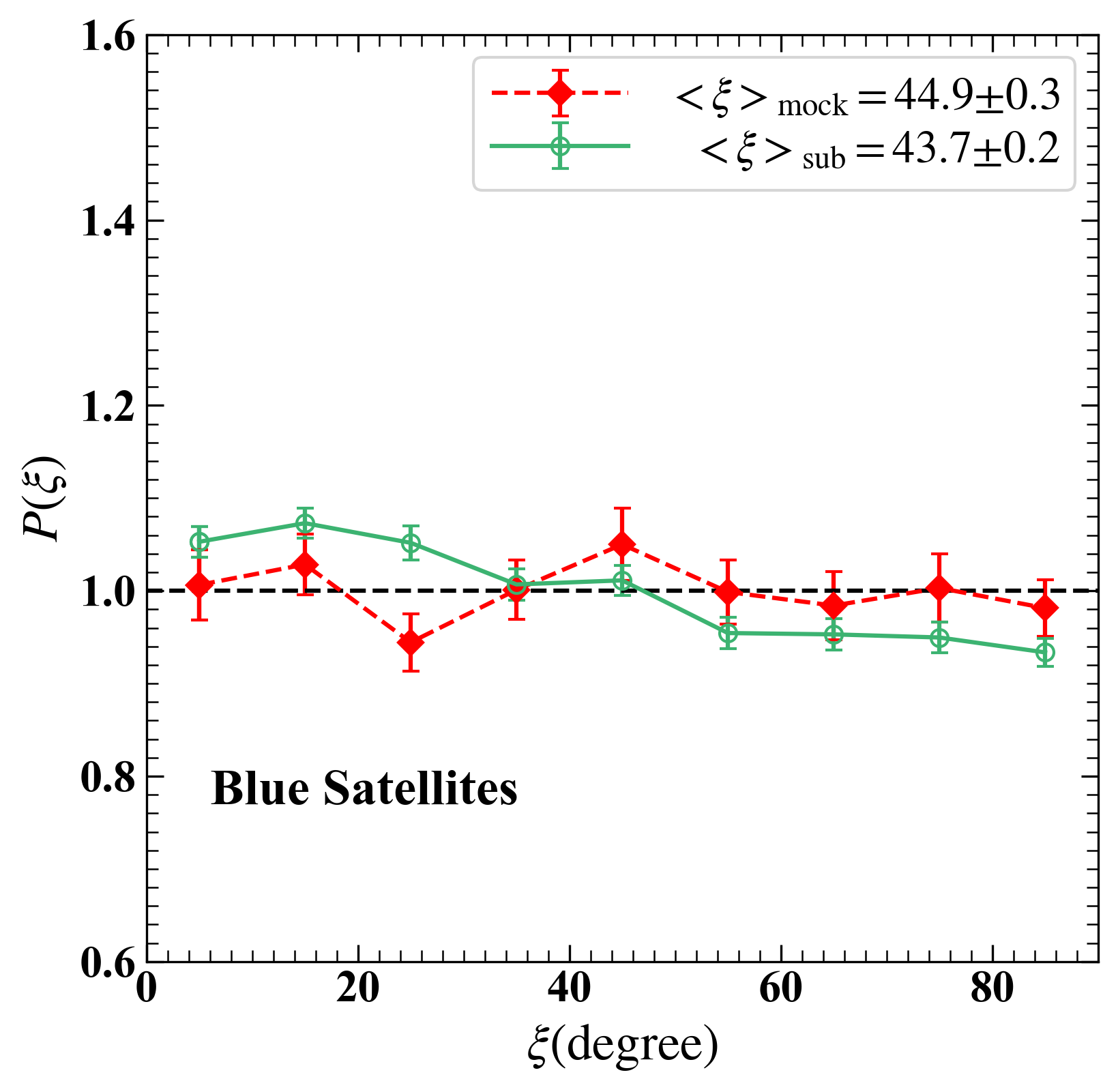}
\includegraphics[width=4.2 cm]{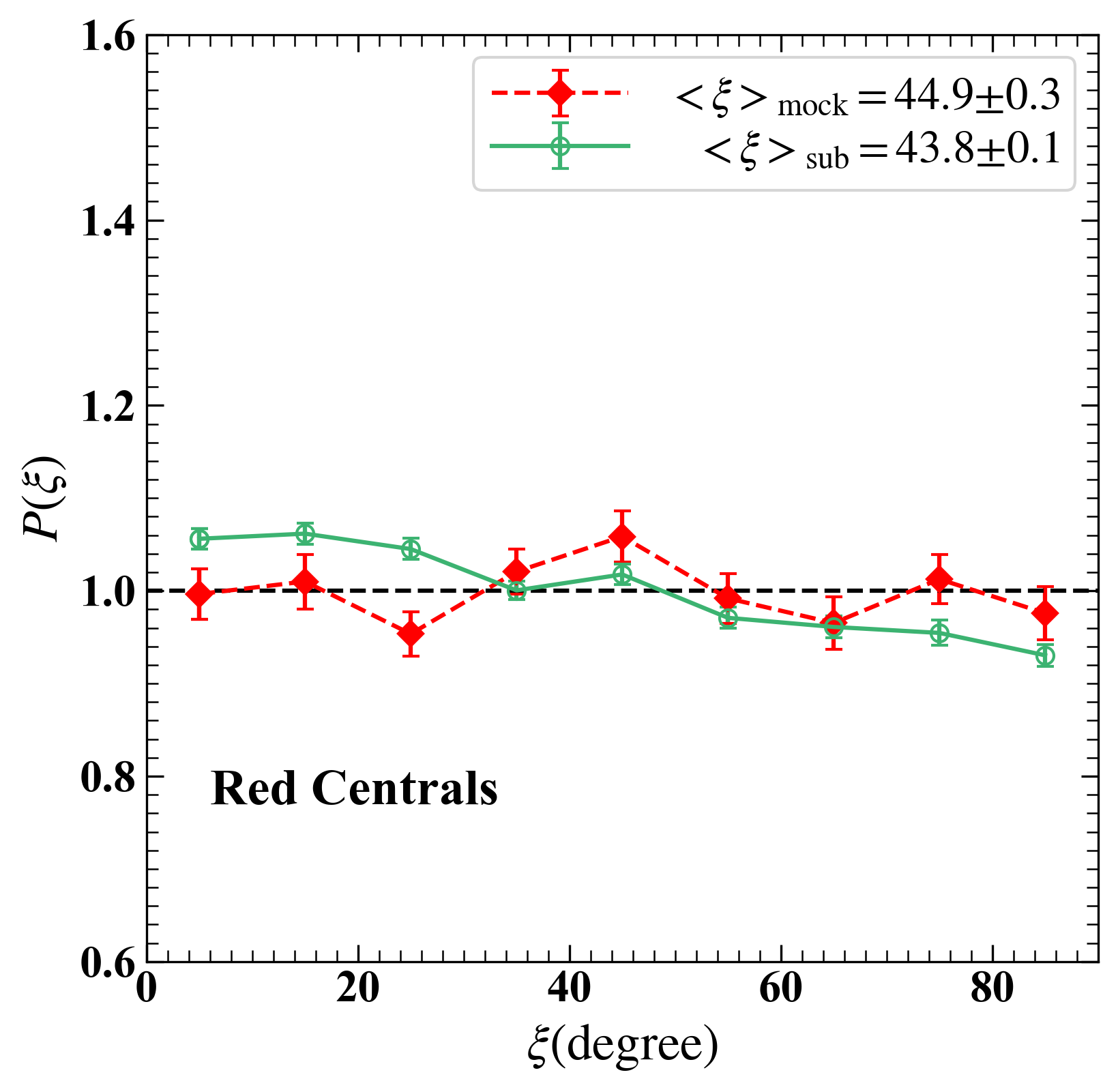}
\includegraphics[width=4.2 cm]{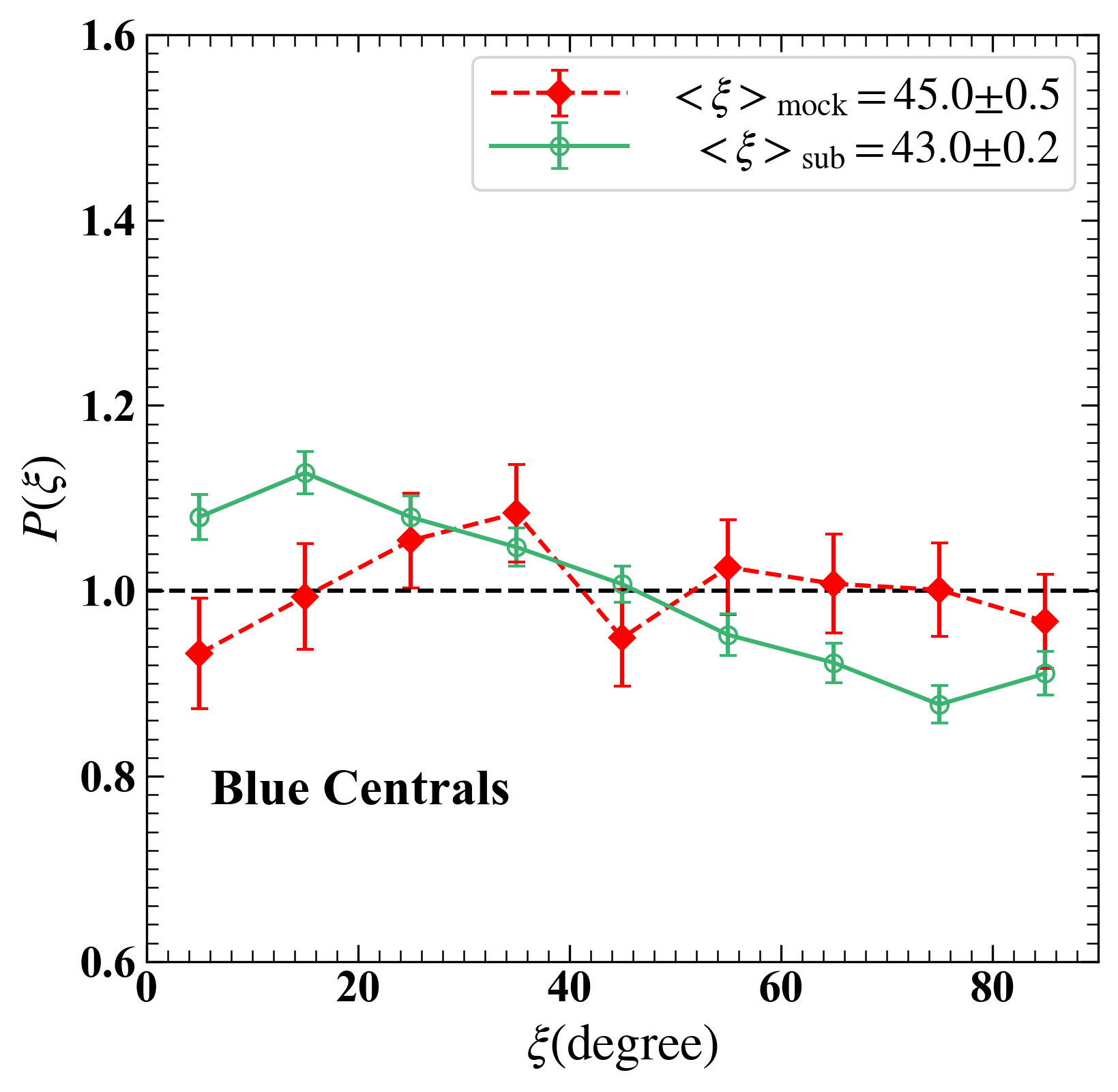}
\caption{Same as Figure \ref{figure theta_All_Ms}, but for subsamples with different colors.
The panels in the rows from top to bottom are for satellite-central, radial, and direct alignment respectively, while the panels in the column from left to right are for ``Red Satellites,'' ``Blue Satellites,'' ``Red Centrals,'' and ``Blue Centrals'' subsample, respectively.}
\label{figure color subsample}
\end{figure*}

The results for color dependence are displayed in Figure \ref{figure color subsample}. The panels are organized from top to bottom for satellite-central alignment, radial alignment, and direct alignment, respectively. The columns from left to right represent the four subsamples, as denoted in the lower corner of each panel. The symbols, lines, and annotations used in the figure are consistent with those in Figure \ref{figure theta_All_Ms}. Upon initial inspection, it is evident that the alignment signals for substructures (represented by the green solid lines) are notably stronger than those for the single-color subgroups, akin to the trends observed in Figure \ref{figure theta_All_Ms}.

As indicated in the legends of the top panels, our predicted average satellite-central alignment angles for ``Red Satellites'' and ``Red Centrals'' ($40^\circ.8 \pm 0^\circ.3/41^\circ.2 \pm 0^\circ.3$) are a bit smaller than the values derived from observational data, specifically $41^\circ.5 \pm 0^\circ.3/41^\circ.5 \pm 0^\circ.2$ \citep{Yang2006} and $42^\circ.0\pm 0^\circ.6/42^\circ.2 \pm 0^\circ.6$ \citep{Rodriguez2022}.
However, the lines representing predictions and observations for red subsamples are nearly indistinguishable, indicating a high level of agreement.
On the other hand, the predicted average angles for blue subsamples, particularly ``Blue Centrals,'' remain notably smaller compared to the observed values, with our prediction of $43^\circ.1 \pm 0^\circ.5$, contrasting with the observed values of $44^\circ.5\pm 0^\circ.5$ \citep{Yang2006} and $45^\circ.8 \pm 0^\circ.6$ \citep{Rodriguez2022}.
The contrasting trends in color dependence predictions between our current study and the previous work by \cite{Tang2020} are indeed intriguing.
The present predicted satellite-central alignment strengths for `Red Satellites', `Red Centrals', `Blue Satellites', and `Blue Centrals' exhibit a clear trend from strong to weak, which aligns well with the observed color dependence trend reported in \cite{Yang2006} and \cite{Rodriguez2022}.
This is the first time that the observed color dependence of satellite-central alignment has been recovered properly by predictions.
This consistency further validates our predictions and highlights the importance of using advanced algorithms and simulations for studying galaxy satellite-central alignment.

The middle panels of the analysis show that the radial alignment signals for all mock subsamples are indeed weak but noticeable. There is a tendency for excessive or deficient alignment at small or large angles, with average angles of $44^\circ.4\pm 0^\circ.3/44^\circ.3\pm 0^\circ.3$ for the ``Red Satellites'' and ``Red Centrals'' subsamples and $44^\circ.3\pm 0v.3$/$44^\circ.4\pm 0^\circ.5$ for the ``Blue Satellites'' and ``Blue Centrals'' subsamples, respectively.
These results indicate that there is a consistent but subtle radial alignment signal present across different subsamples, regardless of galaxy color.
The obvious variations in the average angles for different color subsamples of substructures implies suggest that there might be some differences in the alignment patterns between red and blue counterparts, although the overall alignment signals for mock subsamples remain weak.

The analysis of direct alignment in the bottom panels reveals no distinct signal predicted from any of the single-color subgroups. This is evidenced by the almost flat curves in the normalized probability distribution, with $P(\xi)$ remaining approximately constant at 1, and the average alignment angle $\langle\xi\rangle$ hovering around $45^{\circ}$. However, the color subsamples for substructures have shown significant direct alignment signals.

\begin{figure*}[ht!]
\centering
\includegraphics[width=5.9 cm]{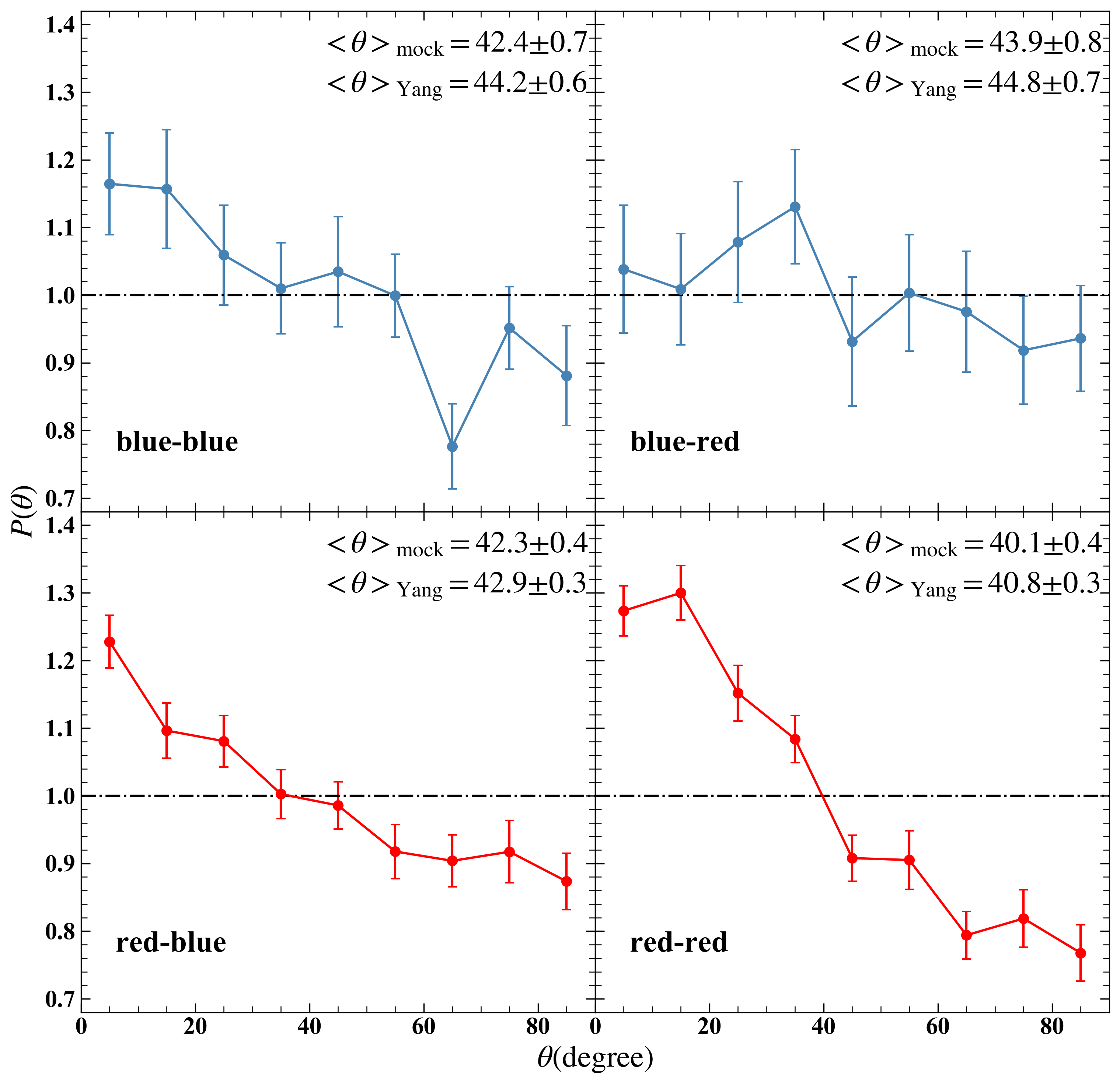}
\includegraphics[width=5.9 cm]{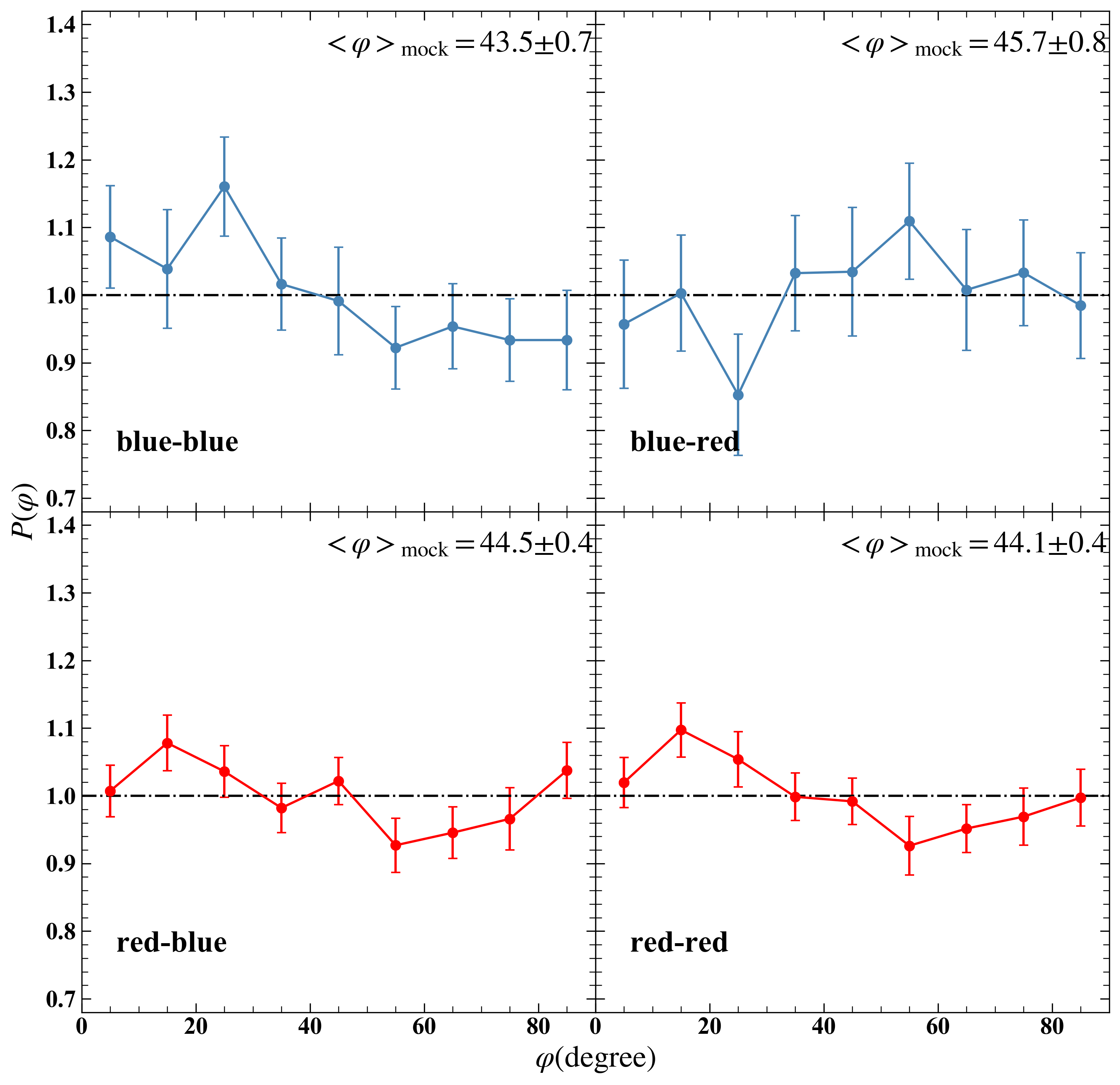}
\includegraphics[width=5.9 cm]{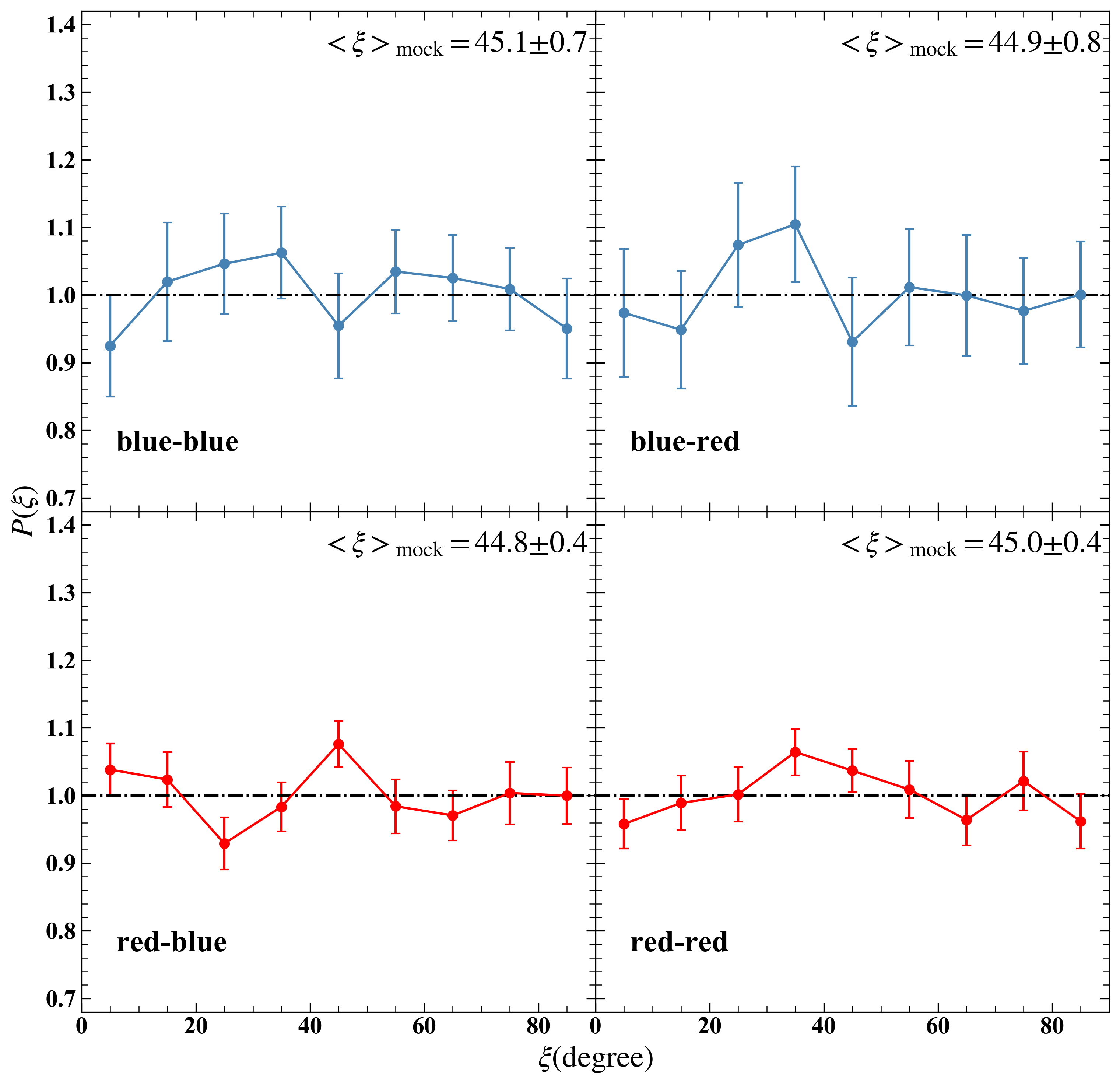}
\caption{Same as Figure \ref{figure theta_All_Ms}, except that here the mock sample has been split further  according to the colors of both the central and the satellite galaxies, as indicated in the low-left corner of each panel. The panels in the column from left to right represent satellite-central, radial, and direct alignment, respectively.}
\label{figure color subsubsample}
\end{figure*}

The analysis conducted by \cite{Yang2006} highlights the importance of considering the color properties of both central and satellite galaxies within a halo, particularly due to the observed
``galactic-conformity'' phenomenon where the color of satellites tends to be correlated with the color of the central galaxy in the same halo \citep[e.g.,][]{Weinmann2006}.
To further explore this color dependence, the mock sample can be divided into double-color subsamples based on the colors of both central and satellite galaxies.
For example, a subsample with red central galaxies and red satellites would be referred to as the red-red subsample, following the convention of naming subsamples based on the colors of centrals and satellites. By creating these double-color subsamples, it becomes possible to investigate how the color properties of both centrals and satellites within a halo influence galaxy alignments and orientations.
The number of pairs for these double-color subsamples can be found in Table  \ref{table1}.

The results presented in the left two columns of panels in Figure \ref{figure color subsubsample} show a comparison between the predictions for satellite-central alignment from different double-color subsamples and the observational results from Figure 4 in \cite{Yang2006} (refer to the average angles labeled ``Yang''). Interestingly, the predictions from the red-red, red-blue, and blue-red subsamples exhibit a good agreement with the observational data.
The prediction from the blue-red subsample (${\langle \theta \rangle}_{mock}=43^\circ.9\pm 0^\circ.8$) closely matches the observed value of $44^\circ.8\pm 0^\circ.7$.
This is surprising, considering the apparent inconsistency illustrated in the top-right panel of Figure \ref{figure color subsample} for ``Blue Centrals''.
On the other hand, for the blue-blue subsample, the predicted alignment angle (${\langle \theta \rangle}_{mock}=42^\circ.4\pm 0^\circ.7$) deviates significantly from the observational result ($44^\circ.2\pm 0^\circ.6$).
These results indicate that the discrepancy between our predictions and observations is minimal and is mainly caused by the excessive anisotropic distribution of blue satellites around blue central galaxies in the simulations, which is the same reason for the too strong alignment signal found in \citet{Tang2021}.

Using the same subsamples, we also predict color dependence for radial and direct alignment, with results shown in the other four columns of the panels. For radial alignment, only the blue-blue subsamples exhibit strong alignment signals, while the other three double-color subsamples show no significant alignment. In contrast, direct alignment shows almost no color dependence, as illustrated in the right panel of Figure \ref{figure color subsubsample}.

Finally, it should be noted that predictions from the double-color subsamples can be used to infer those from single-color subsamples.

\begin{deluxetable*}{ccccc}
\centering
\tablenum{1}
\tablecaption{Number of galaxy pairs for color subsamples\label{table1}}
\tablewidth{0pt}
\tablehead{
\colhead{Type of subsamples} & \colhead{Red Satellites} & \colhead{Blue Satellites} & \colhead{Red Centrals} & \colhead{Blue Centrals}
}
\startdata
Number of pairs  & 6243 & 6744 & 10,338 & 2649 \\
\toprule
Type of subsamples & red-red & red-blue & blue-red & blue-blue \\
Number of pairs  & 5130 & 5208 & 1113 & 1536 \\
\enddata
\label{table}
\end{deluxetable*}


\subsubsection{halo mass and BGG mass}
\label{section_mass}

\begin{figure*}[ht!]
\centering
\includegraphics[width=5.5 cm]{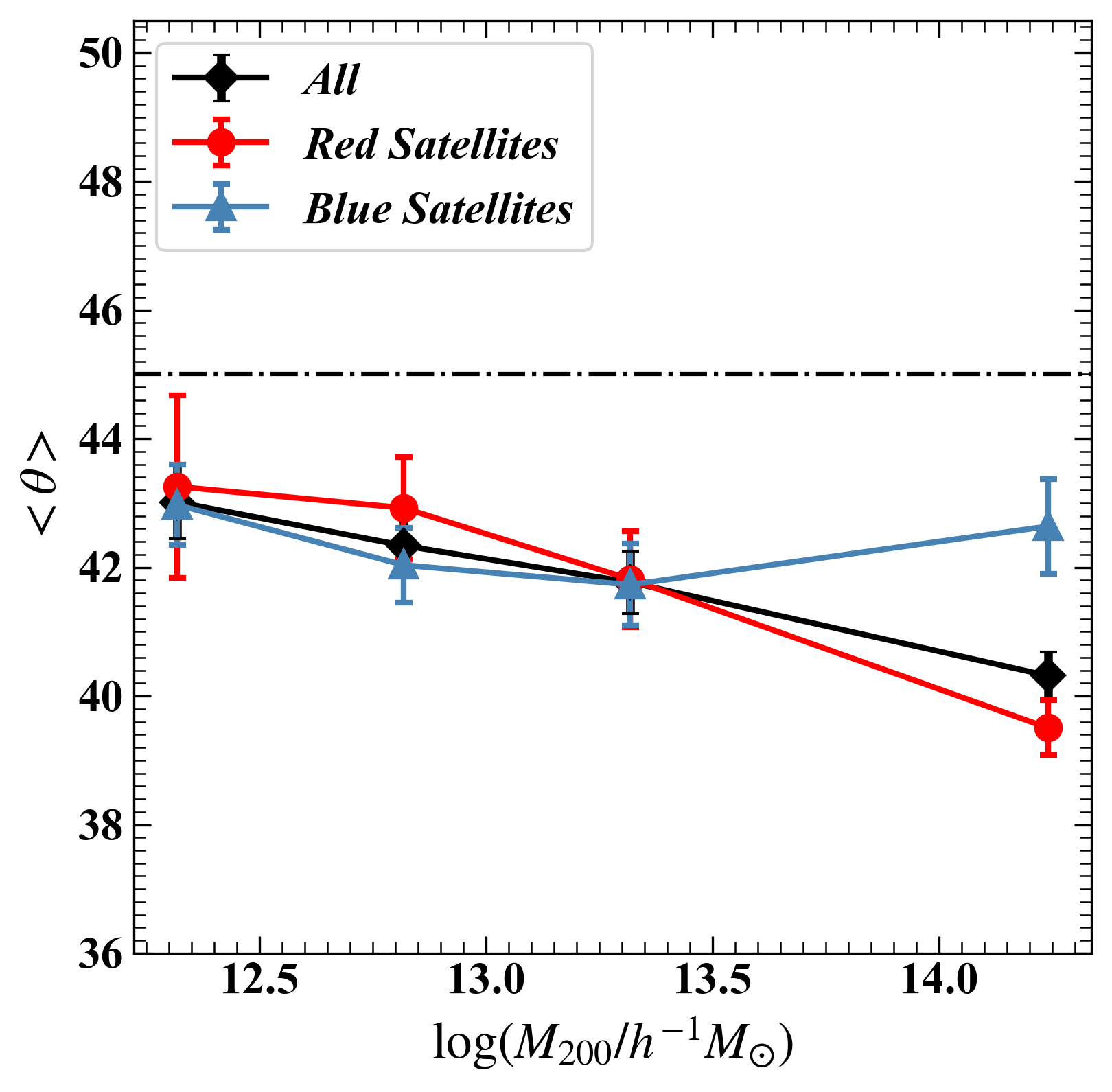}  
\includegraphics[width=5.5 cm]{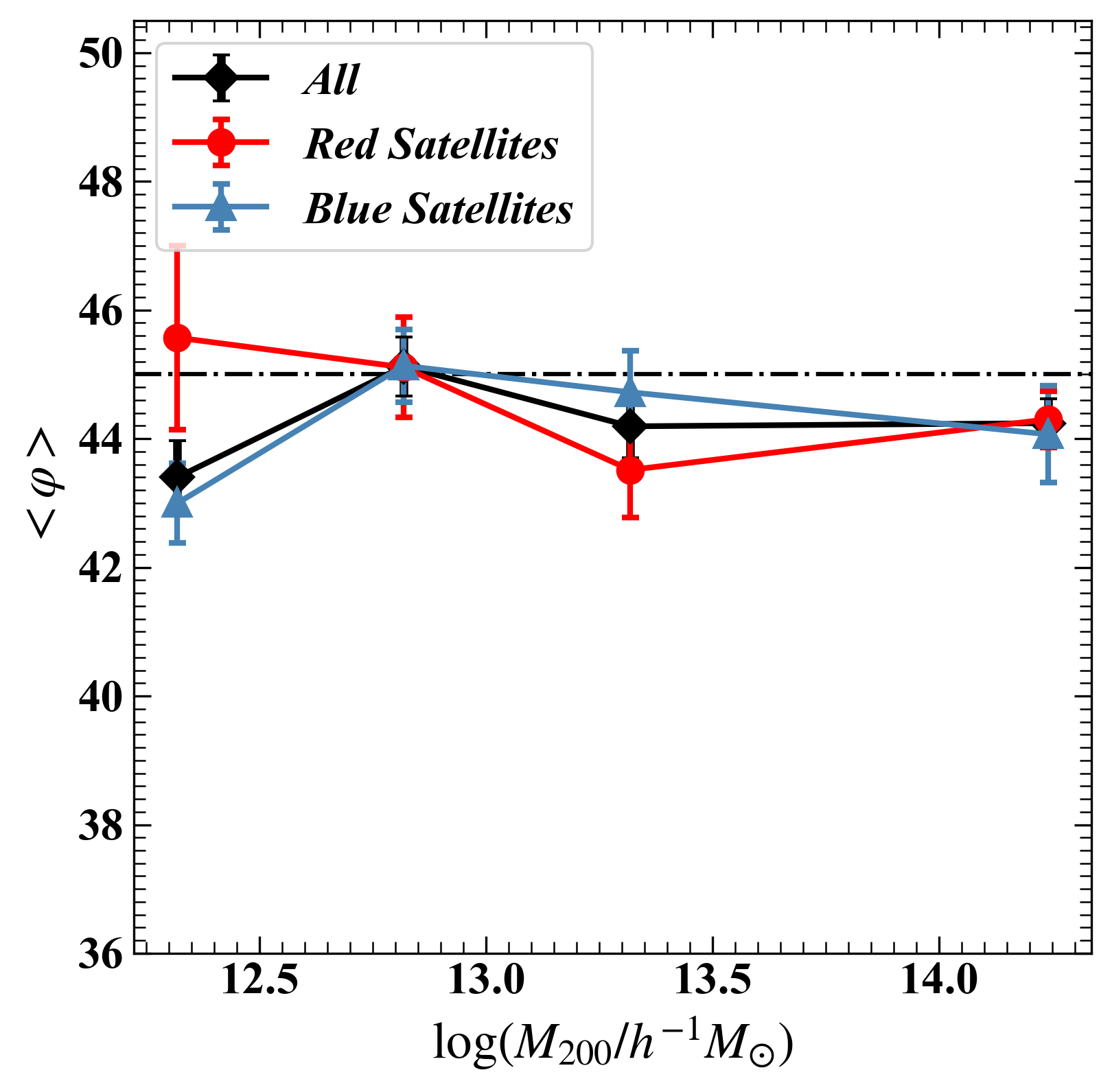}
\includegraphics[width=5.5 cm]{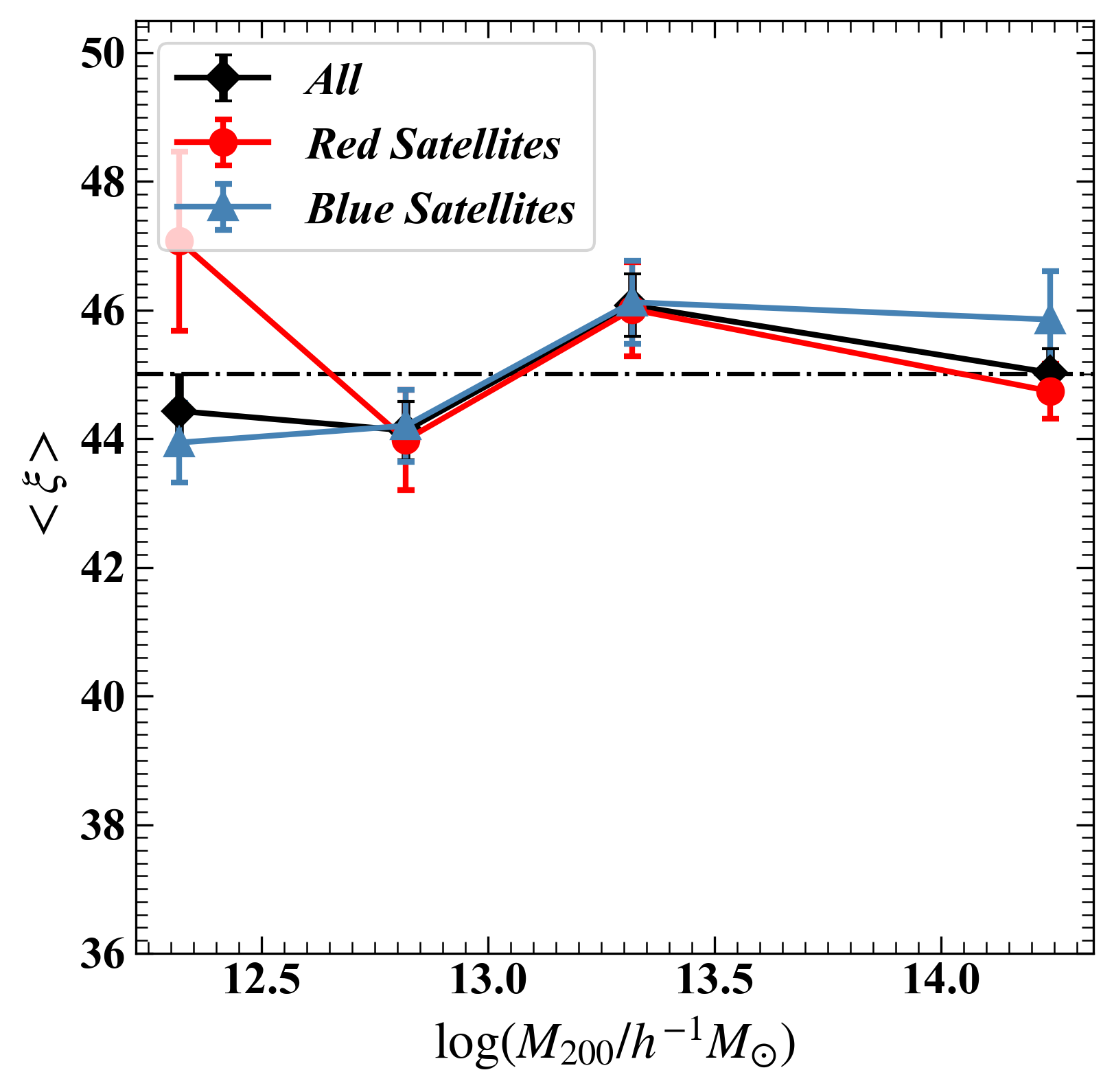}
\includegraphics[width=5.5 cm]{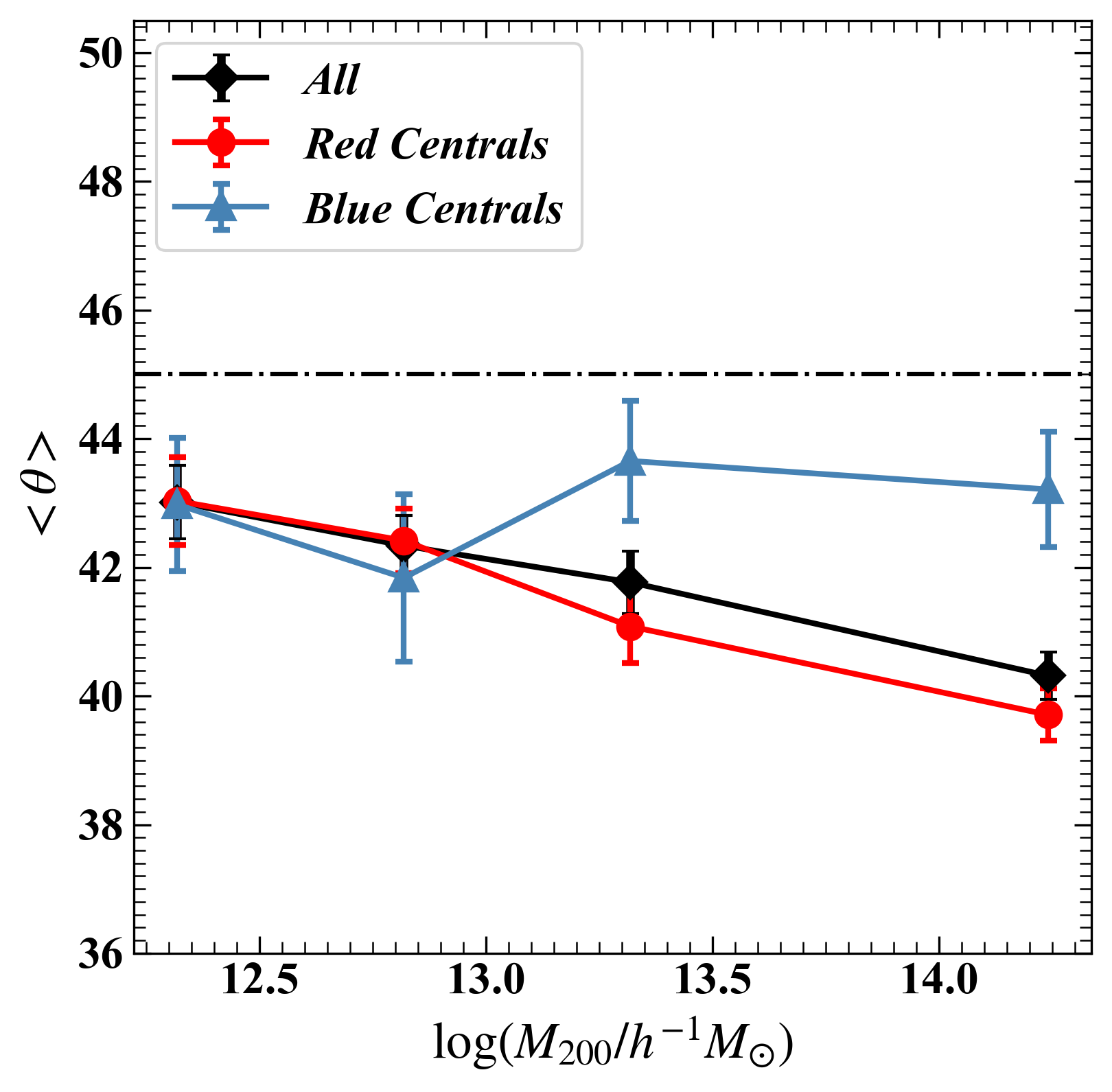}  
\includegraphics[width=5.5 cm]{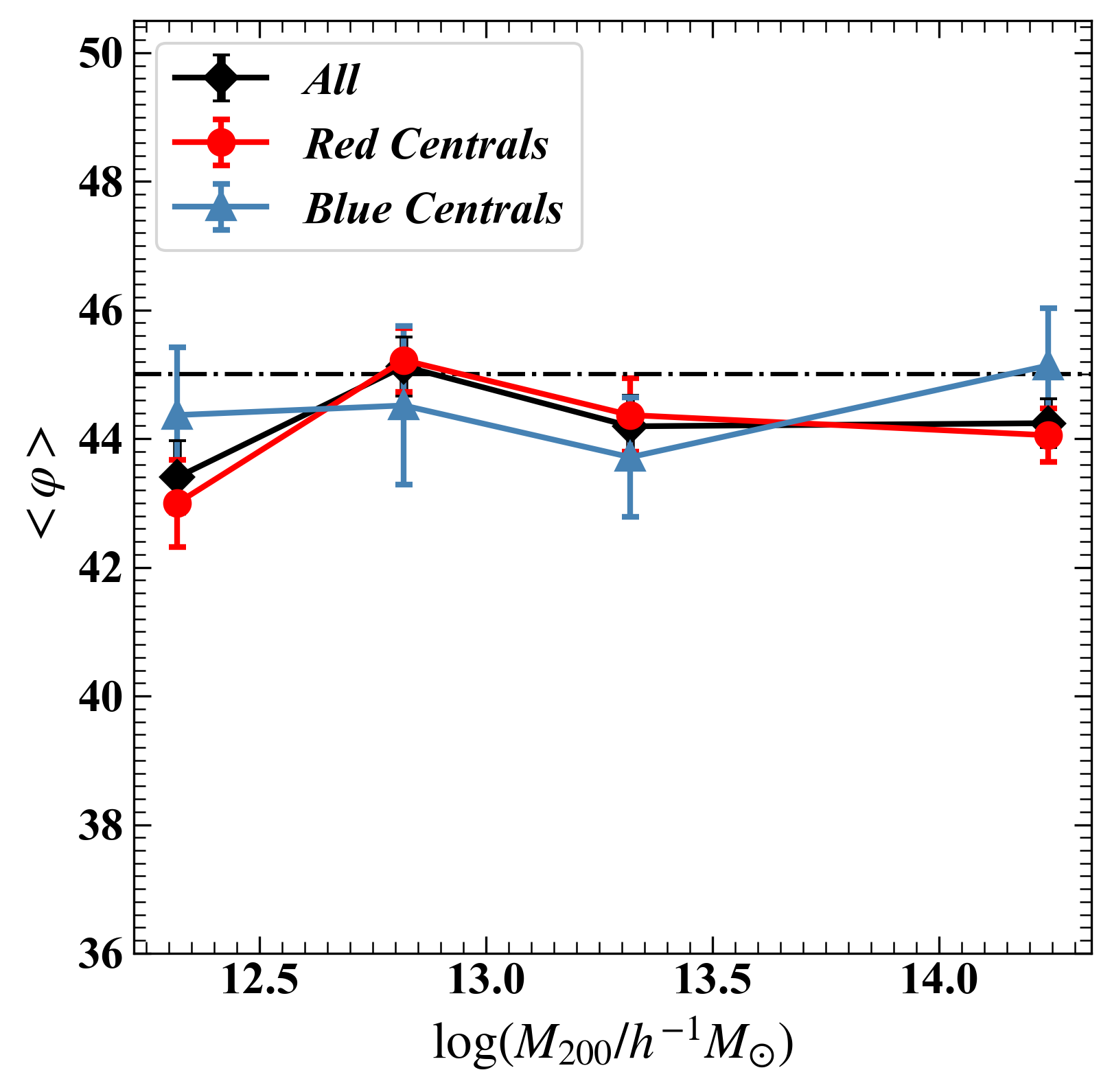}
\includegraphics[width=5.5 cm]{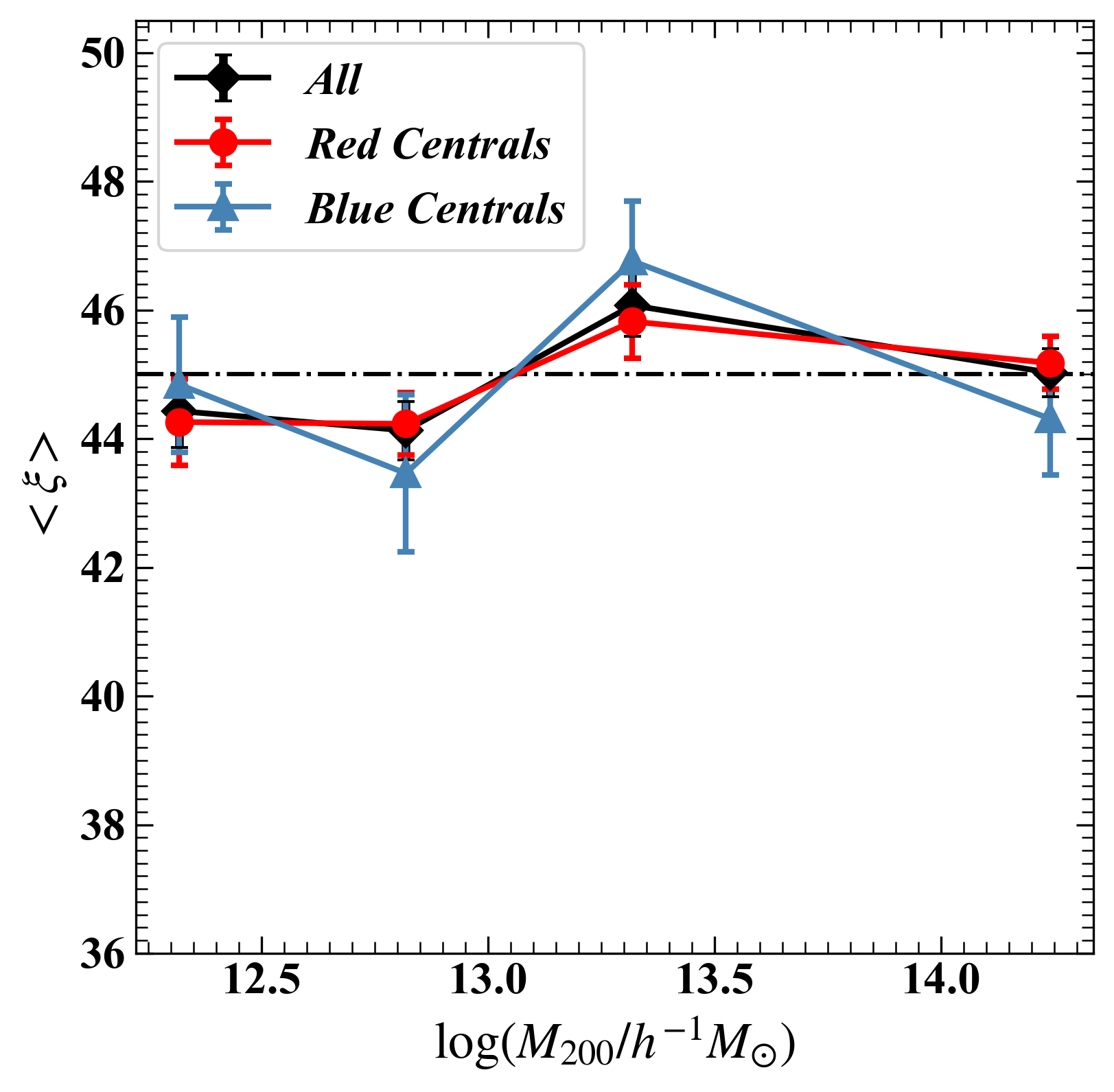}
\caption{The average angle $\langle \theta \rangle$, $\langle \varphi \rangle$ and $\langle \xi \rangle$ as a function of the mass of dark matter halo. The panels in each row, from top to bottom, represent satellites single-color (`Red Satellites', `Blue Satellites') and centrals single-color (`Red Centrals', `Blue Centrals') subsample, respectively. The panels in each column, from left to right, correspond to satellite-central alignment, radial alignment, and direct alignment, respectively.}
\label{figure averangle_halomass}
\end{figure*}

\begin{figure*}[ht!]
\centering
\includegraphics[width=5.5 cm]{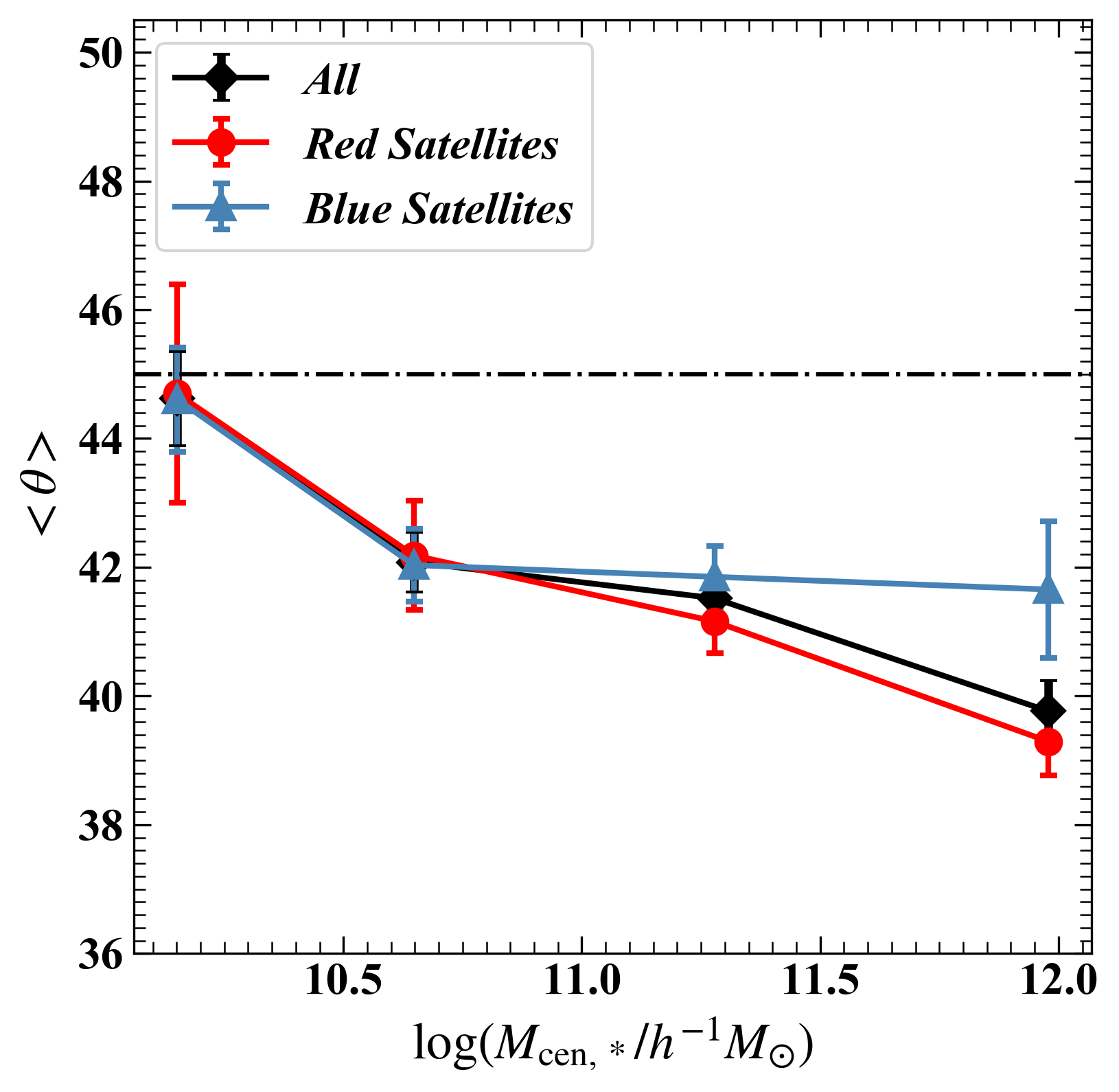}   
\includegraphics[width=5.5 cm]{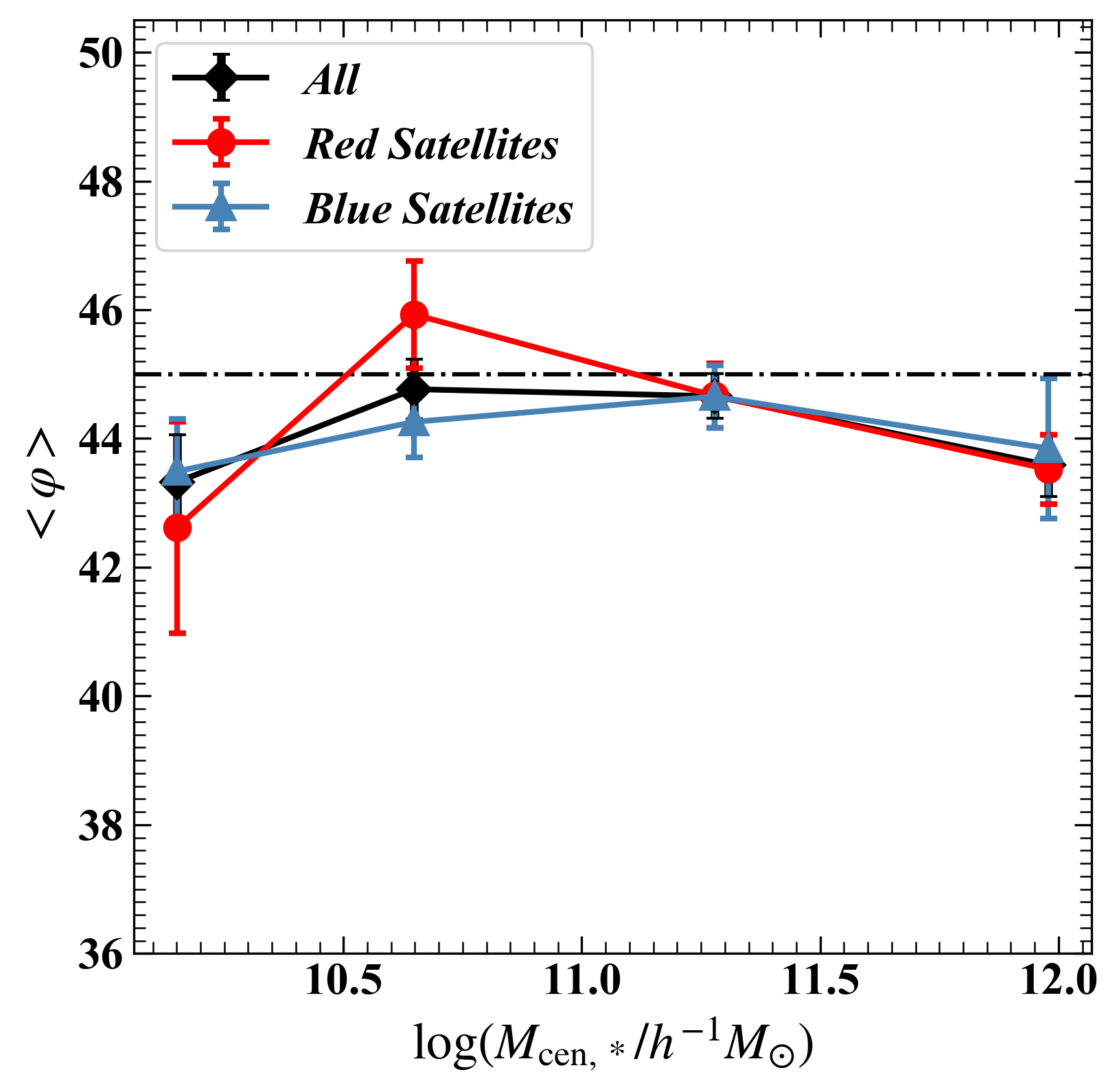}
\includegraphics[width=5.5 cm]{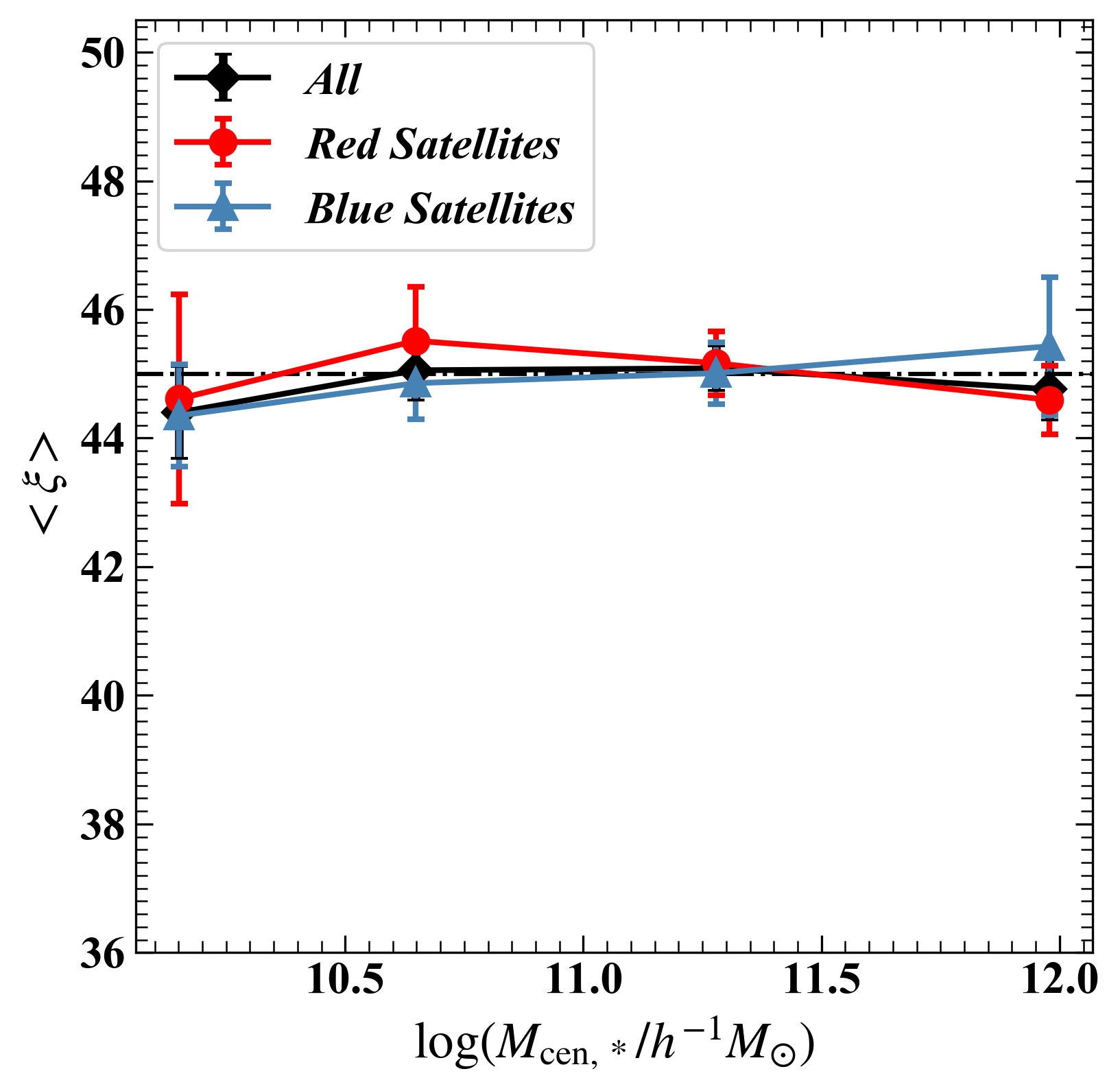}
\includegraphics[width=5.5 cm]{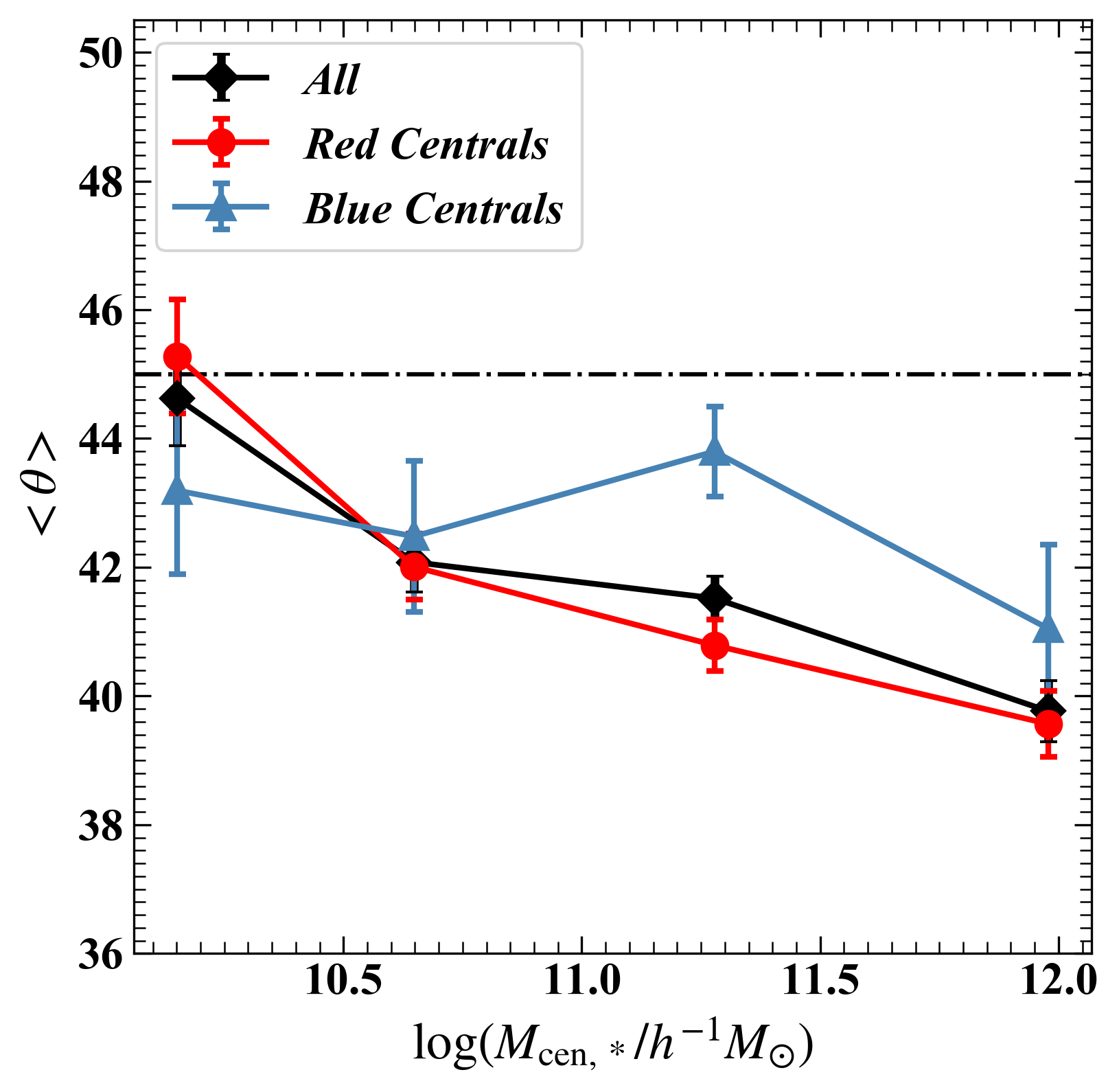}   
\includegraphics[width=5.5 cm]{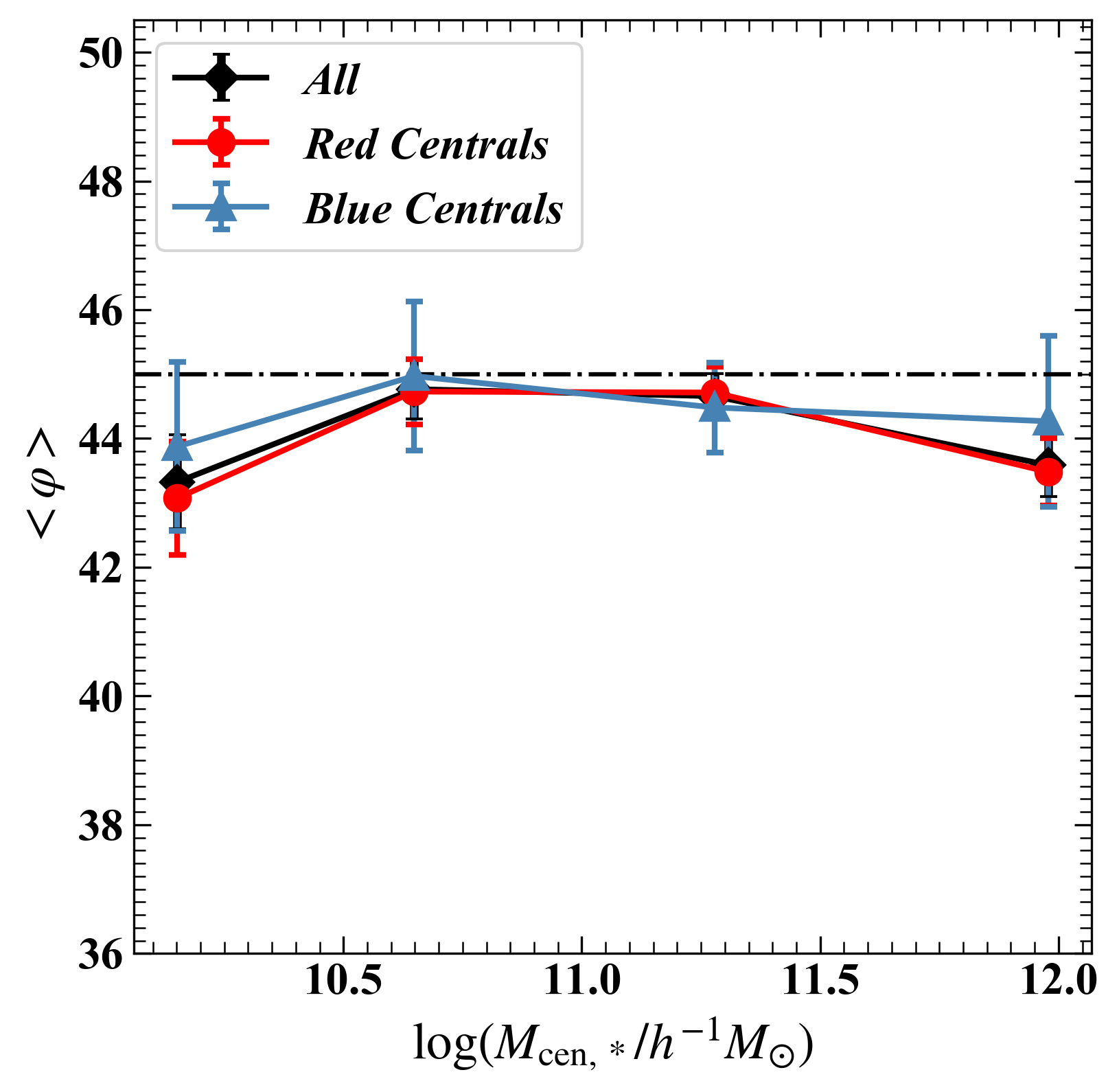}
\includegraphics[width=5.5 cm]{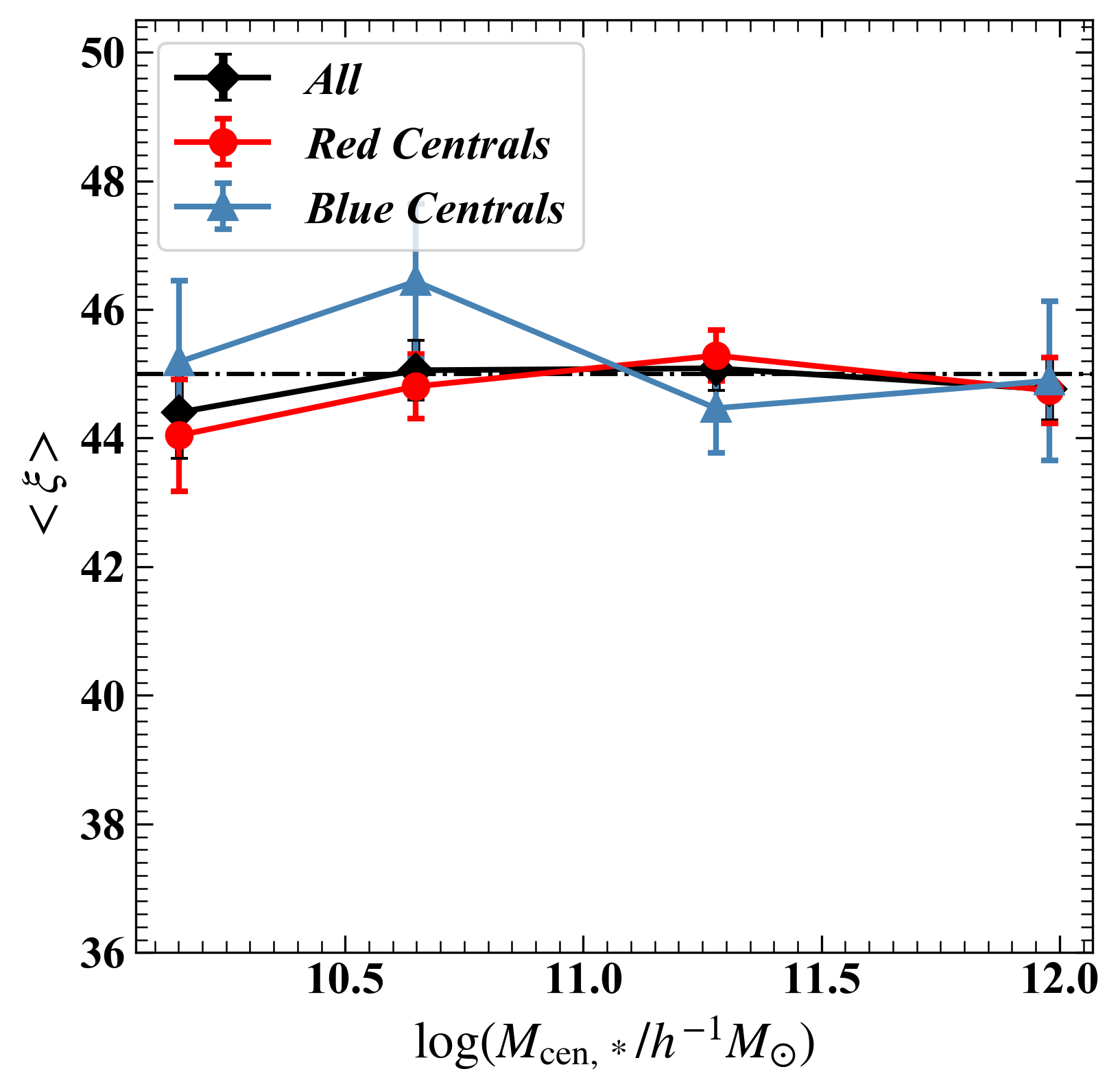}
\caption{Same as Figure \ref{figure averangle_halomass}, but for the dependence on BGG mass.}
\label{figure averangle_stellarmass}
\end{figure*}

There has been evidence that the alignment signal increases with the mass of host halos and BGGs \citep[e.g.,][]{Yang2006, Dong2014,Brainerd2019}.
Figure \ref{figure averangle_halomass} and \ref{figure averangle_stellarmass} have been drawn to reproduce this dependency.

The panels in Figure \ref{figure averangle_halomass} display the average angles $\langle \theta \rangle$, $\langle \varphi \rangle$, and $\langle \xi \rangle$ as functions of the mass of the dark matter halo within a radius of 200 times the mean density of the Universe, $M_{200}$.
The lines represent results for the entire mock sample and the four color subsamples, as indicated in the legends.

The left panels of Figure \ref{figure averangle_halomass} indicate that for the blue subsamples (``Blue Satellites'' and ``Blue Centrals''), $\langle \theta \rangle$ hovers around $43^{\circ}$,
while the alignment strengths of the entire sample (``All'' , the same in the following) and the red subsamples (``Red Satellites'' and ``Red Centrals'') show a positive correlation with halo mass.
This halo mass dependence trend, particularly evident for the entire sample, is consistent with previous predictions \citep[e.g.,][]{Dong2014, Wang2018a}  and observational findings \citep[see, e.g., Figure 5 in][]{Yang2006}.
The pattern of the radial alignment signal variation with halo mass appears somewhat complex, as depicted in the middle column of the panels. It seems that galaxy pairs within less massive halos exhibit a stronger radial alignment signal compared to those in more massive halos, with the exception of the ``Red Satellites'' subsample, which displays large error bars and fluctuations.
This alignment trend aligns with the concept that radial alignment may arise from BGG and satellite interactions, which are more pronounced in relatively small mass halos.
As illustrated in the right column of panels, the average direct alignment angles $\langle \xi \rangle$ exhibit a comparable trend of mass dependence to $\langle \varphi \rangle$, albeit with generally weaker alignment signals on average.

The dependence of alignment on the stellar mass of central galaxies has been investigated and is presented in Figure \ref{figure averangle_stellarmass}. Similar to Figure \ref{figure averangle_halomass}, the satellite-BGG pairs have been grouped into four mass bins based on the range of BGG mass.
The settings for points, lines, colors, and legends are the same as those in Figure \ref{figure averangle_halomass}.
Unsurprisingly, the dependencies on BGG mass for the three types of alignment exhibit similarities to those in Figure \ref{figure averangle_halomass}, considering the relationship between BGG mass and halo virial mass in this mass regime, i.e., $\geq 10^{12} M_\odot$ \citep[e.g., ][]{Nelson2019, Tang2021}.

The strengths of radial and direct alignment appear to be more pronounced at the lower end of the mass range. The intricate patterns observed in the relationship between radial and direct alignment and halo/BGG mass are likely influenced by various factors, such as the dynamical state of satellites within host halos.
Further exploration and analysis are necessary to fully understand and clarify the underlying reasons for these trends. Factors such as the assembly history of halos, environmental influences, and the specific mechanisms driving alignment signals in different mass regimes may all contribute to these complexities.

\subsubsection{BGG Age and Metallicity}
\begin{figure*}[ht!]
\centering
\includegraphics[width=5.5 cm]{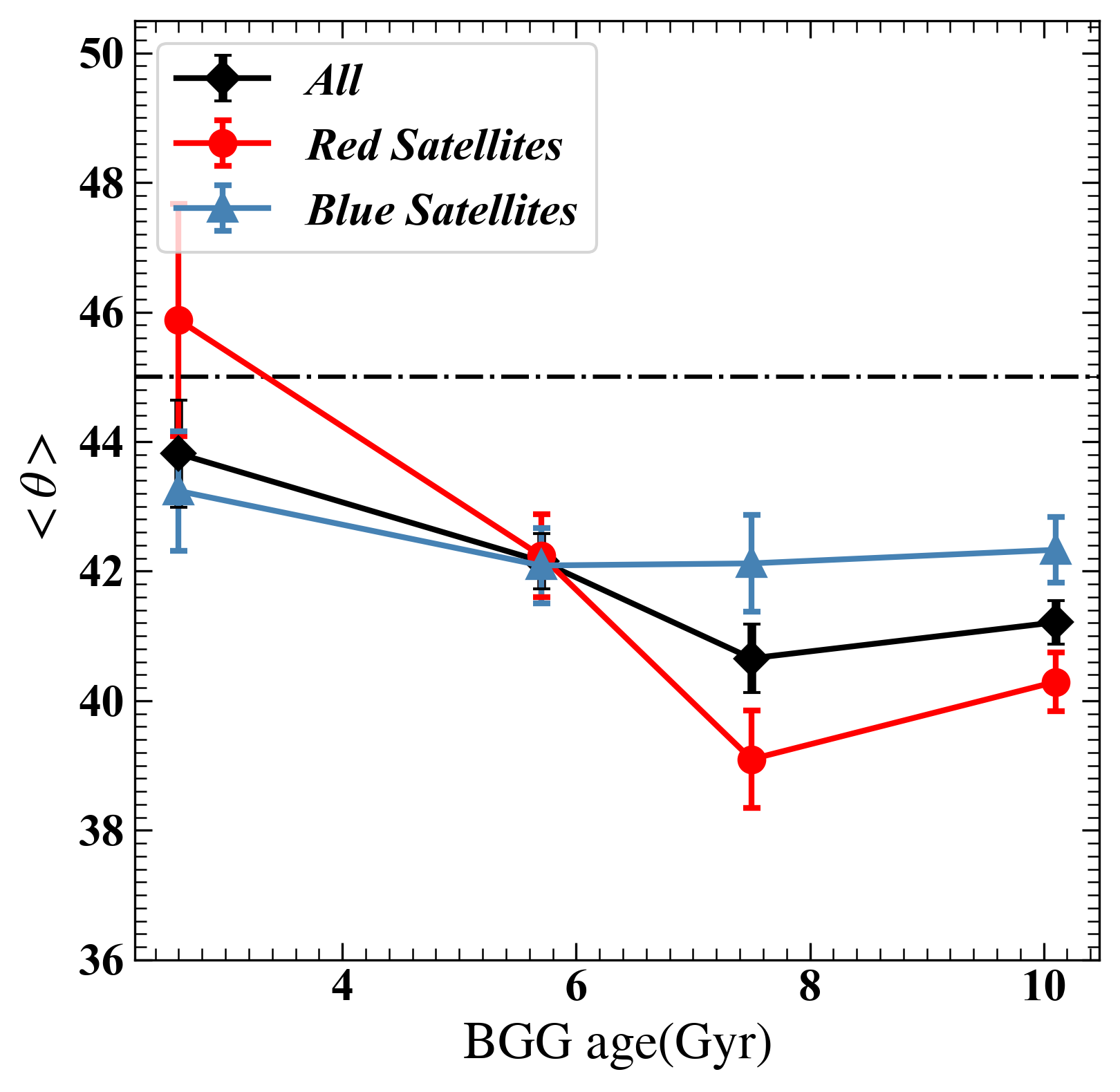}  
\includegraphics[width=5.5 cm]{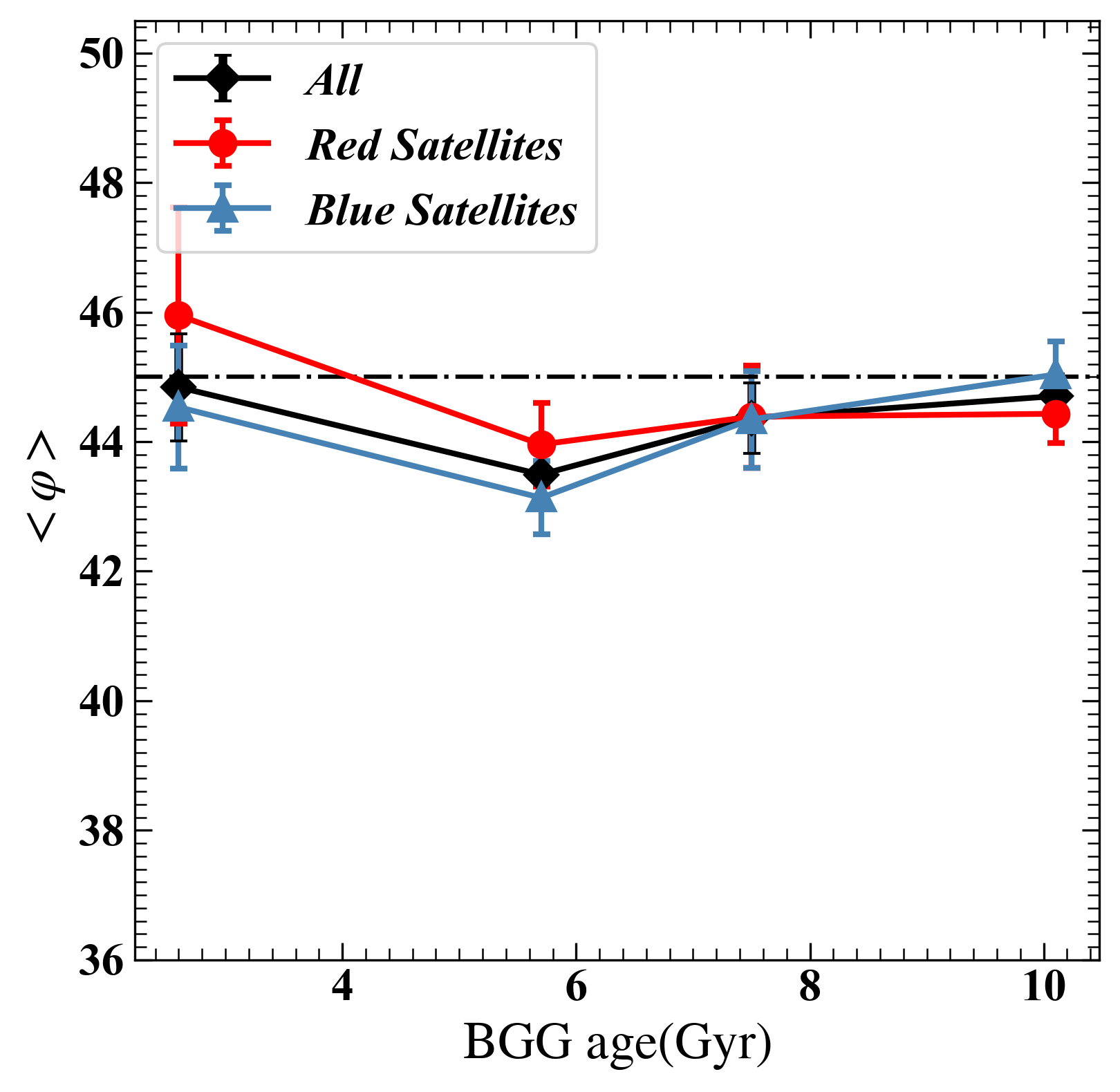}
\includegraphics[width=5.5 cm]{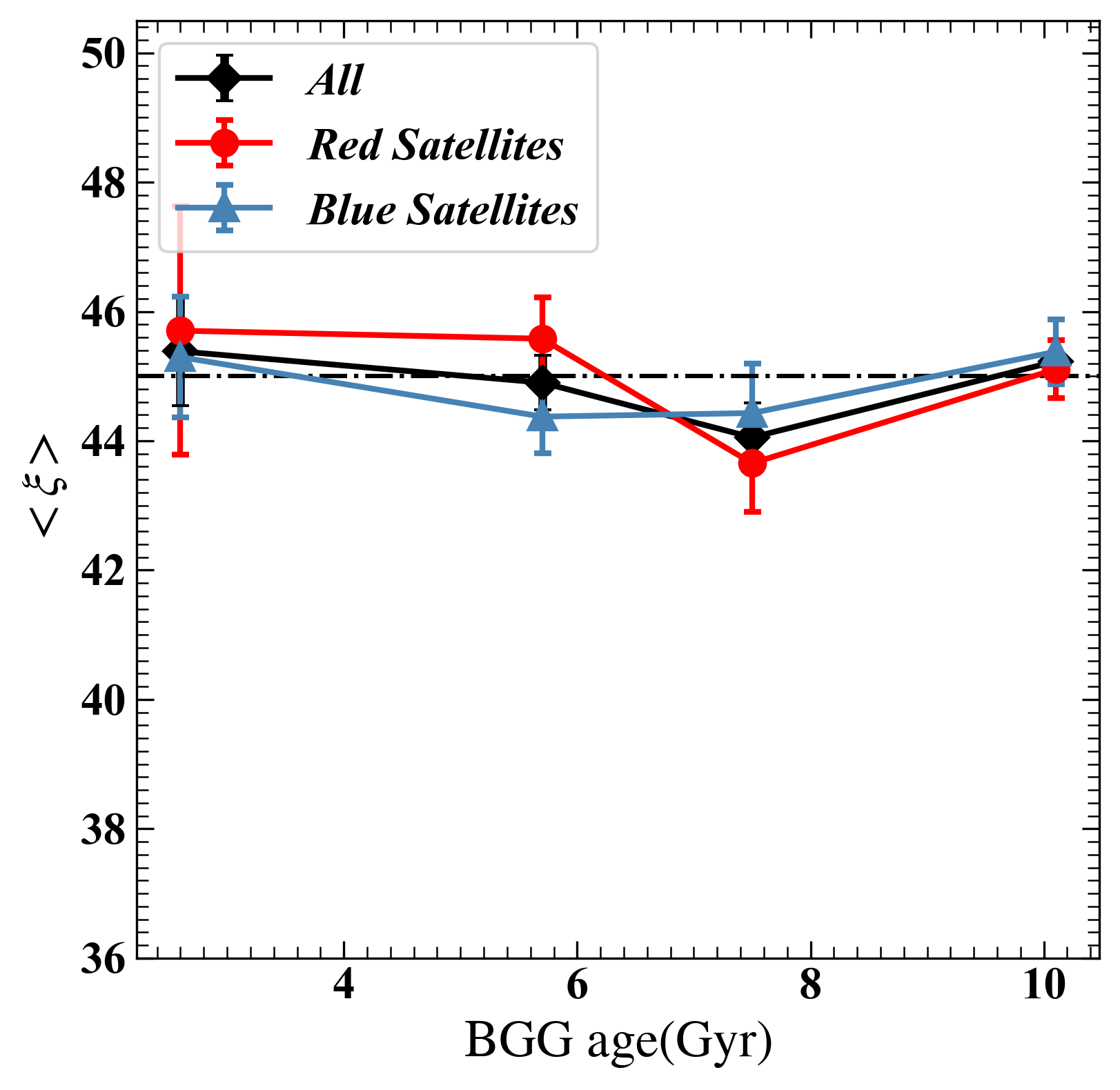}
\includegraphics[width=5.5 cm]{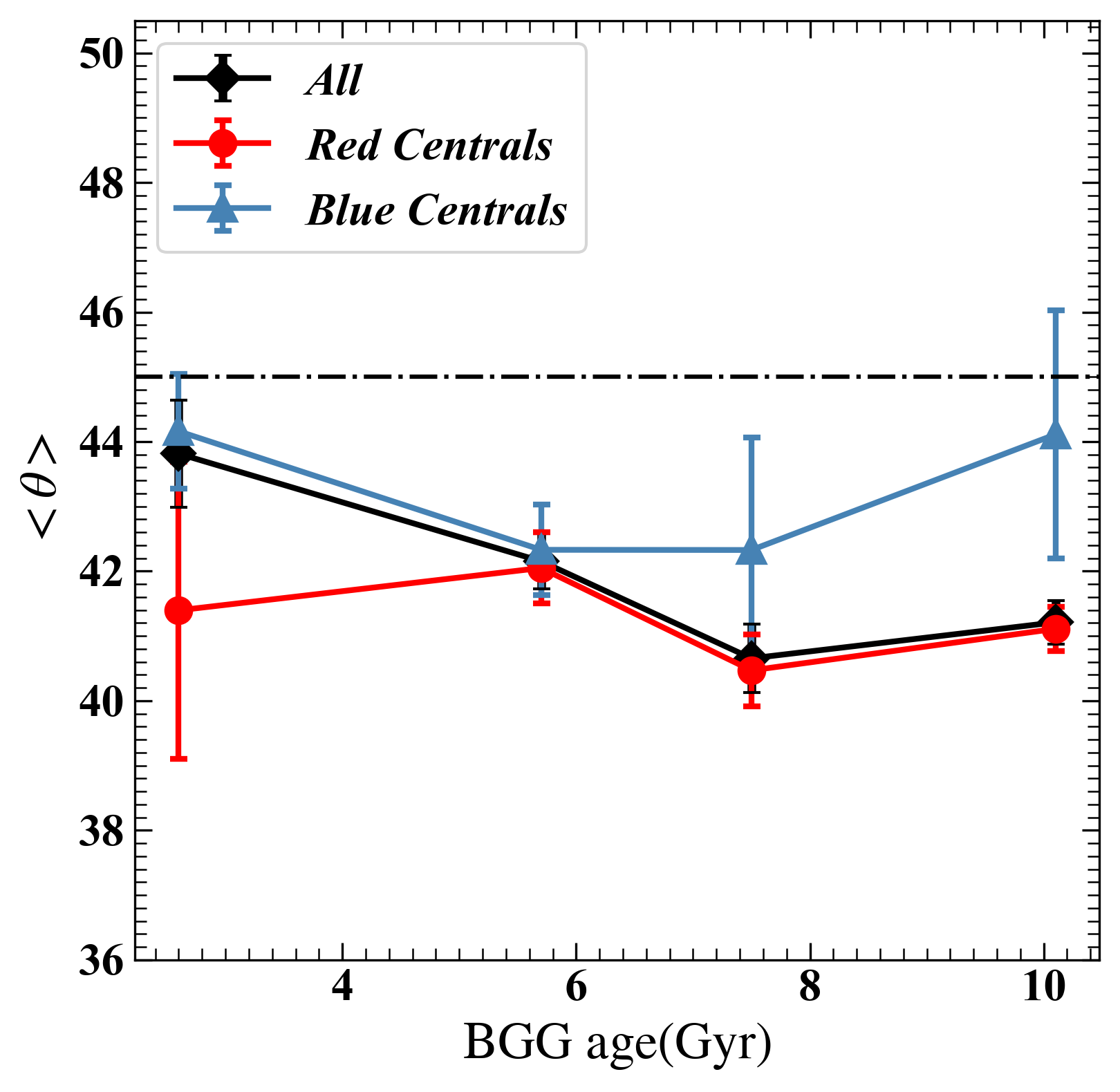}  
\includegraphics[width=5.5 cm]{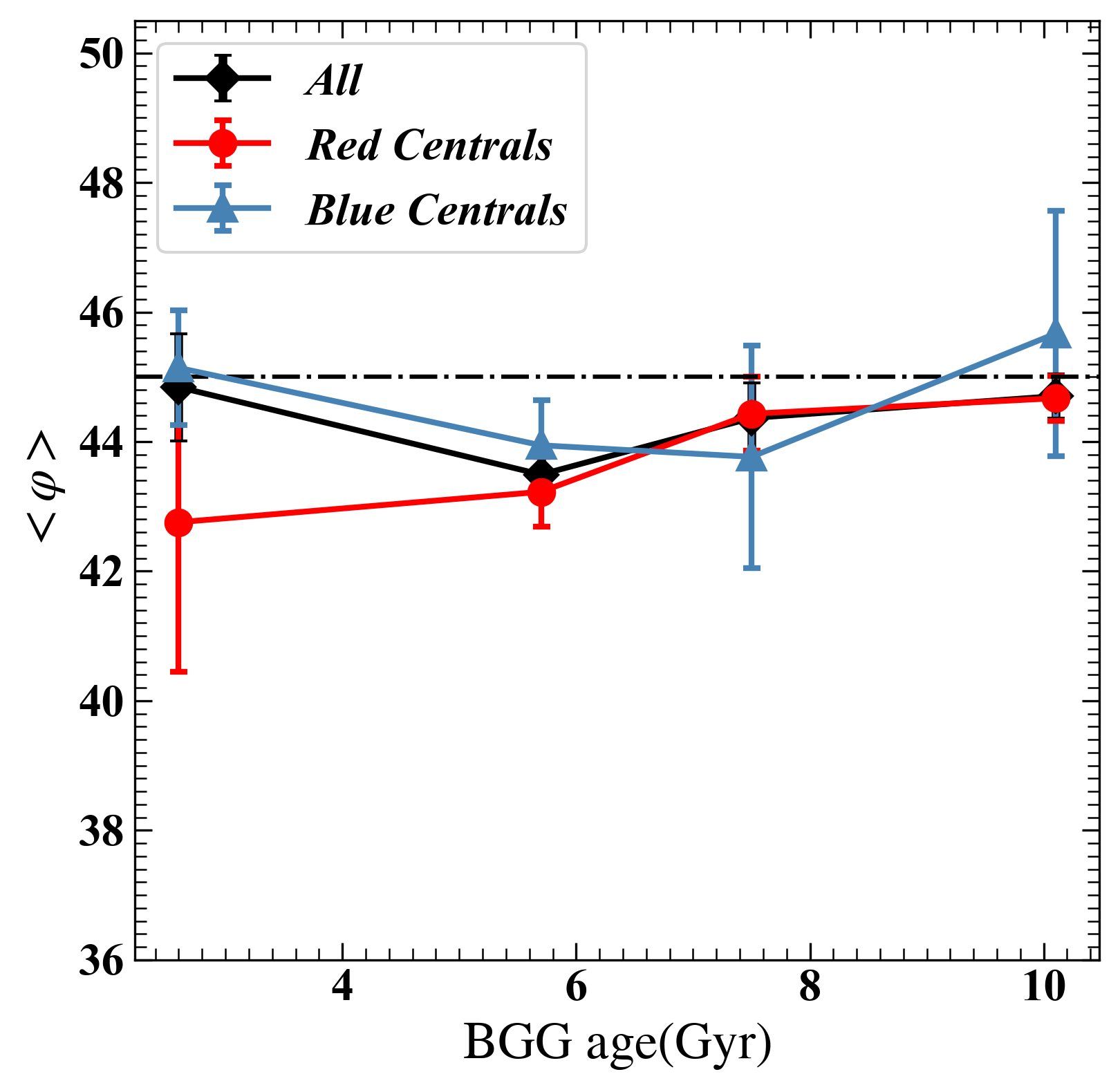}
\includegraphics[width=5.5 cm]{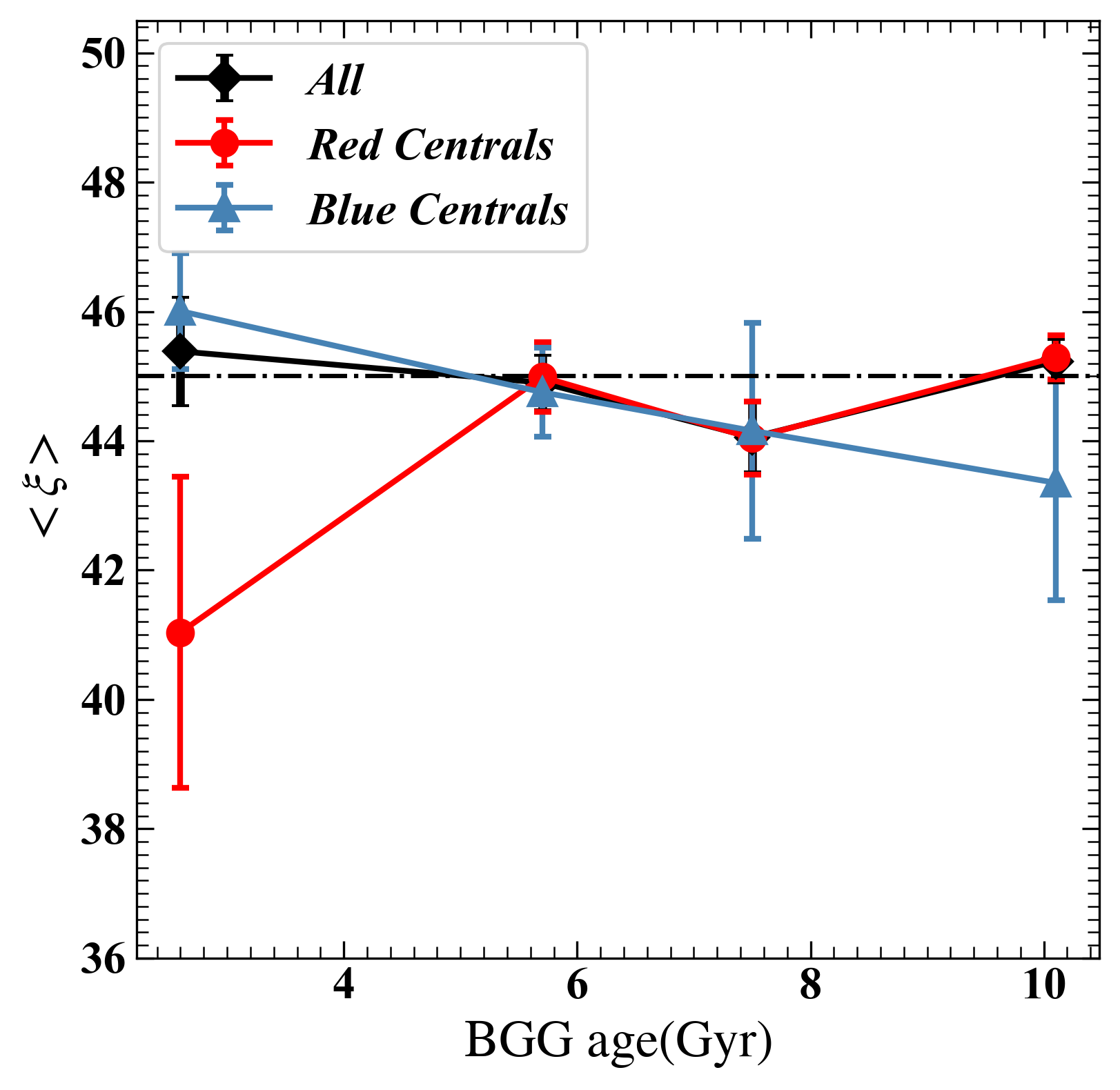}
\caption{Same as Figure \ref{figure averangle_halomass}, but for the dependence on BGG age.}
\label{figure averangle_age}
\end{figure*}

\begin{figure*}[ht!]
\centering
\includegraphics[width=5.5 cm]{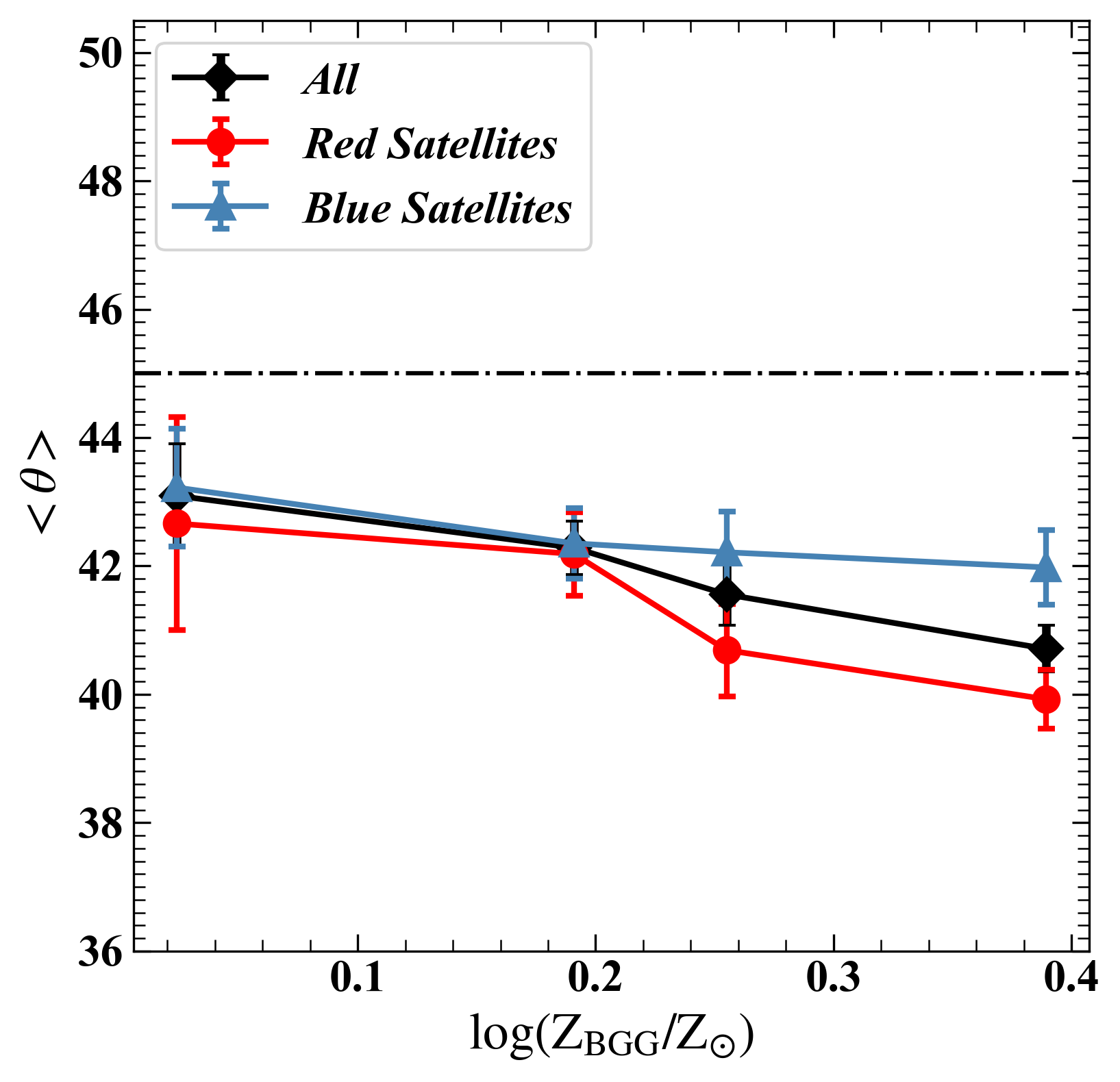}  
\includegraphics[width=5.5 cm]{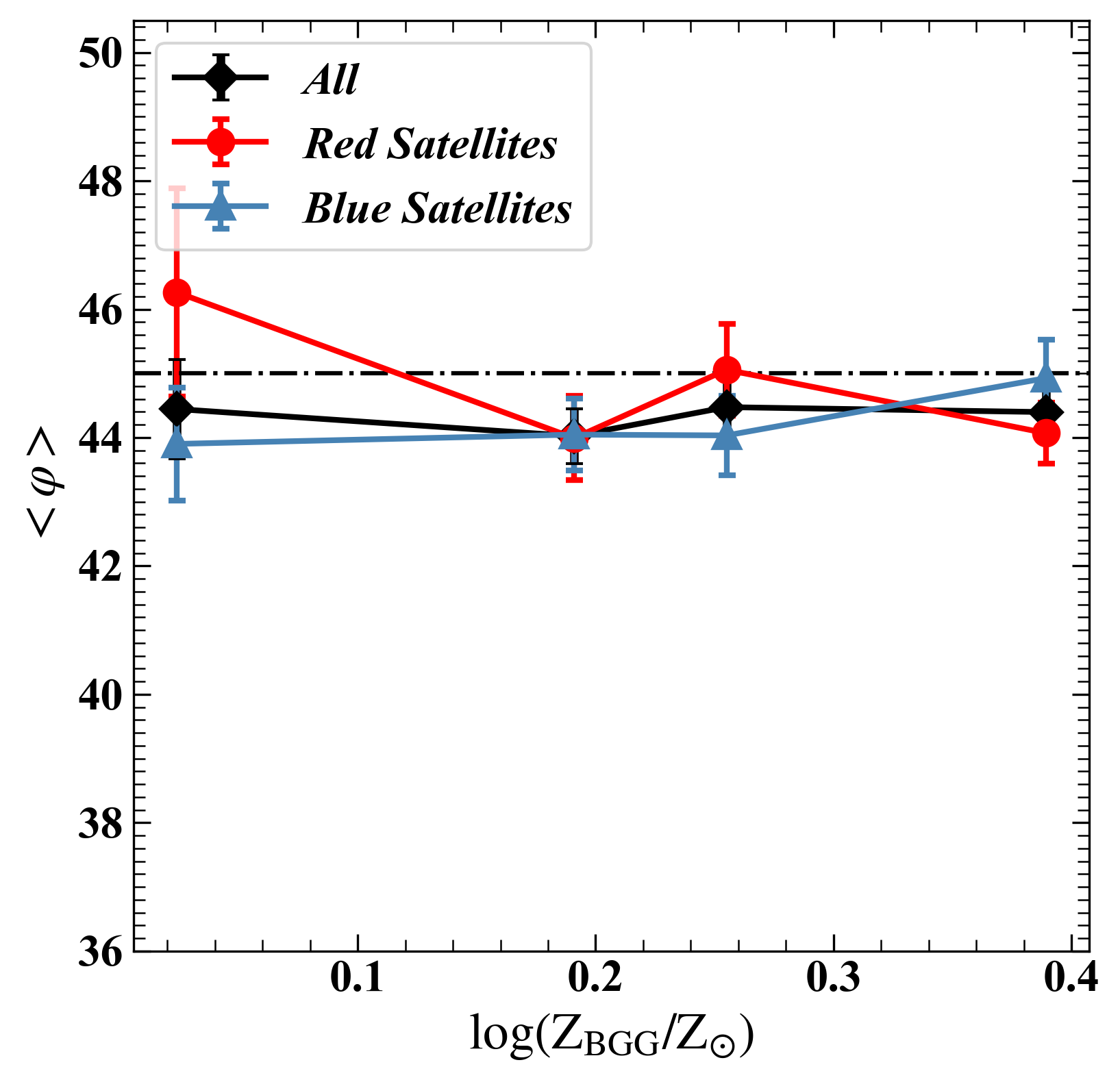}
\includegraphics[width=5.5 cm]{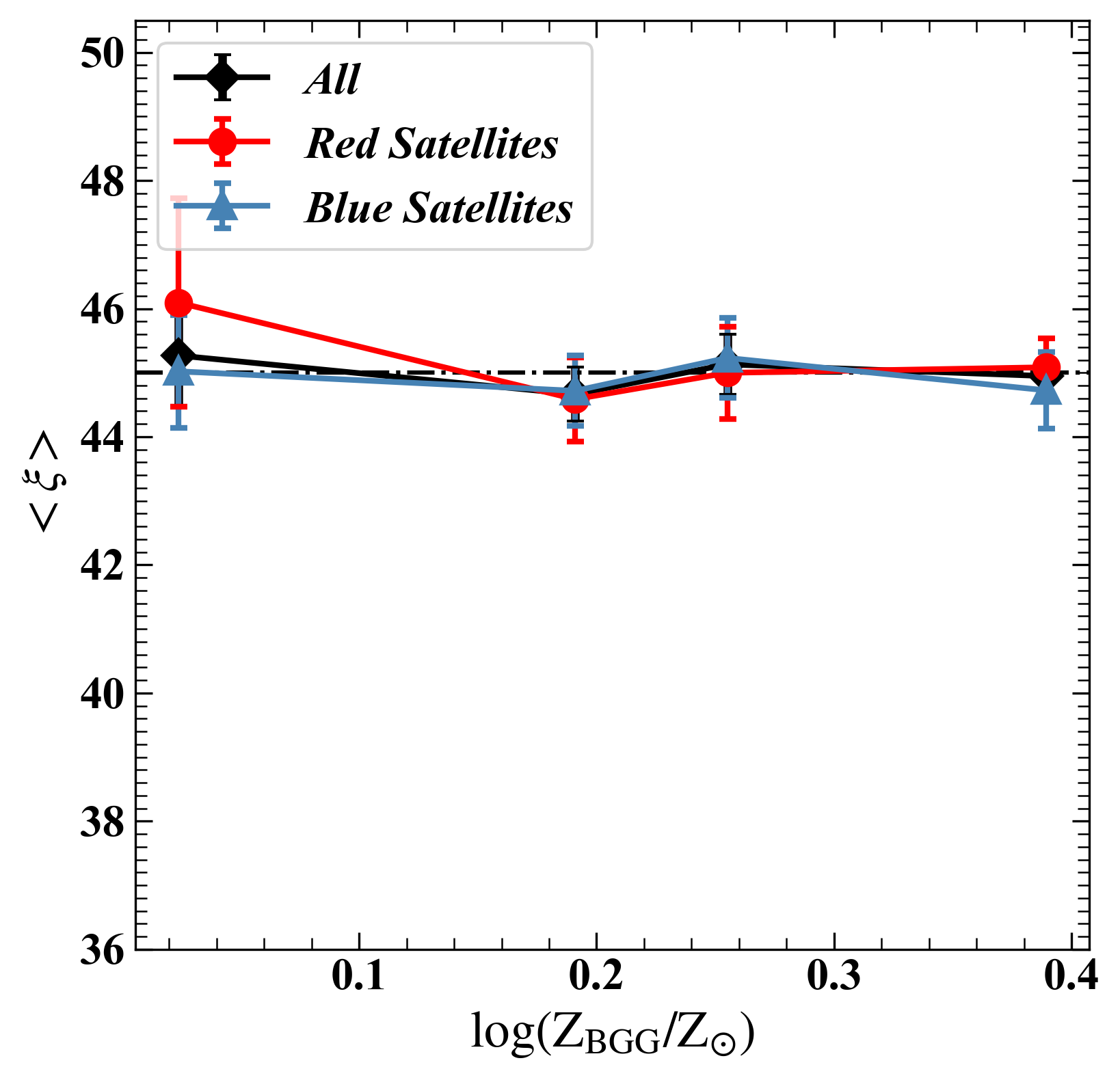}
\includegraphics[width=5.5 cm]{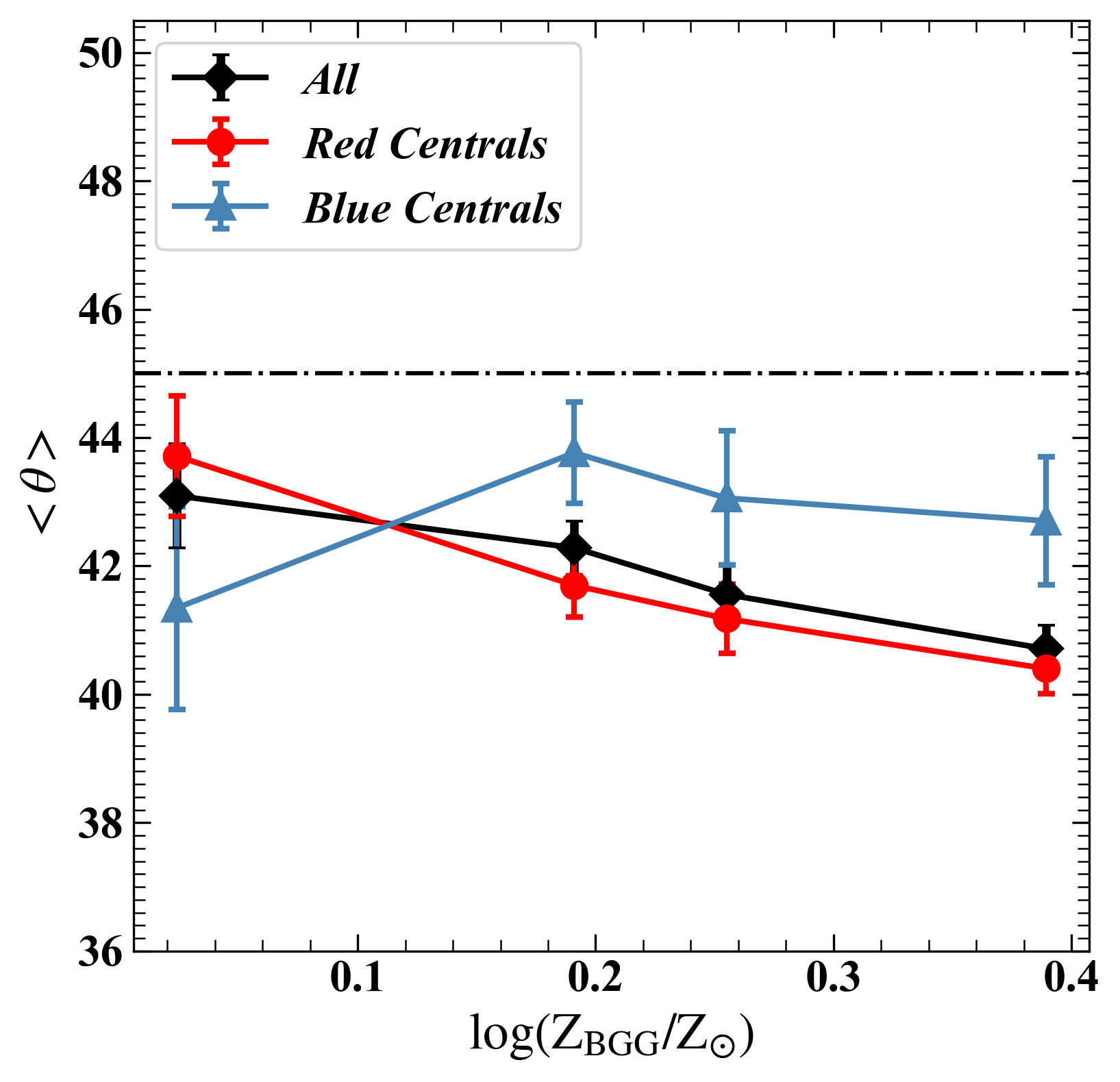}  
\includegraphics[width=5.5 cm]{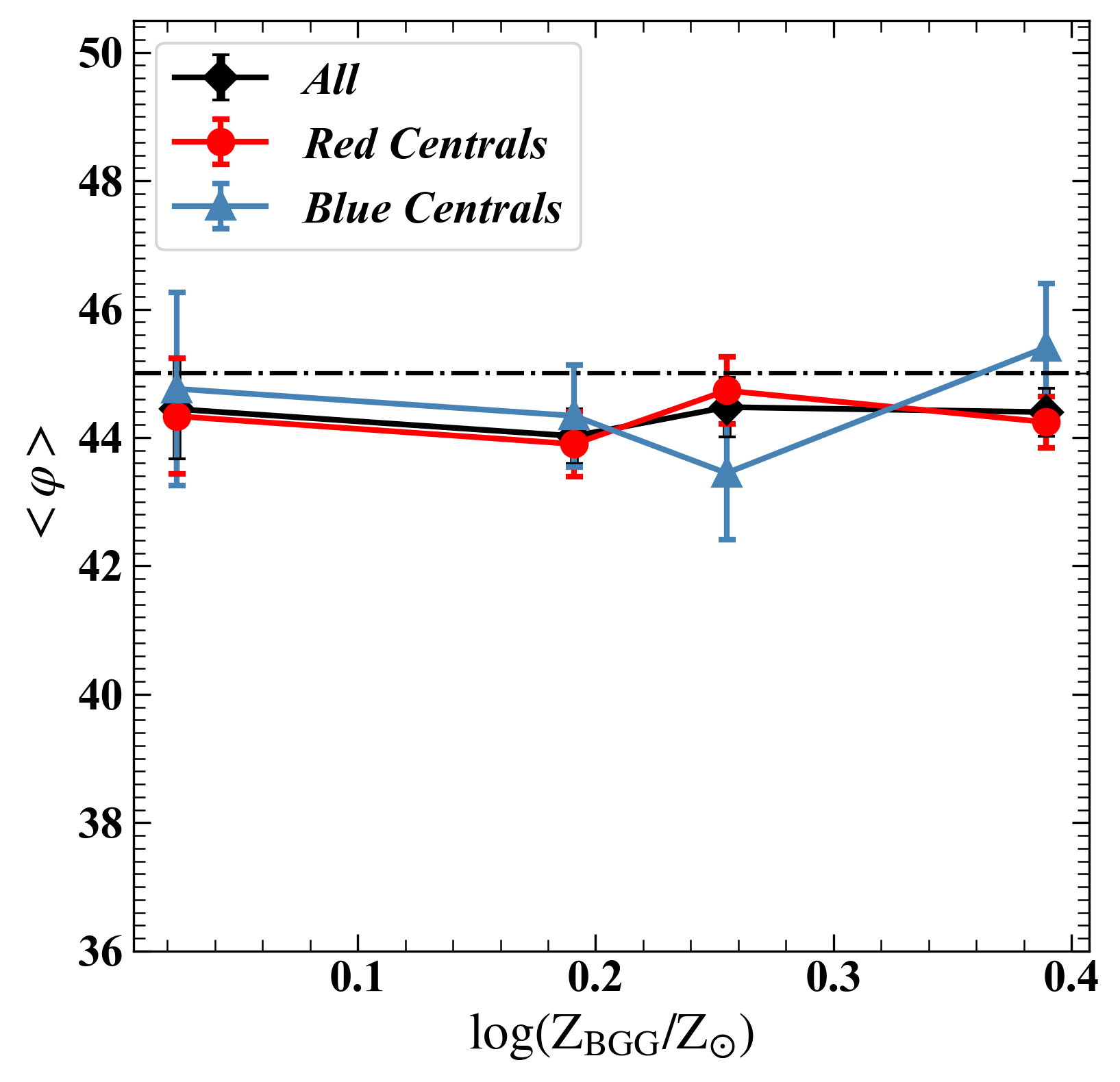}
\includegraphics[width=5.5 cm]{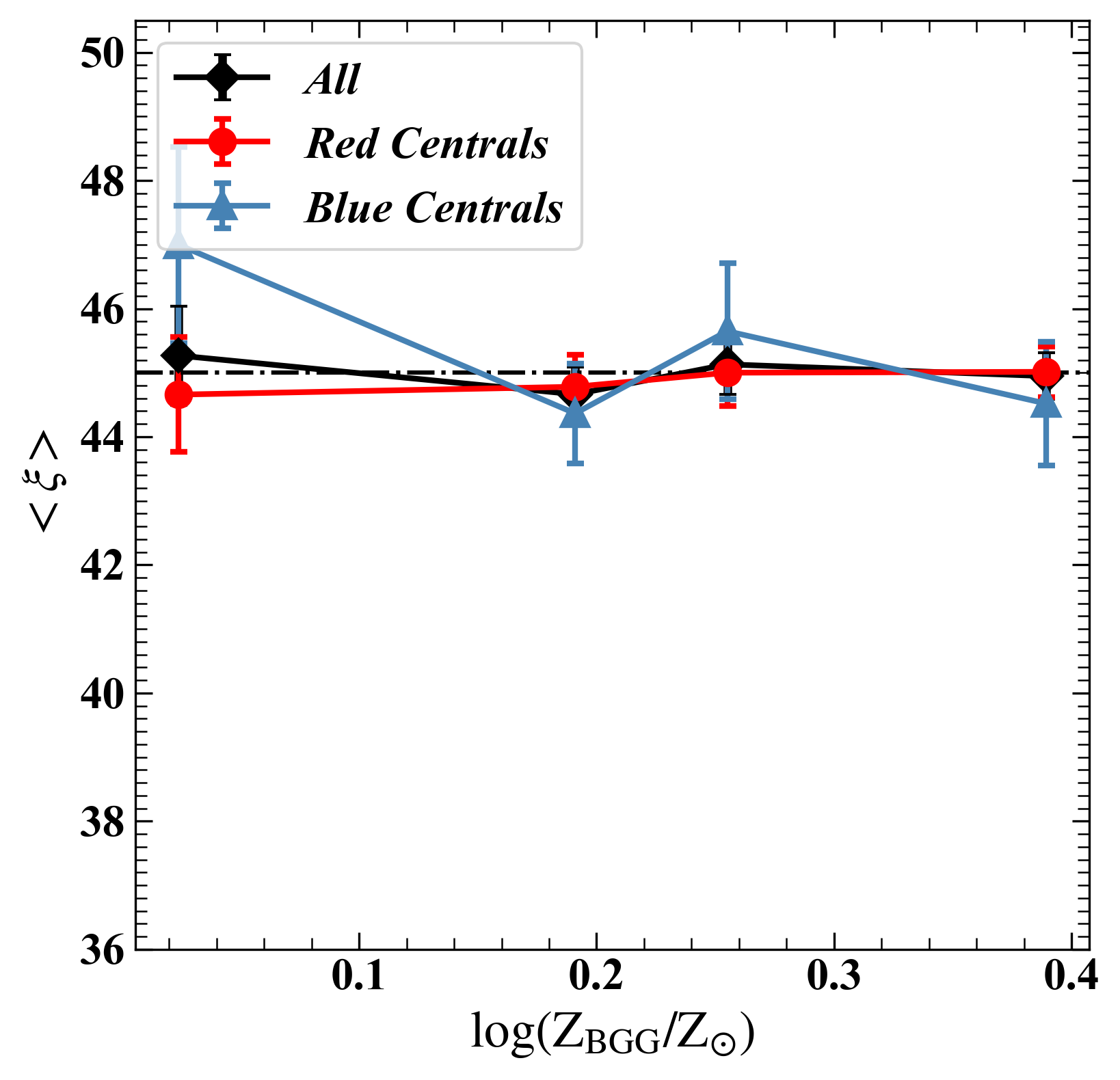}
\caption{Same as Figure \ref{figure averangle_halomass}, but for the dependence on BGG metallicity.}
\label{figure averangle_met}
\end{figure*}

The metallicity and age of galaxies are determined by averaging these properties over all the star particles within the constructed ``galaxy''.
 It is important to note that the metallicity for star particles is obtained from the TNG100-1 snapshots, while the age is calculated based on the formation epoch of the star particles.
To analyze the dependency of alignment on galaxy age and metallicity,
similarly the satellite-BGG pairs will be divided into four consecutive bins with different ranges of BGG age or metallicity.
The results are presented in Figure \ref{figure averangle_age} and \ref{figure averangle_met} for age and metallicity dependence, respectively.

As shown in Figure \ref{figure averangle_age}, on average, for the entire mock sample, the predicted satellite-central alignment strengths exhibit positive correlations with BGG ages. In contrast, the radial and direct alignment strengths show very weak correlations and complex patterns of dependence.
Specifically, for satellite-central alignment in the left-hand side column panels, the blue subsamples demonstrate a complex dependency of alignment on BGG ages, whereas the satellite-central alignment strengths for the red subsamples are positively correlated with BGG ages, particularly for the `Red Satellites' subsample.
However, in the radial/direct alignment analysis shown in the middle-/right-hand side column panels, respectively, the dependence on BGG ages exhibits a high degree of scatter, and no clear trend can be discerned.
In addition, for all three types of alignments, we can see that the BGG age dependence trend of the entire mock sample well align with ``Blue Centrals'' in the small age end and ``Red Centrals'' in the large age end, respectively. This is because in large age regions, central galaxies are dominated by red galaxies, and vice versa.

Figure \ref{figure averangle_met} shows the correlation between alignment and BGG metallicity. The findings suggest that satellite-central alignment patterns resemble those observed with BGG age dependence. Specifically, halos with metal-rich BGGs exhibit more pronounced BGG-satellite alignment compared to those with metal-poor BGGs. This is consistent with the dependence on halo mass or BGG mass, as mass and metallicity are correlated.

Overall, the results exhibit significant scatter in terms of age and metallicity dependence, especially for the blue subsamples, indicating a complex relationship between alignment strength and the properties of BGGs. However, it is crucial to acknowledge that some data points in these plots may lack statistical significance due to the limited number of blue galaxy pairs with old age or high metallicity, and vice versa for red galaxy pairs. To improve statistical robustness, a larger mock sample is required for further analysis.

\subsubsection{radial distance}

\begin{figure*}[ht!]
\centering
\includegraphics[width=5.5 cm]{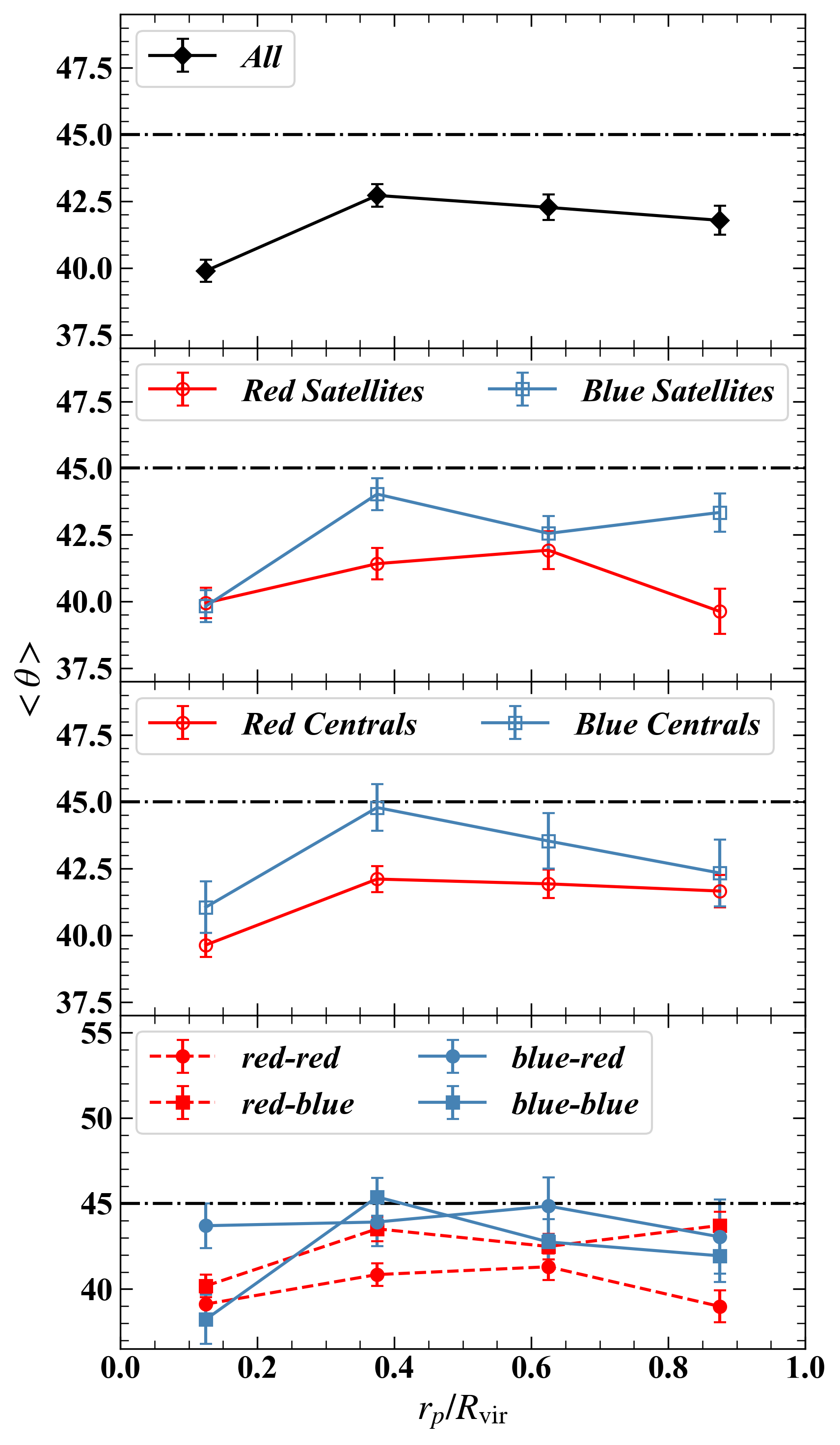}
\includegraphics[width=5.5 cm]{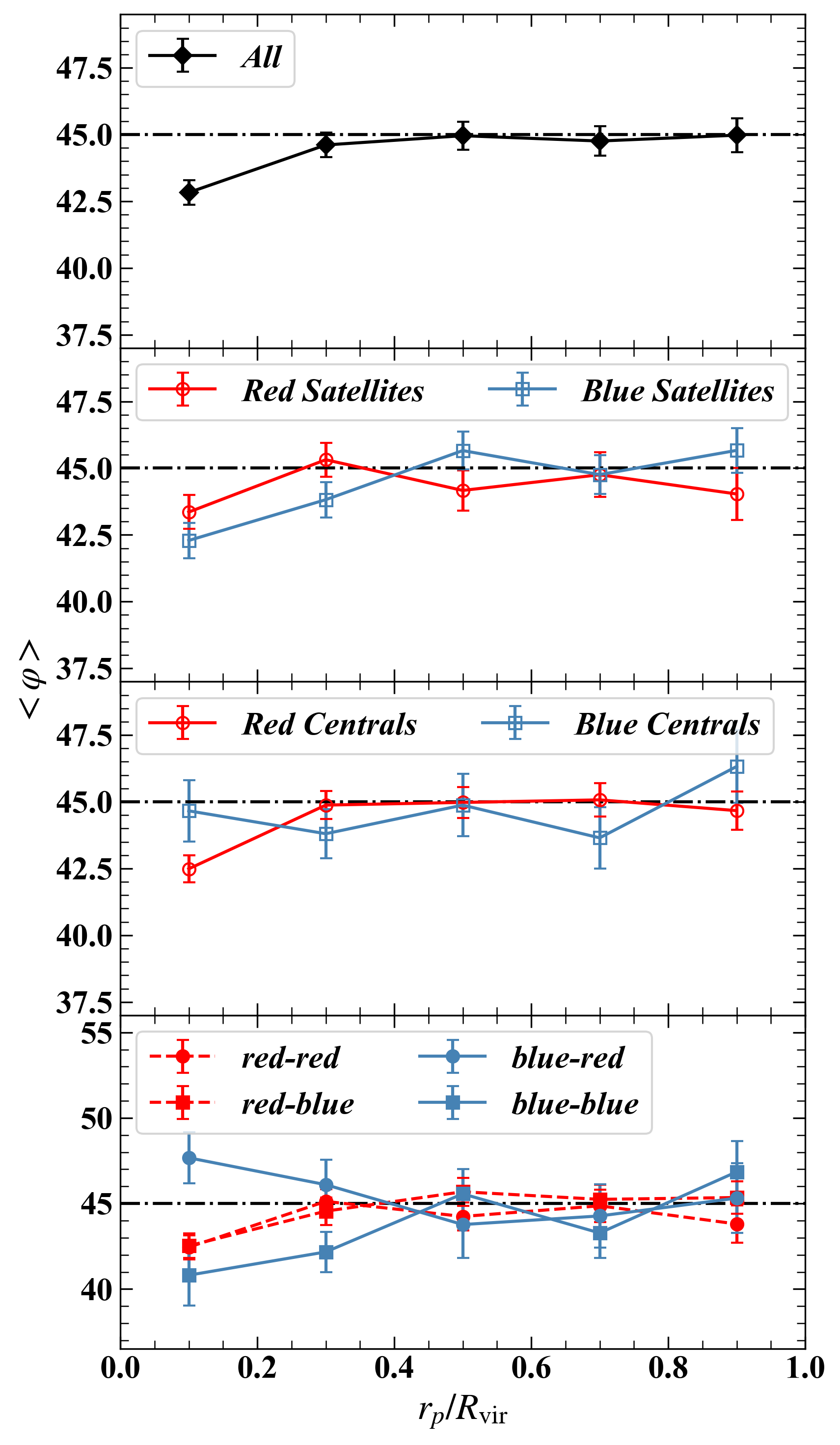}
\includegraphics[width=5.5 cm]{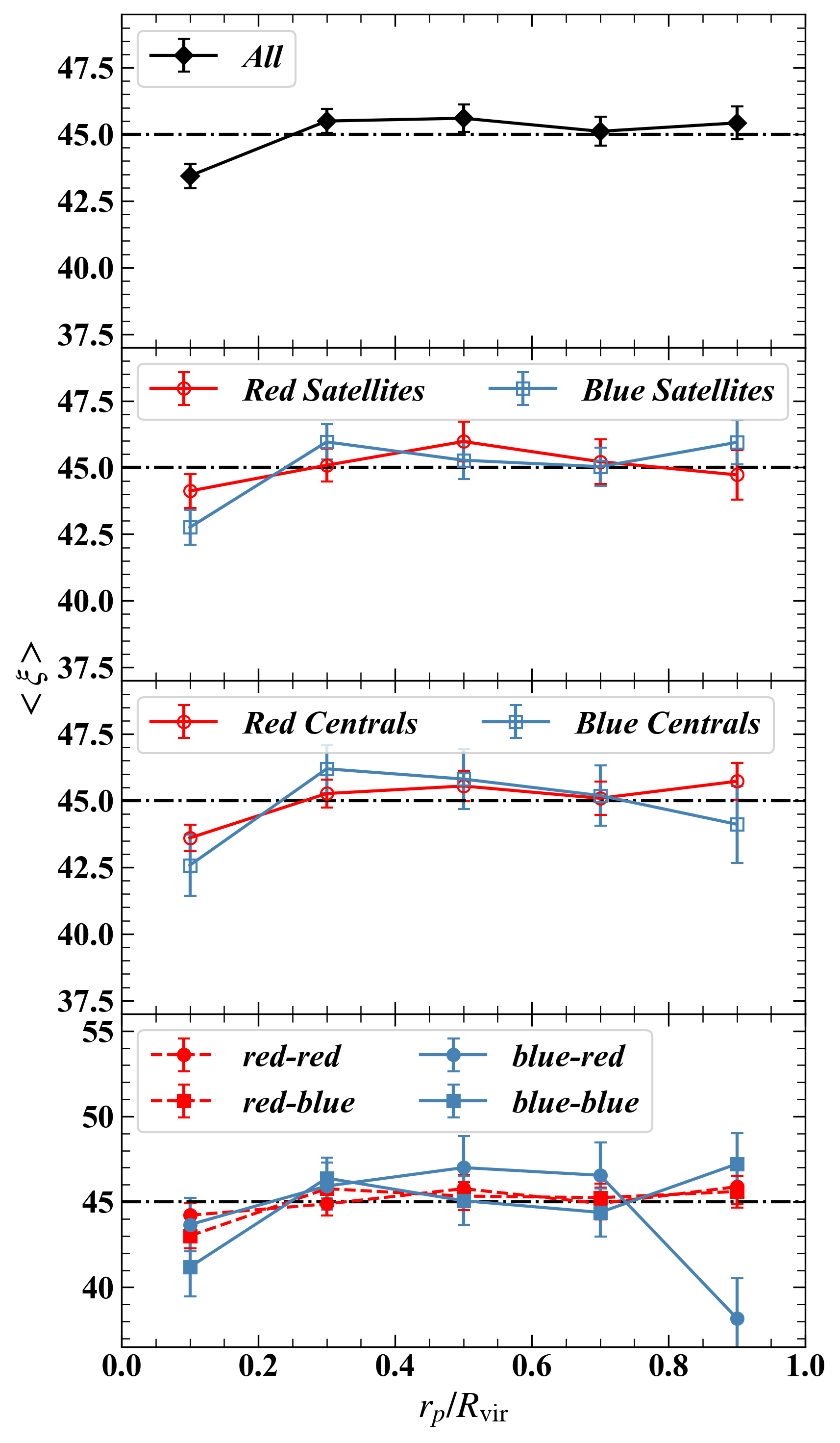}
\caption{Average alignment angles for all satellites and the four color subsamples (indicated in the legends of each panel), as functions of projected distances from the satellites to the center of host halo, normalized by the halo virial radius. The panels in each column, from left to right, correspond to satellite-central alignment, radial alignment, and direct alignment, respectively.}
\label{figure angle rbin0.25_allgalaxy}
\end{figure*}

The dependency of alignment strength on the satellite-centric distance from the BGG center has been extensively studied in the literature \citep[e.g.,][]{Agustsson2006a, Yang2006, Faltenbacher2007, Dong2014, Tang2020}. In this study, further investigation was conducted using the mock sample to examine this relationship.
The average angles for satellite-central, radial, and direct alignment as functions of satellite projected radial distance are shown in the left, middle, and right panels of Figure \ref{figure angle rbin0.25_allgalaxy}, respectively. The projected distances are normalized by the virial radius of the host dark matter halos, and the center of the dark matter halo is calculated by summing the mass-weighted relative coordinates for all types of particles in the FoF group.

The analysis of Figure \ref{figure angle rbin0.25_allgalaxy} reveals that, in general, for the entire mock sample and most of the color subsamples (except for the blue-red subsample), all three types of alignment exhibit a clear radial dependence. Specifically, galaxies located closer to the center of their BGG exhibit stronger alignment compared to those located further away, consistent with previous observational studies \citep{Faltenbacher2007}.
However, the radial dependence for the blue-red subsample appears to differ from the other subsamples, exhibiting abnormal behavior. This discrepancy warrants further investigation to understand the underlying mechanisms driving the alignment patterns observed in this specific color subsample.

In addition, our simulation results highlight the strong significance of satellite-central alignment signals for all satellite galaxies within the entire halo, whereas both radial and direct alignments are particularly noticeable at scales where the projected separation $r_p \lesssim 0.2 R_{\rm{vir}}$.
The results of radial dependence for satellite-central alignment are in broad agreement with observations \citep{Yang2006,Faltenbacher2007} and previous simulation predictions \citep{Dong2014}.

In terms of radial alignment, our predictions show discrepancies compared to the observational findings presented in Figure 3 of \citet{Faltenbacher2007}. The alignment in the ``Blue Satellites'' subsample appears to be more significant than in the ``Red Satellites'' subsample, contrary to the observational results.
Additionally, while the observational conclusion indicates significant signals for red satellites at scales $r_p < 0.7 R_{\rm{vir}}$, our predictions reveal significant alignment signals for both red and blue satellites, but only at small projected separations.

For direct alignment, our predictions are consistent with the observational results on scales where the projected separation $r_p<0.1R_{\rm{vir}}$ \citep[as reported in][]{Faltenbacher2007}.

Interestingly, it has been observed that the ``blue-blue'' subsample consistently exhibits the most significant proximity effect among all three types of alignment.

It is important to note that due to the relatively small size of our mock sample, we did not divide the satellite galaxies into red, green, and blue subsamples as done in the observational study by \cite{Faltenbacher2007}. As a result, caution is advised when interpreting the comparison results above.

\subsection{the influence of surface brightness thresholds}

\begin{figure*}[ht!]
\centering
\includegraphics[width=7. cm]{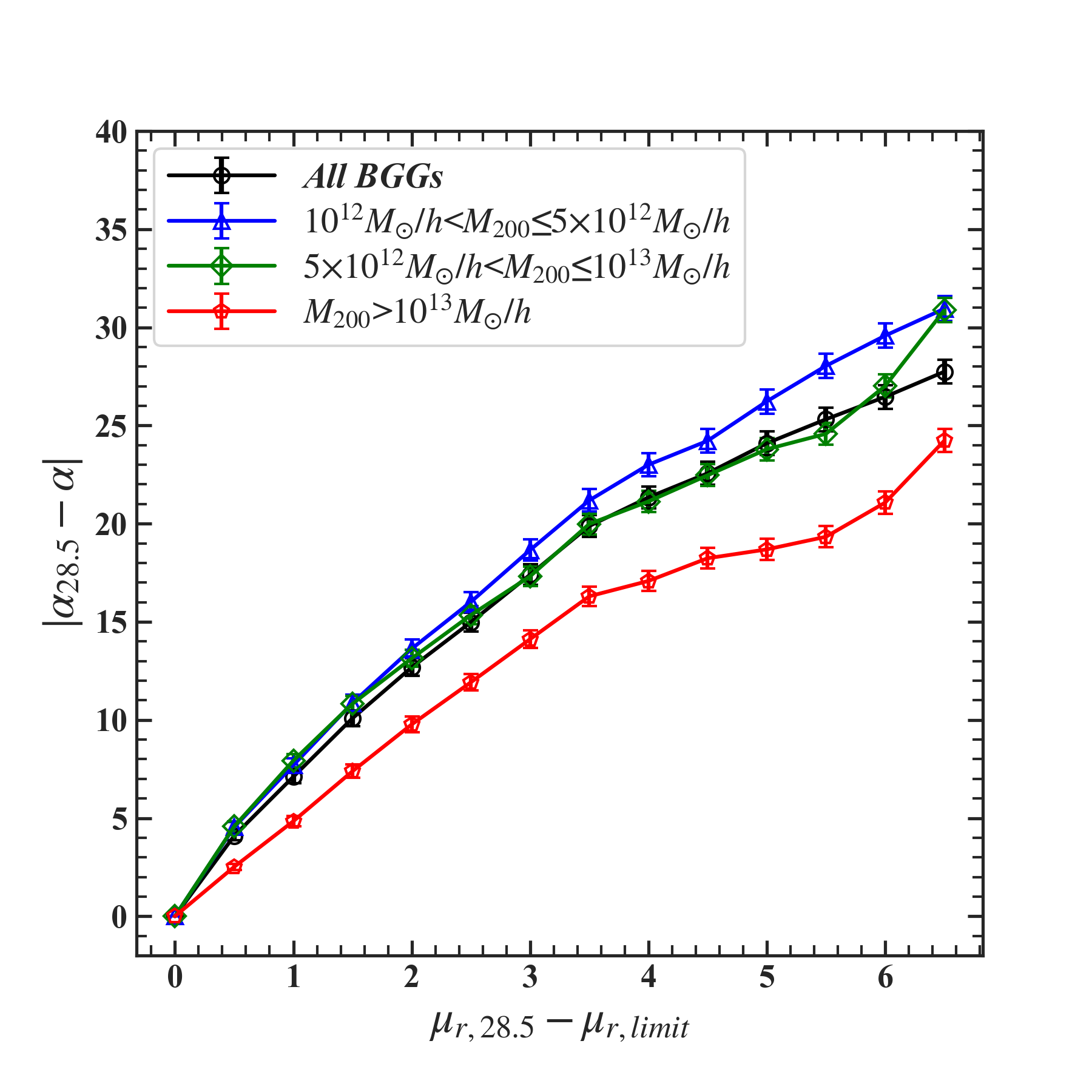}
\includegraphics[width=7. cm]{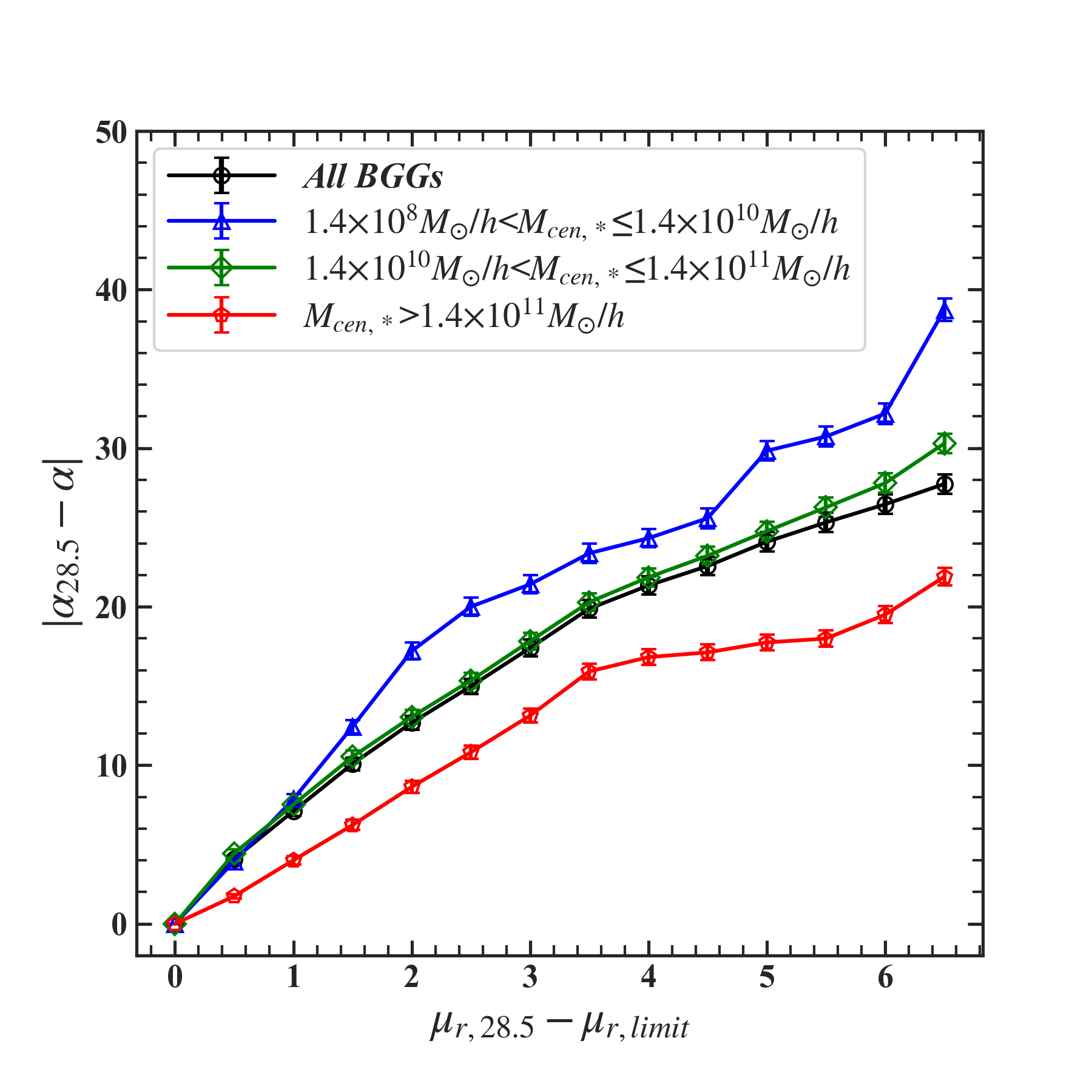}
\caption{The variation of BGG major axes with different surface brightness thresholds. The black solid lines are for all BGGs, while the blue, green and red lines are for  three mass subsamples, as indicated in the upper-left corner of each panel. Here, $\alpha$ represents for the position angle of BGG major axis and the benchmark is set to position angle of BGG major axis at a surface brightness threshold of $\rm \mu_{r}=28.5 \ mag \ arcsec^{-2}$.
The left and right panels shows results for halo and BGG mass, respectively.}
\label{figure abs_delta_Ori}
\end{figure*}

Galaxies in observations are typically defined by the faintest detectable edge determined by the surface brightness depths of surveys like 2dFGRS \citep{Folkes1999}, SDSS \citep{York2000}, and DESI \citep{DESICollaboration2016}, etc. The shape and orientation of galaxies can vary significantly with different surface brightness contours, leading to the ``isophote twist'' phenomenon observed in elliptical \citep[e.g.,][]{Fasano_1989, Nieto_1992, Huang_2013} and spiral galaxies \citep[e.g.,][]{Gong_2023ApJS}.
To ensure a fair comparison with observational results, it is crucial to use the same surface brightness threshold to define mock galaxies in simulations.
We chose a threshold of $25.0 \ mag \ arcsec^{-2}$ in the $r$ band to align with SDSS data \citep{Yang2006,Faltenbacher2007}.
Exploring the influence of surface brightness thresholds on alignment predictions could be a worthwhile research avenue.

The orientation of the BGG can be characterized by the position angle (PA) of its major axis. In our analysis, we calculate the absolute value of the relative PA difference, DPA$=|\alpha_{28.5}-\alpha|$, where $\alpha_{28.5}$ represents the PA of the BGG defined with a surface brightness threshold of $\rm \mu_{r}=28.5 \ mag \ arcsec^{-2}$).
 We have plotted the variation of DPA with different surface brightness thresholds in Figure \ref{figure abs_delta_Ori}. In the figure, the black solid lines represent all BGGs, while the other lines correspond to results categorized by halo mass (left panel) and stellar mass (right panel) in various mass ranges, as indicated in the legends.
This analysis allows us to study how the orientation of BGGs changes with different surface brightness thresholds and how it depends on the properties of the halos and stellar masses.

The analysis of Figure \ref{figure abs_delta_Ori} reveals some key trends regarding the variation of DPA with different surface brightness thresholds and the properties of halo and stellar mass.
We observe that DPA increases significantly with a brighter threshold, $\rm \mu_{r,limit}$, and can reach values up to $\sim 40 ^\circ$.
This suggests that the orientation of BGGs is highly sensitive to the surface brightness threshold used in defining their major axis.
Furthermore, we find that the average DPA for BGGs in less massive halos is larger compared to those in more massive halos.
This difference can be attributed to the fact that low-mass BGGs are more likely to be spiral galaxies, while high-mass BGGs tend to be elliptical galaxies.
Spiral galaxies typically exhibit larger angular momentum and distinctive arm structures, leading to a more pronounced rotation pattern compared to elliptical galaxies.

The variation trend of DPA with BGG stellar mass parallels that of halo mass, given the direct correlation between BGG mass and halo mass. These findings emphasize the necessity of meticulously selecting the surface brightness threshold when determining BGG orientations, particularly in comparisons with observational data. Considering these trends enables researchers to more accurately interpret and analyze the orientation properties of BGGs in galaxy clusters.

Furthermore, as the surface brightness thresholds for defining galaxies become dimmer, the alignment signals tend to strengthen, as shown in Table \ref{table2}. This is because dimmer thresholds encompass more peripheral regions, potentially including tidal tails or wrapped structures, which leads to varied orientations and stronger alignment signals. Notably, as the alignment signals for galaxies with fainter thresholds strengthen, they approach those for stellar substructures (see results in Figure \ref{figure theta_All_Ms}), indicating a closer morphological match. This underscores the importance of surface brightness thresholds in analyzing alignment signals and morphology within galaxy clusters.
We suggest conducting deep image surveys to reveal the radial and direct alignment phenomena.

\begin{deluxetable}{ccccD}
\centering
\tablenum{2}
\tablecaption{$\langle \theta \rangle$,$\langle \varphi \rangle$,$\langle \xi \rangle$ for mock samples with different surface brightness thresholds\label{table2}}
\tablewidth{0pt}
\tablehead{
\colhead{SB Thresholds} & \colhead{$\langle\theta\rangle$} & \colhead{$\langle \varphi \rangle$} & \colhead{$\langle \xi \rangle$} & \colhead{number of pairs}
\\ \colhead{(mag arcsec$^{-2}$)} & \colhead{(degree)} & \colhead{(degree)} & \colhead{(degree)} & \colhead{}
}
\startdata
24.5 & 41.6 & 44.9 & 45.5 & 8371 \\
25.0   & 41.6 & 44.3 & 44.9 & 12,987 \\
25.5 & 41.2 & 44.0 & 44.6 & 18,041 \\
26.0   & 41.1 & 44.2 & 44.5 & 23,518 \\
26.5 & 40.8 & 44.0 & 44.6 & 29,511 \\
27.0   & 40.6 & 43.9 & 44.7 & 36,562 \\
27.5 & 40.5 & 43.7 & 44.6 & 43,760 \\
28.0   & 40.7 & 43.2 & 44.6 & 51,599\\
28.5 & 40.7 & 42.7 & 44.6 & 59,728 \\
\enddata
\label{table}
\end{deluxetable}

\section{Conclusions and Discussion}

In this paper, the TNG100-1 simulations have been utilized to explore three types of galaxy alignment and how they relate to galaxy or halo properties by applying mock observation techniques.

First of all, the simulation predictions accurately replicate the satellite-central and radial alignment effects for the entire mock sample (i.e., all BGG-satellite pairs), aligning well with observational findings \citep{Yang2006,Rodriguez2022}. 
However, no significant signal is detected for direct alignment. Specifically, satellite galaxies tend to align along the major axis of their host BGG and orient their major axes toward the center of their central galaxy. Nevertheless, there is minimal correlation between the orientation of satellite galaxies and the orientation of their host BGG.

Our study reveals that most predictions on galaxy alignment and its dependence on halo or galaxy properties are in line with previous research, with some slight discrepancies.
The main findings on satellite-central alignment, radial alignment, and direct alignment will be summarized and discussed separately.

For satellite-central alignment, predictions using single-color and double-color subsamples demonstrate alignment strengths and color-dependent trends in good agreement with observational results. 
 Our results compellingly demonstrate that the strength of satellite-central alignment increases for single-color subsamples in the following order: ``Blue Centrals,'' ``Blue Satellites,'' ``Red Centrals,'' and ``Red Satellites''. Consequently, we argue that this study is the first to accurately predict the trend of color dependence in satellite-central alignment. It is noteworthy that \cite{Kang2007} used a semi-analytical model to predict the same color dependence and found a different order: ``Blue Satellites,'' ``Red Centrals,'' ``Blue Centrals,'' and ``Red Satellites.'' They also identified an alternative order: ``Blue Satellites,'' ``Blue Centrals,'' ``Red Centrals,'' and ``Red Satellites'' when considering the spin axes as the major axes of central galaxies. None of these previous predictions yielded the correct color dependence.
Nevertheless, it should be noted that in our prediction there are still exceptions where the `Blue Centrals' and blue-blue subsamples exhibit significantly pronounced alignment signals \citep{Yang2006, Rodriguez2022}.
The successful fit of predictions from the blue-red subsample with observational results suggests that the excessive satellite-central alignment observed in the ``Blue Centrals'' subsample can be attributed to the abnormal behavior of the blue-blue subsample. Additionally, the exceptional alignment signals on small scales indicate that the discrepancy is primarily due to the overly strong anisotropy distribution of blue satellites around blue central galaxies in the simulations. This suggests a potential issue with the color assignments for blue satellites in the simulations.
Furthermore, on average, there is a positive correlation between alignment strengths and halo mass, which aligns well with the observed dependence \citep{Yang2006} and previous predictions made using Gadget-2 simulations \citep{Dong2014, Tang2020}.
Predictions were made for the alignment strengths of halos based on BGG mass, age, and metallicity, revealing positive correlations with these properties, mirroring the mass-dependence trends. The metallicity-dependence results resemble predictions for satellites \citep{Dong2014}  but lack observational validation.
Furthermore, alignment strengths across all color subsamples exhibited radial dependence or proximity effects.
The strength of satellite-central alignment signals is significantly influenced by the proximity of satellite galaxies to the central galaxy, with closer satellites yielding stronger alignment signals. These findings align well with both observational studies \citep{Yang2006,Faltenbacher2007} and previous predictions \citep{Dong2014} regarding radial dependence. The dependencies on halo/BGG mass, BGG age, and metallicity can be understood within the framework of cosmic structure and galaxy formation.
On one hand, it is widely recognized that central galaxies in more massive halos tend to be more massive themselves, possess older average stellar ages, and exhibit higher average stellar metallicities. However, these relationships are complex, nonlinear, and have significant scatter.
On the other hand, studies have shown that satellite galaxies often follow an asymmetric infall pattern and are preferentially accreted along the major axis of host halos with higher mass \citep[e.g.,][]{WangHY2005,libeskind2015planes,kubik2017universal,morinaga2020impact}. This anisotropic distribution of satellite galaxies leads to a significant satellite-central alignment.
Consequently, it is reasonable to expect positive correlations between satellite-central alignment strength and halo mass, BGG mass, BGG age, and BGG metallicity.
Nevertheless, it has been observed that the predicted alignment strengths for the blue subsample exhibit significant variability based on BGG properties. This variability can be attributed to the complexity of blue galaxies, as they are challenging to characterize without additional information such as merger history and dynamical stage.
Further investigations are needed.

In terms of radial alignment, it has been observed that there is almost no color dependence within single-color subsamples, although weak but noticeable signals have been detected. On the other hand, when considering double-color subsamples, significant radial alignment signals are present in the red-red and blue-blue subsamples, while the red-blue and blue-red subsamples show minimal radial alignment. These results can be explained by the galactic conformity phenomenon and the influence of tidal stripping on galaxy alignment.
When analyzing all color subsamples, the trends in dependence on halo mass or BGG mass appear to be similar. Specifically, the radial alignment signal tends to be stronger at the lower mass end. However, it is found that the relationship between radial alignment and BGG age and metallicity is weak and complex for the entire sample and all subsamples, with significant scatter due to the small number statistics.
Furthermore, a clear proximity effect is observed, as the predicted alignment signals from the entire mock sample and subsamples for central galaxies are significant at scales where the projected distance is less than 0.2 times the virial radius.
There have been found discrepancies between the predicted dependence of proximity effect on satellite colors and observational results \citep{Faltenbacher2007}.
However, it is important to note that the definition of BGG colors in this study may differ from that used in the study by \citet{Faltenbacher2007}. Further investigation and refinement of methodologies may be necessary to reconcile these differences and enhance our understanding of radial alignment phenomena in galaxy systems.

For direct alignment, no signals were detected in the entire mock sample or in any of the single-color and double-color subsamples. There was also no apparent dependence on halo/BGG mass, BGG age, or metallicity.
However, there has been found significant alignment on scales of $r_p \lesssim 0.2 R_{\rm{vir}}$, generally consistent with observations \citep{Faltenbacher2007}.

Moreover, it is also revealed that galaxy orientations are affected by surface brightness contours, emphasizing the importance of adopting consistent thresholds for a fair comparison with observations. Galaxy alignment, essential for understanding galaxy properties, depends on factors such as surface brightness thresholds, PSF effects, and redshift distribution, etc.
The importance of selection effects in mock observations using simulations, as emphasized in \cite{Tang2021}, stresses the necessity of accounting for these factors to ensure proper and accurate comparisons.

The scenario of galaxy formation within the context of cosmic structure formation provides a comprehensive understanding of the various alignment signals observed in satellite galaxies.
The satellite-central alignment, driven by the large-scale structure of the Universe, sets the stage for the subsequent radial alignment phenomenon influenced by tidal forces \citep[e.g.,][]{Ciotti1994,Usami1997}.
As satellite galaxies move toward the center of the host dark matter halo along the principal axis of the central galaxy, they experience increasing tidal forces that gradually align their shapes and orientations toward the central galaxy. This process leads to a stronger radial alignment signal in satellite galaxies closer to the center of the dark matter halo, reflecting the intensity of the tidal forces exerted on them.
In contrast, the direct alignment signals, which represent the difference between satellite-central alignment and radial alignment angles, are relatively weak and without clear theoretical prediction \citep[see][]{Faltenbacher2007}. This suggests that the direct alignment may be a more subtle or complex alignment mechanism compared to the satellite-central and radial alignments.
Overall, the combined effects of satellite-central alignment, radial alignment due to tidal forces, and the potential presence of direct alignment paint a nuanced picture of how satellite galaxies align with their host dark matter halos and central galaxies during the cosmic evolution of structures.

In addition, the comparison between the alignment signals obtained from mock observations and those derived from stellar substructures using the SUBFIND algorithm reveals interesting insights into the accuracy and reliability of different methods in studying satellite galaxy alignments.
While the mock observational galaxies generated through galaxy reconstruction techniques offer a closer resemblance to the observed satellite distributions and orientations, indicating a higher level of accuracy in capturing the alignment signals, it is essential to note that the signals obtained from stellar substructures are still remarkably significant for all three types of alignment -- satellite-central alignment, radial alignment, and direct alignment.
This suggests that while the mock observational galaxies may provide a more realistic representation of the observed satellite galaxy alignments, the signals derived from stellar substructures using algorithms like SUBFIND still hold considerable value in understanding the alignment phenomena in cosmic structures. The combination of both approaches can offer a more comprehensive and robust analysis of satellite galaxy alignments, helping us gain deeper insights into the complex interplay between galaxies, dark matter halos, and the large-scale structure of the Universe.
It is suggested to take sufficiently deep galaxy images to find the outer faint ingredients and recover the prominent alignment signals derived from substructures \citep[see, e.g.,][for estimation and discussions]{Johnston2001}.

In summary, while the explanations for satellite-central alignment are well-understood and dependent on certain properties, the reasons behind radial and direct alignment are more complex and not fully elucidated due to various factors influencing the alignment patterns. Our analysis underscores the importance of galaxy properties and positions in influencing satellite alignment signals.
These results offer valuable contributions to unraveling the mechanisms governing galaxy alignment and shed light on the intricate interplay between galaxy properties and their distribution within dark matter halos. By delving deeper into these alignment phenomena, we can enhance our understanding of the intricate processes shaping the cosmic web and the distribution of galaxies within it.
Furthermore, precise quantification of galaxy intrinsic alignment is crucial for mitigating its impact on weak lensing analyses \citep[e.g.,][]{Heymans2004,Kirk2015, Edo2017, Zhang2022, zhou2023intrinsic}.
Given the challenges in simulating galaxy formation due to the complexity and costs involved, there is a need to further uncover the detailed physical mechanisms underlying satellite-central alignment. Addressing the discrepancies identified in this study will require advanced simulations and innovative methodologies in the future.
By advancing our understanding of satellite-central alignment and developing improved simulation techniques, we can refine our models and enhance the accuracy of weak lensing analyses. This will ultimately contribute to more reliable interpretations of observational data and deepen our knowledge of the complex interplay between galaxies and dark matter halos in the Universe.

\begin{acknowledgments}
The authors thank the anonymous referee for useful suggestions and Dr. Dandan Xu for helping download part of the TNG public data.
We acknowledges the support from the NSFC grants (Nos.12073089, 12003079).
L.T.  is also supported by the Natural Science Foundation of Sichuan Province (No. 2022NSFSC1842), the Fundamental Research Funds of China West Normal University (CWNU, No. 21E029) and the Sichuan Youth Science and Technology Innovation Research Team (21CXTD0038).
The calculations were carried out on the Kunlun HPC in SPA, SYSU.
\end{acknowledgments}

\bibliography{sample63}{}

\begin{thebibliography}{}
\expandafter\ifx\csname natexlab\endcsname\relax\def\natexlab#1{#1}\fi
\providecommand{\url}[1]{\href{#1}{#1}}
\providecommand{\dodoi}[1]{doi:~\href{http://doi.org/#1}{\nolinkurl{#1}}}
\providecommand{\doeprint}[1]{\href{http://ascl.net/#1}{\nolinkurl{http://ascl.net/#1}}}
\providecommand{\doarXiv}[1]{\href{https://arxiv.org/abs/#1}{\nolinkurl{https://arxiv.org/abs/#1}}}

\bibitem[{{Agustsson} \& {Brainerd}(2006{\natexlab{a}})}]{Agustsson2006b}
{Agustsson}, I., \& {Brainerd}, T.~G. 2006{\natexlab{a}}, \apj, 650, 550,
  \dodoi{10.1086/507084}

\bibitem[{{Agustsson} \& {Brainerd}(2006{\natexlab{b}})}]{Agustsson2006a}
---. 2006{\natexlab{b}}, \apjl, 644, L25, \dodoi{10.1086/505465}

\bibitem[{{Agustsson} \& {Brainerd}(2010)}]{Agustsson&Brainerd2010}
---. 2010, \apj, 709, 1321, \dodoi{10.1088/0004-637X/709/2/1321}

\bibitem[{{Baes} {et~al.}(2011){Baes}, {Verstappen}, {De Looze}, {Fritz},
  {Saftly}, {Vidal P{\'e}rez}, {Stalevski}, \& {Valcke}}]{Baes2011}
{Baes}, M., {Verstappen}, J., {De Looze}, I., {et~al.} 2011, \apjs, 196, 22,
  \dodoi{10.1088/0067-0049/196/2/22}

\bibitem[{{Bahl} \& {Baumgardt}(2014)}]{Bahl2014}
{Bahl}, H., \& {Baumgardt}, H. 2014, \mnras, 438, 2916,
  \dodoi{10.1093/mnras/stt2399}

\bibitem[{{Bailin} {et~al.}(2008){Bailin}, {Power}, {Norberg}, {Zaritsky}, \&
  {Gibson}}]{Bailin2008}
{Bailin}, J., {Power}, C., {Norberg}, P., {Zaritsky}, D., \& {Gibson}, B.~K.
  2008, \mnras, 390, 1133, \dodoi{10.1111/j.1365-2966.2008.13828.x}

\bibitem[{Bernardi {et~al.}(2013)Bernardi, Meert, Sheth, Vikram,
  Huertas-Company, Mei, \& Shankar}]{Bernardi2013}
Bernardi, M., Meert, A., Sheth, R.~K., {et~al.} 2013, Monthly Notices of the
  Royal Astronomical Society, 436, 697, \dodoi{10.1093/mnras/stt1607}

\bibitem[{Blanton \& Roweis(2007)}]{blanton2007k}
Blanton, M.~R., \& Roweis, S. 2007, The Astronomical Journal, 133, 734

\bibitem[{{Brainerd}(2005)}]{Brainerd2005}
{Brainerd}, T.~G. 2005, \apjl, 628, L101, \dodoi{10.1086/432713}

\bibitem[{{Brainerd} \& {Yamamoto}(2019)}]{Brainerd2019}
{Brainerd}, T.~G., \& {Yamamoto}, M. 2019, \mnras, 489, 459,
  \dodoi{10.1093/mnras/stz2102}

\bibitem[{{Bruzual} \& {Charlot}(2003)}]{bruzual2003stellar}
{Bruzual}, G., \& {Charlot}, S. 2003, \mnras, 344, 1000,
  \dodoi{10.1046/j.1365-8711.2003.06897.x}

\bibitem[{{Ciotti} \& {Dutta}(1994)}]{Ciotti1994}
{Ciotti}, L., \& {Dutta}, S.~N. 1994, \mnras, 270, 390,
  \dodoi{10.1093/mnras/270.2.390}

\bibitem[{{Davis} {et~al.}(1985){Davis}, {Efstathiou}, {Frenk}, \&
  {White}}]{Davis1985}
{Davis}, M., {Efstathiou}, G., {Frenk}, C.~S., \& {White}, S.~D.~M. 1985, \apj,
  292, 371, \dodoi{10.1086/163168}

\bibitem[{{Debattista} {et~al.}(2015){Debattista}, {van den Bosch},
  {Ro{\v{s}}kar}, {Quinn}, {Moore}, \& {Cole}}]{Debattista2015}
{Debattista}, V.~P., {van den Bosch}, F.~C., {Ro{\v{s}}kar}, R., {et~al.} 2015,
  \mnras, 452, 4094, \dodoi{10.1093/mnras/stv1563}

\bibitem[{{DESI Collaboration} {et~al.}(2016){DESI Collaboration}, {Aghamousa},
  {Aguilar}, {Ahlen}, {Alam}, {Allen}, {Allende Prieto}, {Annis}, {Bailey},
  {Balland}, {Ballester}, {Baltay}, {Beaufore}, {Bebek}, {Beers}, {Bell},
  {Bernal}, {Besuner}, {Beutler}, {Blake}, {Bleuler}, {Blomqvist}, {Blum},
  {Bolton}, {Briceno}, {Brooks}, {Brownstein}, {Buckley-Geer}, {Burden},
  {Burtin}, {Busca}, {Cahn}, {Cai}, {Cardiel-Sas}, {Carlberg}, {Carton},
  {Casas}, {Castander}, {Cervantes-Cota}, {Claybaugh}, {Close}, {Coker},
  {Cole}, {Comparat}, {Cooper}, {Cousinou}, {Crocce}, {Cuby}, {Cunningham},
  {Davis}, {Dawson}, {de la Macorra}, {De Vicente}, {Delubac}, {Derwent},
  {Dey}, {Dhungana}, {Ding}, {Doel}, {Duan}, {Ealet}, {Edelstein},
  {Eftekharzadeh}, {Eisenstein}, {Elliott}, {Escoffier}, {Evatt}, {Fagrelius},
  {Fan}, {Fanning}, {Farahi}, {Farihi}, {Favole}, {Feng}, {Fernandez},
  {Findlay}, {Finkbeiner}, {Fitzpatrick}, {Flaugher}, {Flender}, {Font-Ribera},
  {Forero-Romero}, {Fosalba}, {Frenk}, {Fumagalli}, {Gaensicke}, {Gallo},
  {Garcia-Bellido}, {Gaztanaga}, {Pietro Gentile Fusillo}, {Gerard},
  {Gershkovich}, {Giannantonio}, {Gillet}, {Gonzalez-de-Rivera},
  {Gonzalez-Perez}, {Gott}, {Graur}, {Gutierrez}, {Guy}, {Habib}, {Heetderks},
  {Heetderks}, {Heitmann}, {Hellwing}, {Herrera}, {Ho}, {Holland}, {Honscheid},
  {Huff}, {Hutchinson}, {Huterer}, {Hwang}, {Illa Laguna}, {Ishikawa},
  {Jacobs}, {Jeffrey}, {Jelinsky}, {Jennings}, {Jiang}, {Jimenez}, {Johnson},
  {Joyce}, {Jullo}, {Juneau}, {Kama}, {Karcher}, {Karkar}, {Kehoe}, {Kennamer},
  {Kent}, {Kilbinger}, {Kim}, {Kirkby}, {Kisner}, {Kitanidis}, {Kneib},
  {Koposov}, {Kovacs}, {Koyama}, {Kremin}, {Kron}, {Kronig}, {Kueter-Young},
  {Lacey}, {Lafever}, {Lahav}, {Lambert}, {Lampton}, {Landriau}, {Lang},
  {Lauer}, {Le Goff}, {Le Guillou}, {Le Van Suu}, {Lee}, {Lee}, {Leitner},
  {Lesser}, {Levi}, {L'Huillier}, {Li}, {Liang}, {Lin}, {Linder}, {Loebman},
  {Luki{\'c}}, {Ma}, {MacCrann}, {Magneville}, {Makarem}, {Manera}, {Manser},
  {Marshall}, {Martini}, {Massey}, {Matheson}, {McCauley}, {McDonald},
  {McGreer}, {Meisner}, {Metcalfe}, {Miller}, {Miquel}, {Moustakas}, {Myers},
  {Naik}, {Newman}, {Nichol}, {Nicola}, {Nicolati da Costa}, {Nie}, {Niz},
  {Norberg}, {Nord}, {Norman}, {Nugent}, {O'Brien}, {Oh}, {Olsen}, {Padilla},
  {Padmanabhan}, {Padmanabhan}, {Palanque-Delabrouille}, {Palmese},
  {Pappalardo}, {P{\^a}ris}, {Park}, {Patej}, {Peacock}, {Peiris}, {Peng},
  {Percival}, {Perruchot}, {Pieri}, {Pogge}, {Pollack}, {Poppett}, {Prada},
  {Prakash}, {Probst}, {Rabinowitz}, {Raichoor}, {Ree}, {Refregier}, {Regal},
  {Reid}, {Reil}, {Rezaie}, {Rockosi}, {Roe}, {Ronayette}, {Roodman}, {Ross},
  {Ross}, {Rossi}, {Rozo}, {Ruhlmann-Kleider}, {Rykoff}, {Sabiu}, {Samushia},
  {Sanchez}, {Sanchez}, {Schlegel}, {Schneider}, {Schubnell}, {Secroun},
  {Seljak}, {Seo}, {Serrano}, {Shafieloo}, {Shan}, {Sharples}, {Sholl},
  {Shourt}, {Silber}, {Silva}, {Sirk}, {Slosar}, {Smith}, {Smoot}, {Som},
  {Song}, {Sprayberry}, {Staten}, {Stefanik}, {Tarle}, {Sien Tie}, {Tinker},
  {Tojeiro}, {Valdes}, {Valenzuela}, {Valluri}, {Vargas-Magana}, {Verde},
  {Walker}, {Wang}, {Wang}, {Weaver}, {Weaverdyck}, {Wechsler}, {Weinberg},
  {White}, {Yang}, {Yeche}, {Zhang}, {Zhao}, {Zheng}, {Zhou}, {Zhou}, {Zhu},
  {Zou}, \& {Zu}}]{DESICollaboration2016}
{DESI Collaboration}, {Aghamousa}, A., {Aguilar}, J., {et~al.} 2016, arXiv
  e-prints, arXiv:1611.00036, \dodoi{10.48550/arXiv.1611.00036}

\bibitem[{{Dolag} {et~al.}(2009){Dolag}, {Borgani}, {Murante}, \&
  {Springel}}]{dolag2009}
{Dolag}, K., {Borgani}, S., {Murante}, G., \& {Springel}, V. 2009, \mnras, 399,
  497, \dodoi{10.1111/j.1365-2966.2009.15034.x}

\bibitem[{{Dong} {et~al.}(2014){Dong}, {Lin}, {Kang}, {Ocean Wang}, {Dutton},
  \& {Macci{\`o}}}]{Dong2014}
{Dong}, X.~C., {Lin}, W.~P., {Kang}, X., {et~al.} 2014, \apjl, 791, L33,
  \dodoi{10.1088/2041-8205/791/2/L33}

\bibitem[{{Faltenbacher} {et~al.}(2008){Faltenbacher}, {Jing}, {Li}, {Mao},
  {Mo}, {Pasquali}, \& {van den Bosch}}]{Faltenbacher2008}
{Faltenbacher}, A., {Jing}, Y.~P., {Li}, C., {et~al.} 2008, \apj, 675, 146,
  \dodoi{10.1086/525243}

\bibitem[{{Faltenbacher} {et~al.}(2007){Faltenbacher}, {Li}, {Mao}, {van den
  Bosch}, {Yang}, {Jing}, {Pasquali}, \& {Mo}}]{Faltenbacher2007}
{Faltenbacher}, A., {Li}, C., {Mao}, S., {et~al.} 2007, \apjl, 662, L71,
  \dodoi{10.1086/519683}

\bibitem[{{Faltenbacher} {et~al.}(2009){Faltenbacher}, {Li}, {White}, {Jing},
  {Shu-DeMao}, \& {Wang}}]{Faltenbacher2009}
{Faltenbacher}, A., {Li}, C., {White}, S. D.~M., {et~al.} 2009, Research in
  Astronomy and Astrophysics, 9, 41, \dodoi{10.1088/1674-4527/9/1/004}

\bibitem[{{Fasano} \& {Bonoli}(1989)}]{Fasano_1989}
{Fasano}, G., \& {Bonoli}, C. 1989, \aaps, 79, 291

\bibitem[{{Ferrari} {et~al.}(2006){Ferrari}, {Arnaud}, {Ettori},
  {Maurogordato}, \& {Rho}}]{Ferrari2006}
{Ferrari}, C., {Arnaud}, M., {Ettori}, S., {Maurogordato}, S., \& {Rho}, J.
  2006, \aap, 446, 417, \dodoi{10.1051/0004-6361:20053946}

\bibitem[{{Folkes} {et~al.}(1999){Folkes}, {Ronen}, {Price}, {Lahav},
  {Colless}, {Maddox}, {Deeley}, {Glazebrook}, {Bland-Hawthorn}, {Cannon},
  {Cole}, {Collins}, {Couch}, {Driver}, {Dalton}, {Efstathiou}, {Ellis},
  {Frenk}, {Kaiser}, {Lewis}, {Lumsden}, {Peacock}, {Peterson}, {Sutherland},
  \& {Taylor}}]{Folkes1999}
{Folkes}, S., {Ronen}, S., {Price}, I., {et~al.} 1999, \mnras, 308, 459,
  \dodoi{10.1046/j.1365-8711.1999.02721.x}

\bibitem[{{Ghigna} {et~al.}(2000){Ghigna}, {Moore}, {Governato}, {Lake},
  {Quinn}, \& {Stadel}}]{Ghigna2000}
{Ghigna}, S., {Moore}, B., {Governato}, F., {et~al.} 2000, \apj, 544, 616,
  \dodoi{10.1086/317221}

\bibitem[{Gong {et~al.}(2023)Gong, Mao, Gao, \& Yu}]{Gong_2023ApJS}
Gong, J.-Y., Mao, Y.-W., Gao, H., \& Yu, S.-Y. 2023, The Astrophysical Journal
  Supplement Series, 267, 26, \dodoi{10.3847/1538-4365/acd554}

\bibitem[{{Hawley} \& {Peebles}(1975)}]{Hawley&Peebles1975}
{Hawley}, D.~L., \& {Peebles}, P.~J.~E. 1975, \aj, 80, 477,
  \dodoi{10.1086/111768}

\bibitem[{{Heymans} {et~al.}(2004){Heymans}, {Brown}, {Heavens},
  {Meisenheimer}, {Taylor}, \& {Wolf}}]{Heymans2004}
{Heymans}, C., {Brown}, M., {Heavens}, A., {et~al.} 2004, \mnras, 347, 895,
  \dodoi{10.1111/j.1365-2966.2004.07264.x}

\bibitem[{{Holmberg}(1969)}]{Holmberg1969}
{Holmberg}, E. 1969, Arkiv for Astronomi, 5, 305

\bibitem[{Huang {et~al.}(2013)Huang, Ho, Peng, Li, \& Barth}]{Huang_2013}
Huang, S., Ho, L.~C., Peng, C.~Y., Li, Z.-Y., \& Barth, A.~J. 2013, The
  Astrophysical Journal, 766, 47, \dodoi{10.1088/0004-637X/766/1/47}

\bibitem[{{Jing} {et~al.}(2006){Jing}, {Zhang}, {Lin}, {Gao}, \&
  {Springel}}]{Jing2006ApJ}
{Jing}, Y.~P., {Zhang}, P., {Lin}, W.~P., {Gao}, L., \& {Springel}, V. 2006,
  \apjl, 640, L119, \dodoi{10.1086/503547}

\bibitem[{{Joachimi} {et~al.}(2015){Joachimi}, {Cacciato}, {Kitching},
  {Leonard}, {Mandelbaum}, {Sch{\"a}fer}, {Sif{\'o}n}, {Hoekstra}, {Kiessling},
  {Kirk}, \& {Rassat}}]{Joachimi2015}
{Joachimi}, B., {Cacciato}, M., {Kitching}, T.~D., {et~al.} 2015, \ssr, 193, 1,
  \dodoi{10.1007/s11214-015-0177-4}

\bibitem[{{Johnston} {et~al.}(2001){Johnston}, {Sackett}, \&
  {Bullock}}]{Johnston2001}
{Johnston}, K.~V., {Sackett}, P.~D., \& {Bullock}, J.~S. 2001, \apj, 557, 137,
  \dodoi{10.1086/321644}

\bibitem[{{Kang} {et~al.}(2007){Kang}, {van den Bosch}, {Yang}, {Mao}, {Mo},
  {Li}, \& {Jing}}]{Kang2007}
{Kang}, X., {van den Bosch}, F.~C., {Yang}, X., {et~al.} 2007, \mnras, 378,
  1531, \dodoi{10.1111/j.1365-2966.2007.11902.x}

\bibitem[{{Kiessling} {et~al.}(2015){Kiessling}, {Cacciato}, {Joachimi},
  {Kirk}, {Kitching}, {Leonard}, {Mandelbaum}, {Sch{\"a}fer}, {Sif{\'o}n},
  {Brown}, \& {Rassat}}]{Kiessling2015}
{Kiessling}, A., {Cacciato}, M., {Joachimi}, B., {et~al.} 2015, \ssr, 193, 67,
  \dodoi{10.1007/s11214-015-0203-6}

\bibitem[{{Kirk} {et~al.}(2015){Kirk}, {Brown}, {Hoekstra}, {Joachimi},
  {Kitching}, {Mandelbaum}, {Sif{\'o}n}, {Cacciato}, {Choi}, {Kiessling},
  {Leonard}, {Rassat}, \& {Sch{\"a}fer}}]{Kirk2015}
{Kirk}, D., {Brown}, M.~L., {Hoekstra}, H., {et~al.} 2015, \ssr, 193, 139,
  \dodoi{10.1007/s11214-015-0213-4}

\bibitem[{{Knebe} {et~al.}(2008){Knebe}, {Yahagi}, {Kase}, {Lewis}, \&
  {Gibson}}]{Knebe2008}
{Knebe}, A., {Yahagi}, H., {Kase}, H., {Lewis}, G., \& {Gibson}, B.~K. 2008,
  \mnras, 388, L34, \dodoi{10.1111/j.1745-3933.2008.00495.x}

\bibitem[{{Knebe} {et~al.}(2020){Knebe}, {G{\'a}mez-Mar{\'\i}n}, {Pearce},
  {Cui}, {Hoffmann}, {De Petris}, {Power}, {Haggar}, \&
  {Mostoghiu}}]{Knebe2020}
{Knebe}, A., {G{\'a}mez-Mar{\'\i}n}, M., {Pearce}, F.~R., {et~al.} 2020,
  \mnras, 495, 3002, \dodoi{10.1093/mnras/staa1407}

\bibitem[{{Kubik} {et~al.}(2017){Kubik}, {Libeskind}, {Knebe}, {Courtois},
  {Yepes}, {Gottl{\"o}ber}, \& {Hoffman}}]{kubik2017universal}
{Kubik}, B., {Libeskind}, N.~I., {Knebe}, A., {et~al.} 2017, \mnras, 472, 4099,
  \dodoi{10.1093/mnras/stx2263}

\bibitem[{{Lacey} \& {Cole}(1993)}]{Lacey&Cole1993}
{Lacey}, C., \& {Cole}, S. 1993, \mnras, 262, 627,
  \dodoi{10.1093/mnras/262.3.627}

\bibitem[{{Li} {et~al.}(2013){Li}, {Jing}, {Faltenbacher}, \& {Wang}}]{Li2013}
{Li}, C., {Jing}, Y.~P., {Faltenbacher}, A., \& {Wang}, J. 2013, \apjl, 770,
  L12, \dodoi{10.1088/2041-8205/770/1/L12}

\bibitem[{{Libeskind} {et~al.}(2015){Libeskind}, {Hoffman}, {Tully},
  {Courtois}, {Pomar{\`e}de}, {Gottl{\"o}ber}, \&
  {Steinmetz}}]{libeskind2015planes}
{Libeskind}, N.~I., {Hoffman}, Y., {Tully}, R.~B., {et~al.} 2015, \mnras, 452,
  1052, \dodoi{10.1093/mnras/stv1302}

\bibitem[{{Lin} {et~al.}(2006){Lin}, {Jing}, {Mao}, {Gao}, \&
  {McCarthy}}]{Lin2006ApJ}
{Lin}, W.~P., {Jing}, Y.~P., {Mao}, S., {Gao}, L., \& {McCarthy}, I.~G. 2006,
  \apj, 651, 636, \dodoi{10.1086/508052}

\bibitem[{{Mandelbaum} {et~al.}(2006){Mandelbaum}, {Hirata}, {Ishak}, {Seljak},
  \& {Brinkmann}}]{Mandelbaum2006}
{Mandelbaum}, R., {Hirata}, C.~M., {Ishak}, M., {Seljak}, U., \& {Brinkmann},
  J. 2006, \mnras, 367, 611, \dodoi{10.1111/j.1365-2966.2005.09946.x}

\bibitem[{Mo {et~al.}(2010)Mo, Van~den Bosch, \& White}]{mo2010galaxy}
Mo, H., Van~den Bosch, F., \& White, S. 2010, Galaxy formation and evolution
  (Cambridge University Press)

\bibitem[{{Moffat}(1969)}]{Moffat1969}
{Moffat}, A.~F.~J. 1969, \aap, 3, 455

\bibitem[{{Morinaga} \& {Ishiyama}(2020)}]{morinaga2020impact}
{Morinaga}, Y., \& {Ishiyama}, T. 2020, \mnras, 495, 502,
  \dodoi{10.1093/mnras/staa1180}

\bibitem[{Nelson {et~al.}(2017)Nelson, Pillepich, Springel, Weinberger,
  Hernquist, Pakmor, Genel, Torrey, Vogelsberger, Kauffmann, Marinacci, \&
  Naiman}]{Nelson2017}
Nelson, D., Pillepich, A., Springel, V., {et~al.} 2017, Monthly Notices of the
  Royal Astronomical Society, 475, 624, \dodoi{10.1093/mnras/stx3040}

\bibitem[{{Nelson} {et~al.}(2019){Nelson}, {Springel}, {Pillepich},
  {Rodriguez-Gomez}, {Torrey}, {Genel}, {Vogelsberger}, {Pakmor}, {Marinacci},
  {Weinberger}, {Kelley}, {Lovell}, {Diemer}, \& {Hernquist}}]{Nelson2019}
{Nelson}, D., {Springel}, V., {Pillepich}, A., {et~al.} 2019, Computational
  Astrophysics and Cosmology, 6, 2, \dodoi{10.1186/s40668-019-0028-x}

\bibitem[{{Nieto} {et~al.}(1992){Nieto}, {Bender}, {Poulain}, \&
  {Surma}}]{Nieto_1992}
{Nieto}, J.~L., {Bender}, R., {Poulain}, P., \& {Surma}, P. 1992, \aap, 257, 97

\bibitem[{{Pereira} {et~al.}(2008){Pereira}, {Bryan}, \& {Gill}}]{Pereira2008}
{Pereira}, M.~J., {Bryan}, G.~L., \& {Gill}, S. P.~D. 2008, \apj, 672, 825,
  \dodoi{10.1086/523830}

\bibitem[{{Pereira} \& {Kuhn}(2005)}]{Pereira&Kuhn2005}
{Pereira}, M.~J., \& {Kuhn}, J.~R. 2005, \apjl, 627, L21,
  \dodoi{10.1086/432089}

\bibitem[{{Pillepich} {et~al.}(2018{\natexlab{a}}){Pillepich}, {Springel},
  {Nelson}, {Genel}, {Naiman}, {Pakmor}, {Hernquist}, {Torrey}, {Vogelsberger},
  {Weinberger}, \& {Marinacci}}]{pillepich2018simulating}
{Pillepich}, A., {Springel}, V., {Nelson}, D., {et~al.} 2018{\natexlab{a}},
  \mnras, 473, 4077, \dodoi{10.1093/mnras/stx2656}

\bibitem[{{Pillepich} {et~al.}(2018{\natexlab{b}}){Pillepich}, {Nelson},
  {Hernquist}, {Springel}, {Pakmor}, {Torrey}, {Weinberger}, {Genel}, {Naiman},
  {Marinacci}, \& {Vogelsberger}}]{Pillepich2018}
{Pillepich}, A., {Nelson}, D., {Hernquist}, L., {et~al.} 2018{\natexlab{b}},
  \mnras, 475, 648, \dodoi{10.1093/mnras/stx3112}

\bibitem[{{Planck Collaboration} {et~al.}(2016){Planck Collaboration}, {Ade},
  {Aghanim}, {Arnaud}, {Ashdown}, {Aumont}, {Baccigalupi}, {Banday},
  {Barreiro}, {Bartlett}, {Bartolo}, {Battaner}, {Battye}, {Benabed},
  {Beno{\^\i}t}, {Benoit-L{\'e}vy}, {Bernard}, {Bersanelli}, {Bielewicz},
  {Bock}, {Bonaldi}, {Bonavera}, {Bond}, {Borrill}, {Bouchet}, {Boulanger},
  {Bucher}, {Burigana}, {Butler}, {Calabrese}, {Cardoso}, {Catalano},
  {Challinor}, {Chamballu}, {Chary}, {Chiang}, {Chluba}, {Christensen},
  {Church}, {Clements}, {Colombi}, {Colombo}, {Combet}, {Coulais}, {Crill},
  {Curto}, {Cuttaia}, {Danese}, {Davies}, {Davis}, {de Bernardis}, {de Rosa},
  {de Zotti}, {Delabrouille}, {D{\'e}sert}, {Di Valentino}, {Dickinson},
  {Diego}, {Dolag}, {Dole}, {Donzelli}, {Dor{\'e}}, {Douspis}, {Ducout},
  {Dunkley}, {Dupac}, {Efstathiou}, {Elsner}, {En{\ss}lin}, {Eriksen},
  {Farhang}, {Fergusson}, {Finelli}, {Forni}, {Frailis}, {Fraisse},
  {Franceschi}, {Frejsel}, {Galeotta}, {Galli}, {Ganga}, {Gauthier}, {Gerbino},
  {Ghosh}, {Giard}, {Giraud-H{\'e}raud}, {Giusarma}, {Gjerl{\o}w},
  {Gonz{\'a}lez-Nuevo}, {G{\'o}rski}, {Gratton}, {Gregorio}, {Gruppuso},
  {Gudmundsson}, {Hamann}, {Hansen}, {Hanson}, {Harrison}, {Helou},
  {Henrot-Versill{\'e}}, {Hern{\'a}ndez-Monteagudo}, {Herranz}, {Hildebrandt},
  {Hivon}, {Hobson}, {Holmes}, {Hornstrup}, {Hovest}, {Huang}, {Huffenberger},
  {Hurier}, {Jaffe}, {Jaffe}, {Jones}, {Juvela}, {Keih{\"a}nen}, {Keskitalo},
  {Kisner}, {Kneissl}, {Knoche}, {Knox}, {Kunz}, {Kurki-Suonio}, {Lagache},
  {L{\"a}hteenm{\"a}ki}, {Lamarre}, {Lasenby}, {Lattanzi}, {Lawrence}, {Leahy},
  {Leonardi}, {Lesgourgues}, {Levrier}, {Lewis}, {Liguori}, {Lilje},
  {Linden-V{\o}rnle}, {L{\'o}pez-Caniego}, {Lubin}, {Mac{\'\i}as-P{\'e}rez},
  {Maggio}, {Maino}, {Mandolesi}, {Mangilli}, {Marchini}, {Maris}, {Martin},
  {Martinelli}, {Mart{\'\i}nez-Gonz{\'a}lez}, {Masi}, {Matarrese}, {McGehee},
  {Meinhold}, {Melchiorri}, {Melin}, {Mendes}, {Mennella}, {Migliaccio},
  {Millea}, {Mitra}, {Miville-Desch{\^e}nes}, {Moneti}, {Montier}, {Morgante},
  {Mortlock}, {Moss}, {Munshi}, {Murphy}, {Naselsky}, {Nati}, {Natoli},
  {Netterfield}, {N{\o}rgaard-Nielsen}, {Noviello}, {Novikov}, {Novikov},
  {Oxborrow}, {Paci}, {Pagano}, {Pajot}, {Paladini}, {Paoletti}, {Partridge},
  {Pasian}, {Patanchon}, {Pearson}, {Perdereau}, {Perotto}, {Perrotta},
  {Pettorino}, {Piacentini}, {Piat}, {Pierpaoli}, {Pietrobon}, {Plaszczynski},
  {Pointecouteau}, {Polenta}, {Popa}, {Pratt}, {Pr{\'e}zeau}, {Prunet},
  {Puget}, {Rachen}, {Reach}, {Rebolo}, {Reinecke}, {Remazeilles}, {Renault},
  {Renzi}, {Ristorcelli}, {Rocha}, {Rosset}, {Rossetti}, {Roudier},
  {Rouill{\'e} d'Orfeuil}, {Rowan-Robinson}, {Rubi{\~n}o-Mart{\'\i}n},
  {Rusholme}, {Said}, {Salvatelli}, {Salvati}, {Sandri}, {Santos},
  {Savelainen}, {Savini}, {Scott}, {Seiffert}, {Serra}, {Shellard}, {Spencer},
  {Spinelli}, {Stolyarov}, {Stompor}, {Sudiwala}, {Sunyaev}, {Sutton},
  {Suur-Uski}, {Sygnet}, {Tauber}, {Terenzi}, {Toffolatti}, {Tomasi},
  {Tristram}, {Trombetti}, {Tucci}, {Tuovinen}, {T{\"u}rler}, {Umana},
  {Valenziano}, {Valiviita}, {Van Tent}, {Vielva}, {Villa}, {Wade}, {Wandelt},
  {Wehus}, {White}, {White}, {Wilkinson}, {Yvon}, {Zacchei}, \&
  {Zonca}}]{Planck2016}
{Planck Collaboration}, {Ade}, P.~A.~R., {Aghanim}, N., {et~al.} 2016, \aap,
  594, A13, \dodoi{10.1051/0004-6361/201525830}

\bibitem[{{Plionis} {et~al.}(2003){Plionis}, {Benoist}, {Maurogordato},
  {Ferrari}, \& {Basilakos}}]{Plionis2003}
{Plionis}, M., {Benoist}, C., {Maurogordato}, S., {Ferrari}, C., \&
  {Basilakos}, S. 2003, \apj, 594, 144, \dodoi{10.1086/376892}

\bibitem[{{Rodriguez} {et~al.}(2022){Rodriguez}, {Merch{\'a}n}, \&
  {Artale}}]{Rodriguez2022}
{Rodriguez}, F., {Merch{\'a}n}, M., \& {Artale}, M.~C. 2022, \mnras, 514, 1077,
  \dodoi{10.1093/mnras/stac1428}

\bibitem[{{Rong} {et~al.}(2015){Rong}, {Yi}, {Zhang}, \& {Tu}}]{Rong2015}
{Rong}, Y., {Yi}, S.-X., {Zhang}, S.-N., \& {Tu}, H. 2015, \mnras, 451, 2536,
  \dodoi{10.1093/mnras/stv865}

\bibitem[{{Sales} \& {Lambas}(2004)}]{Sales&Lambas2004}
{Sales}, L., \& {Lambas}, D.~G. 2004, \mnras, 348, 1236,
  \dodoi{10.1111/j.1365-2966.2004.07443.x}

\bibitem[{{Sastry}(1968)}]{Sastry1968}
{Sastry}, G.~N. 1968, \pasp, 80, 252, \dodoi{10.1086/128626}

\bibitem[{{Schneider} {et~al.}(2013){Schneider}, {Cole}, {Frenk}, {Kelvin},
  {Mandelbaum}, {Norberg}, {Bland-Hawthorn}, {Brough}, {Driver}, {Hopkins},
  {Liske}, {Loveday}, \& {Robotham}}]{Schneider2013}
{Schneider}, M.~D., {Cole}, S., {Frenk}, C.~S., {et~al.} 2013, \mnras, 433,
  2727, \dodoi{10.1093/mnras/stt855}

\bibitem[{{Shen} {et~al.}(2003){Shen}, {Mo}, {White}, {Blanton}, {Kauffmann},
  {Voges}, {Brinkmann}, \& {Csabai}}]{Shen2003}
{Shen}, S., {Mo}, H.~J., {White}, S. D.~M., {et~al.} 2003, \mnras, 343, 978,
  \dodoi{10.1046/j.1365-8711.2003.06740.x}

\bibitem[{{Springel} {et~al.}(2001){Springel}, {White}, {Tormen}, \&
  {Kauffmann}}]{Springel2001}
{Springel}, V., {White}, S. D.~M., {Tormen}, G., \& {Kauffmann}, G. 2001,
  \mnras, 328, 726, \dodoi{10.1046/j.1365-8711.2001.04912.x}

\bibitem[{{Springel} {et~al.}(2018){Springel}, {Pakmor}, {Pillepich},
  {Weinberger}, {Nelson}, {Hernquist}, {Vogelsberger}, {Genel}, {Torrey},
  {Marinacci}, \& {Naiman}}]{springel2018first}
{Springel}, V., {Pakmor}, R., {Pillepich}, A., {et~al.} 2018, \mnras, 475, 676,
  \dodoi{10.1093/mnras/stx3304}

\bibitem[{{Tang} {et~al.}(2020){Tang}, {Lin}, \& {Wang}}]{Tang2020}
{Tang}, L., {Lin}, W., \& {Wang}, Y. 2020, \apj, 893, 87,
  \dodoi{10.3847/1538-4357/ab8292}

\bibitem[{{Tang} {et~al.}(2021){Tang}, {Lin}, {Wang}, \&
  {Napolitano}}]{Tang2021}
{Tang}, L., {Lin}, W., {Wang}, Y., \& {Napolitano}, N.~R. 2021, \mnras, 508,
  3321, \dodoi{10.1093/mnras/stab2722}

\bibitem[{{Taylor} {et~al.}(2015){Taylor}, {Hopkins}, {Baldry},
  {Bland-Hawthorn}, {Brown}, {Colless}, {Driver}, {Norberg}, {Robotham},
  {Alpaslan}, {Brough}, {Cluver}, {Gunawardhana}, {Kelvin}, {Liske},
  {Conselice}, {Croom}, {Foster}, {Jarrett}, {Lara-Lopez}, \&
  {Loveday}}]{Taylor2015}
{Taylor}, E.~N., {Hopkins}, A.~M., {Baldry}, I.~K., {et~al.} 2015, \mnras, 446,
  2144, \dodoi{10.1093/mnras/stu1900}

\bibitem[{{Tenneti} {et~al.}(2021){Tenneti}, {Kitching}, {Joachimi}, \& {Di
  Matteo}}]{Tenneti2021}
{Tenneti}, A., {Kitching}, T.~D., {Joachimi}, B., \& {Di Matteo}, T. 2021,
  \mnras, 501, 5859, \dodoi{10.1093/mnras/staa3934}

\bibitem[{{Usami} \& {Fujimoto}(1997)}]{Usami1997}
{Usami}, M., \& {Fujimoto}, M. 1997, \apj, 487, 489, \dodoi{10.1086/304624}

\bibitem[{{van Uitert} {et~al.}(2017){van Uitert}, {Hoekstra}, {Joachimi},
  {Schneider}, {Bland-Hawthorn}, {Choi}, {Erben}, {Heymans}, {Hildebrandt},
  {Hopkins}, {Klaes}, {Kuijken}, {Nakajima}, {Napolitano}, {Schrabback},
  {Valentijn}, \& {Viola}}]{Edo2017}
{van Uitert}, E., {Hoekstra}, H., {Joachimi}, B., {et~al.} 2017, Monthly
  Notices of the Royal Astronomical Society, 467, 4131,
  \dodoi{10.1093/mnras/stx344}

\bibitem[{{Velliscig} {et~al.}(2015){Velliscig}, {Cacciato}, {Schaye},
  {Hoekstra}, {Bower}, {Crain}, {van Daalen}, {Furlong}, {McCarthy},
  {Schaller}, \& {Theuns}}]{Velliscig2015}
{Velliscig}, M., {Cacciato}, M., {Schaye}, J., {et~al.} 2015, \mnras, 454,
  3328, \dodoi{10.1093/mnras/stv2198}

\bibitem[{{Wang} {et~al.}(2005){Wang}, {Jing}, {Mao}, \& {Kang}}]{WangHY2005}
{Wang}, H.~Y., {Jing}, Y.~P., {Mao}, S., \& {Kang}, X. 2005, \mnras, 364, 424,
  \dodoi{10.1111/j.1365-2966.2005.09543.x}

\bibitem[{{Wang} {et~al.}(2019){Wang}, {Guo}, {Libeskind}, {Tempel}, {Wei}, \&
  {Kang}}]{Wang2019}
{Wang}, P., {Guo}, Q., {Libeskind}, N.~I., {et~al.} 2019, \mnras, 484, 4325,
  \dodoi{10.1093/mnras/stz285}

\bibitem[{{Wang} \& {Kang}(2018)}]{Wang2018b}
{Wang}, P., \& {Kang}, X. 2018, \mnras, 473, 1562,
  \dodoi{10.1093/mnras/stx2466}

\bibitem[{{Wang} {et~al.}(2021){Wang}, {Libeskind}, {Tempel}, {Kang}, \&
  {Guo}}]{Wang2021}
{Wang}, P., {Libeskind}, N.~I., {Tempel}, E., {Kang}, X., \& {Guo}, Q. 2021,
  Nature Astronomy, 5, 1077, \dodoi{10.1038/s41550-021-01443-8}

\bibitem[{{Wang} {et~al.}(2020){Wang}, {Libeskind}, {Tempel}, {Pawlowski},
  {Kang}, \& {Guo}}]{Wang2020}
{Wang}, P., {Libeskind}, N.~I., {Tempel}, E., {et~al.} 2020, \apj, 900, 129,
  \dodoi{10.3847/1538-4357/aba6ea}

\bibitem[{{Wang} {et~al.}(2018){Wang}, {Luo}, {Kang}, {Libeskind}, {Wang},
  {Zhang}, {Tempel}, \& {Guo}}]{Wang2018a}
{Wang}, P., {Luo}, Y., {Kang}, X., {et~al.} 2018, \apj, 859, 115,
  \dodoi{10.3847/1538-4357/aabe2b}

\bibitem[{{Wang} {et~al.}(2014){Wang}, {Lin}, {Kang}, {Dutton}, {Yu}, \&
  {Macci{\`o}}}]{Wang2014}
{Wang}, Y.~O., {Lin}, W.~P., {Kang}, X., {et~al.} 2014, \apj, 786, 8,
  \dodoi{10.1088/0004-637X/786/1/8}

\bibitem[{{Weinberger} {et~al.}(2017){Weinberger}, {Springel}, {Hernquist},
  {Pillepich}, {Marinacci}, {Pakmor}, {Nelson}, {Genel}, {Vogelsberger},
  {Naiman}, \& {Torrey}}]{weinberger2016simulating}
{Weinberger}, R., {Springel}, V., {Hernquist}, L., {et~al.} 2017, \mnras, 465,
  3291, \dodoi{10.1093/mnras/stw2944}

\bibitem[{{Weinmann} {et~al.}(2006){Weinmann}, {van den Bosch}, {Yang}, \&
  {Mo}}]{Weinmann2006}
{Weinmann}, S.~M., {van den Bosch}, F.~C., {Yang}, X., \& {Mo}, H.~J. 2006,
  \mnras, 366, 2, \dodoi{10.1111/j.1365-2966.2005.09865.x}

\bibitem[{{White} \& {Frenk}(1991)}]{White1991}
{White}, S. D.~M., \& {Frenk}, C.~S. 1991, \apj, 379, 52,
  \dodoi{10.1086/170483}

\bibitem[{{White} \& {Rees}(1978)}]{white1978core}
{White}, S.~D.~M., \& {Rees}, M.~J. 1978, \mnras, 183, 341,
  \dodoi{10.1093/mnras/183.3.341}

\bibitem[{{Yang} {et~al.}(2005){Yang}, {Mo}, {Jing}, \& {van den
  Bosch}}]{yang2005galaxy}
{Yang}, X., {Mo}, H.~J., {Jing}, Y.~P., \& {van den Bosch}, F.~C. 2005, \mnras,
  358, 217, \dodoi{10.1111/j.1365-2966.2005.08801.x}

\bibitem[{{Yang} {et~al.}(2009){Yang}, {Mo}, \& {van den Bosch}}]{Yang2009}
{Yang}, X., {Mo}, H.~J., \& {van den Bosch}, F.~C. 2009, \apj, 695, 900,
  \dodoi{10.1088/0004-637X/695/2/900}

\bibitem[{{Yang} {et~al.}(2006){Yang}, {van den Bosch}, {Mo}, {Mao}, {Kang},
  {Weinmann}, {Guo}, \& {Jing}}]{Yang2006}
{Yang}, X., {van den Bosch}, F.~C., {Mo}, H.~J., {et~al.} 2006, \mnras, 369,
  1293, \dodoi{10.1111/j.1365-2966.2006.10373.x}

\bibitem[{{York} {et~al.}(2000){York}, {Adelman}, {Anderson}, {Anderson},
  {Annis}, {Bahcall}, {Bakken}, {Barkhouser}, {Bastian}, {Berman}, {Boroski},
  {Bracker}, {Briegel}, {Briggs}, {Brinkmann}, {Brunner}, {Burles}, {Carey},
  {Carr}, {Castander}, {Chen}, {Colestock}, {Connolly}, {Crocker}, {Csabai},
  {Czarapata}, {Davis}, {Doi}, {Dombeck}, {Eisenstein}, {Ellman}, {Elms},
  {Evans}, {Fan}, {Federwitz}, {Fiscelli}, {Friedman}, {Frieman}, {Fukugita},
  {Gillespie}, {Gunn}, {Gurbani}, {de Haas}, {Haldeman}, {Harris}, {Hayes},
  {Heckman}, {Hennessy}, {Hindsley}, {Holm}, {Holmgren}, {Huang}, {Hull},
  {Husby}, {Ichikawa}, {Ichikawa}, {Ivezi{\'c}}, {Kent}, {Kim}, {Kinney},
  {Klaene}, {Kleinman}, {Kleinman}, {Knapp}, {Korienek}, {Kron}, {Kunszt},
  {Lamb}, {Lee}, {Leger}, {Limmongkol}, {Lindenmeyer}, {Long}, {Loomis},
  {Loveday}, {Lucinio}, {Lupton}, {MacKinnon}, {Mannery}, {Mantsch}, {Margon},
  {McGehee}, {McKay}, {Meiksin}, {Merelli}, {Monet}, {Munn}, {Narayanan},
  {Nash}, {Neilsen}, {Neswold}, {Newberg}, {Nichol}, {Nicinski}, {Nonino},
  {Okada}, {Okamura}, {Ostriker}, {Owen}, {Pauls}, {Peoples}, {Peterson},
  {Petravick}, {Pier}, {Pope}, {Pordes}, {Prosapio}, {Rechenmacher}, {Quinn},
  {Richards}, {Richmond}, {Rivetta}, {Rockosi}, {Ruthmansdorfer}, {Sandford},
  {Schlegel}, {Schneider}, {Sekiguchi}, {Sergey}, {Shimasaku}, {Siegmund},
  {Smee}, {Smith}, {Snedden}, {Stone}, {Stoughton}, {Strauss}, {Stubbs},
  {SubbaRao}, {Szalay}, {Szapudi}, {Szokoly}, {Thakar}, {Tremonti}, {Tucker},
  {Uomoto}, {Vanden Berk}, {Vogeley}, {Waddell}, {Wang}, {Watanabe},
  {Weinberg}, {Yanny}, {Yasuda}, \& {SDSS Collaboration}}]{York2000}
{York}, D.~G., {Adelman}, J., {Anderson}, John~E., J., {et~al.} 2000, \aj, 120,
  1579, \dodoi{10.1086/301513}

\bibitem[{{Zhang} \& {Wang}(2019)}]{Zhang2019}
{Zhang}, M.-G., \& {Wang}, Y. 2019, Research in Astronomy and Astrophysics, 19,
  181, \dodoi{10.1088/1674-4527/19/12/181}

\bibitem[{Zhang {et~al.}(2022)Zhang, Liu, Wei, Li, Luo, Kang, \&
  Fan}]{Zhang2022}
Zhang, T., Liu, X., Wei, C., {et~al.} 2022, The Astrophysical Journal, 940, 96,
  \dodoi{10.3847/1538-4357/ac9a4c}

\bibitem[{{Zhang} {et~al.}(2021){Zhang}, {Yang}, \& {Guo}}]{Zhang2021}
{Zhang}, Y., {Yang}, X., \& {Guo}, H. 2021, \mnras, 500, 1895,
  \dodoi{10.1093/mnras/staa2356}

\bibitem[{{Zhou} {et~al.}(2023){Zhou}, {Tong}, {Troxel}, {Blazek}, {Lin},
  {Bacon}, {Bleem}, {Chang}, {Costanzi}, {DeRose}, {Dietrich}, {Drlica-Wagner},
  {Gruen}, {Gruendl}, {Hoyle}, {Jarvis}, {MacCrann}, {Mawdsley}, {McClintock},
  {Melchior}, {Prat}, {Pujol}, {Rozo}, {Rykoff}, {Samuroff}, {Sheldon}, {Shin},
  {Rosell}, {Yanny}, {S{\'a}nchez}, {Tucker}, {Sevilla-Noarbe}, {Zuntz},
  {Varga}, {Zhang}, {Alves}, {Amon}, {Bertin}, {Brooks}, {Burke}, {Kind}, {da
  Costa}, {Davis}, {De Vicente}, {Desai}, {Diehl}, {Doel}, {Everett},
  {Ferrero}, {Flaugher}, {Frieman}, {Gerdes}, {Gutierrez}, {Hinton},
  {Hollowood}, {Honscheid}, {James}, {Jeltema}, {Kuehn}, {Lahav}, {Lima},
  {Marshall}, {Mena-Fern{\'a}ndez}, {Menanteau}, {Miquel}, {Palmese},
  {Paz-Chinch{\'o}n}, {Pieres}, {Malag{\'o}n}, {Porredon}, {Raveri}, {Romer},
  {Sanchez}, {Smith}, {Soares-Santos}, {Suchyta}, {Swanson}, {Tarle}, {To},
  {Weaverdyck}, {Weller}, \& {Wiseman}}]{zhou2023intrinsic}
{Zhou}, C., {Tong}, A., {Troxel}, M.~A., {et~al.} 2023, \mnras, 526, 323,
  \dodoi{10.1093/mnras/stad2712}

\end{thebibliography}
\bibliographystyle{aasjournal}

\end{document}